\documentclass[11pt]{article}
\usepackage{color}		
\usepackage{graphicx}	
\usepackage{amsbsy}		
\usepackage{amsfonts}		
\usepackage{amsmath}		
\usepackage{amssymb}		
\usepackage[title]{appendix}
\usepackage{soul}
\usepackage{subcaption}
\usepackage{cite}
\usepackage{dsfont}
\usepackage{mathrsfs}
\usepackage{float}
\usepackage[hypertexnames=false,pdfencoding=auto]{hyperref}
\pdfoutput=1
\makeatletter
\g@addto@macro\bfseries{\boldmath}
\makeatother	

\DeclareTextCommand{\uppercaseDelta}{PU}{\83\224}
\DeclareTextCommand{\lowercasegamma}{PU}{\83\263}
\DeclareMathOperator\sgn{sgn}
\renewcommand{\Re}{\operatorname{Re}}
\renewcommand{\Im}{\operatorname{Im}}

\newcommand\Tstrut{\rule{0pt}{2.9ex}}         
\newcommand\Bstrut{\rule[-1.2ex]{0pt}{0pt}}   
\newcommand\TBstrut{\Tstrut\Bstrut}

\newcommand{\RNum}[1]{\uppercase\expandafter{\romannumeral #1\relax}}
\newenvironment{Eqnarray}%
     {\arraycolsep 0.14em\begin{eqnarray}}{\end{eqnarray}}
\def\beq{\begin{equation}}
\def\eeq{\end{equation}}
\def\beqa{\begin{Eqnarray}}
\def\eeqa{\end{Eqnarray}}

\def\eq#1{eq.~(\ref{#1})}
\def\Eq#1{Eq.~(\ref{#1})}

\def\eqs#1#2{eqs.~(\ref{#1}) and (\ref{#2})}

\def\eqst#1#2{eqs.~(\ref{#1})--(\ref{#2})}
\def\Eqst#1#2{Eqs.~(\ref{#1})--(\ref{#2})}

\def\sbma{s_{\beta-\alpha}}
\def\cbma{c_{\beta-\alpha}}

\def\T{{\mathsf T}}
\def\phaa{\phantom{AA}}
\def\phm{\phantom{-}}
\def\ifmath#1{\relax\ifmmode #1\else $#1$\fi}
\def\ls#1{\ifmath{_{\lower1.5pt\hbox{$\scriptstyle #1$}}}}
\def\rs#1{\ifmath{^{\raise2pt\hbox{$\scriptstyle #1$}}}}

\def\lsim{\mathrel{\raise.3ex\hbox{$<$\kern-.75em\lower1ex\hbox{$\sim$}}}}
\def\gsim{\mathrel{\raise.3ex\hbox{$>$\kern-.75em\lower1ex\hbox{$\sim$}}}}
\def\bra#1{\left\langle #1\right|}
\def\ket#1{\left| #1\right\rangle}

\makeatletter

\newcommand{\Rmnum}[1]{\expandafter\@slowromancap\romannumeral #1@}
\makeatother

\newcommand{\half}{\tfrac{1}{2}}
\newcommand{\nn}{\nonumber}
\def\thefootnote{\fnsymbol{footnote}}
\makeatletter
\let\save@mathaccent\mathaccent
\newcommand*\if@single[3]{%
  \setbox0\hbox{${\mathaccent"0362{#1}}^H$}%
  \setbox2\hbox{${\mathaccent"0362{\kern0pt#1}}^H$}%
  \ifdim\ht0=\ht2 #3\else #2\fi
  }
\newcommand*\rel@kern[1]{\kern#1\dimexpr\macc@kerna}
\newcommand*\widebar[1]{\@ifnextchar^{{\wide@bar{#1}{0}}}{\wide@bar{#1}{1}}}
\newcommand*\wide@bar[2]{\if@single{#1}{\wide@bar@{#1}{#2}{1}}{\wide@bar@{#1}{#2}{2}}}
\newcommand*\wide@bar@[3]{%
  \begingroup
  \def\mathaccent##1##2{%
    \let\mathaccent\save@mathaccent
    \if#32 \let\macc@nucleus\first@char \fi
    \setbox\z@\hbox{$\macc@style{\macc@nucleus}_{}$}%
    \setbox\tw@\hbox{$\macc@style{\macc@nucleus}{}_{}$}%
    \dimen@\wd\tw@
    \advance\dimen@-\wd\z@
    \divide\dimen@ 3
    \@tempdima\wd\tw@
    \advance\@tempdima-\scriptspace
    \divide\@tempdima 10
    \advance\dimen@-\@tempdima
    \ifdim\dimen@>\z@ \dimen@0pt\fi
    \rel@kern{0.6}\kern-\dimen@
    \if#31
      \overline{\rel@kern{-0.6}\kern\dimen@\macc@nucleus\rel@kern{0.4}\kern\dimen@}%
      \advance\dimen@0.4\dimexpr\macc@kerna
      \let\final@kern#2%
      \ifdim\dimen@<\z@ \let\final@kern1\fi
      \if\final@kern1 \kern-\dimen@\fi
    \else
      \overline{\rel@kern{-0.6}\kern\dimen@#1}%
    \fi
  }%
  \macc@depth\@ne
  \let\math@bgroup\@empty \let\math@egroup\macc@set@skewchar
  \mathsurround\z@ \frozen@everymath{\mathgroup\macc@group\relax}%
  \macc@set@skewchar\relax
  \let\mathaccentV\macc@nested@a
  \if#31
    \macc@nested@a\relax111{#1}%
  \else
    \def\gobble@till@marker##1\endmarker{}%
    \futurelet\first@char\gobble@till@marker#1\endmarker
    \ifcat\noexpand\first@char A\else
      \def\first@char{}%
    \fi
    \macc@nested@a\relax111{\first@char}%
  \fi
  \endgroup
}
\makeatother

\begin{document}
\begin{flushright}
\normalsize{
SCIPP-22/02 \\
December, 2022
}
\end{flushright}
\vspace{1cm}

\begin{center}
\Large\bf\boldmath
Accommodating Hints of New Heavy Scalars in the Framework of the
Flavor-Aligned Two-Higgs-Doublet Model

\unboldmath
\end{center}
\vspace{0.4cm}
\begin{center}

Joseph M.~Connell,$^{1,}$\footnote{Electronic address: jomaconn@ucsc.edu}
P.M. Ferreira,$^{2,}$\footnote{Electronic address: pmmferreira@fc.ul.pt}
Howard E.~Haber$^{1,}$\footnote{Electronic address: haber@scipp.ucsc.edu}\\

\vspace{0.4cm}
 {\sl $^1$ Santa Cruz Institute for Particle Physics \\
 University of California, Santa Cruz, CA 95064 USA}\\[0.2cm]
\vspace{0.2cm}

 {\sl $^2$ Centro de Fisica Te\'orica e Computacional, Faculdade de Ci\^encias,\\
 Universidade de Lisboa, Avenida Professor Gama Pinto, 2, 1649-003 Lisboa, Portugal.}\\[0.2cm]
\end{center}
\vspace{0.2cm}

\renewcommand{\thefootnote}{\arabic{footnote}}
\setcounter{footnote}{0}

\begin{abstract}
Searches for new neutral Higgs bosons of an extended Higgs sector at the LHC can be interpreted in the framework of the two-Higgs doublet model.  By employing generic flavor-aligned Higgs--fermion Yukawa couplings, we propose an analysis that uses experimental data to determine whether flavor alignment is a consequence of a symmetry that is either exact or at most softly broken.  We illustrate our proposal in two different scenarios based on a few 3 sigma (local) excesses observed by the ATLAS and CMS Collaborations in their searches for heavy scalars.
In Scenario 1, an excess of events is interpreted as $A\to ZH\to \ell^+\ell^- b\bar{b}$ (where $\ell=e$ or $\mu$), with the CP-odd and CP-even neutral scalar masses given by $m_A=610$~GeV and $m_H=290$~GeV, respectively.  In Scenario~2, an excess of events in the production of $t\bar{t}$ and $\tau^+\tau^-$ final states is interpreted as decays of a CP-odd scalar of mass $m_A=400$~GeV.   Scenario 1 is consistent with Type-I Yukawa interactions, which can arise in a 2HDM subject to a softly-broken $\mathbb{Z}_2$ discrete symmetry.  Scenario 2 is inconsistent with a symmetry-based flavor alignment, but can be consistent with more general flavor-aligned Higgs--fermion Yukawa couplings.

\end{abstract}
\newpage

\section{Introduction}

After ten years of Higgs boson studies, the LHC data show no significant deviations from the predictions of the Standard Model (SM).
The phenomenological profile of the Higgs boson resembles that of the SM with precisions approaching 10\% in some channels~\cite{ATLAS:2022vkf,CMS:2022dwd}.   One may be left wondering
whether we have reached the end of our exploration of the theory of elementary particles and their interactions.

However, the Standard Model is known to be incomplete (e.g., it cannot accommodate dark matter, baryogenesis, and neutrino masses, while providing no explanation for the origin of the electroweak energy scale).
Whether departures will first be revealed at the TeV scale, perhaps in future LHC experiments, or whether physics beyond the SM enters at a much higher energy scale
remains to be seen.  Nevertheless, independently of whether the more profound questions associated with the incompleteness of the SM can be directly addressed
at the LHC, one can pose the following pedestrian query.   Given the nonminimal nature of the matter and gauge multiplets that comprise the SM, should one also expect a nonminimal scalar sector as well?  If yes, is it possible (and perhaps even likely) that
additional particles beyond the SM not yet discovered will eventually emerge from future LHC data?   Examples of such additional states could be new gauge bosons
[implying that the gauge group relevant for TeV scale physics is larger than SU(3)$\times$SU(2)$\times$U(1)], new fermionic states (such as vectorlike quarks and
leptons), new scalar states (e.g., an extended Higgs sector), supersymmetric partners of SM particles, or even more exotic objects such as leptoquarks.

In this work, we focus on the possibility that the scalar sector includes additional color singlet neutral and charged scalars beyond the
SM Higgs boson.   It is certainly an important experimental question to ask whether such states exist in a mass range accessible to the LHC.  Indeed, the ATLAS
and CMS Collaborations have performed numerous searches for such new scalar states using the Run~1 and Run 2 data sets, and such searches will continue and will
be expanded during Run~3 and beyond.

So far, no definitive signals of new scalar states have been announced.  From time to time, small excesses of events emerge in some search channels, as one
would expect based on fluctuations of data from the size of the data samples.   Nevertheless, if new scalar states do exist in Nature in a range that
can be probed by the LHC, then the initial signal of these states will often resemble the excesses due to expected fluctuations in the data.  Of course, increasing
the size of the data samples as more data is collected will reveal which of these two possibilities is the correct interpretation.

In searching for evidence for new scalar states, one often is required to make model assumptions in developing the search strategies and in interpreting the
results.  The more specific the model assumptions are, the less flexible the data analysis.  On the other hand, the more generic the model, the more difficult it is to
focus on specific experimental signatures.   In proposing searches for extended Higgs sector phenomena, we find it convenient to focus on the two Higgs doublet
extension of the Standard Model (2HDM)~\cite{Branco:2011iw}.  This model possesses the main ingredients for new phenomena that one expects in most extended Higgs sectors.
These include charged scalars, CP-odd scalars (if the neutral scalar sector is CP-conserving) or neutral scalars of indefinite CP (if the scalar sector is CP-violating),
and the possibility of Higgs-mediated flavor changing neutral currents (FCNCs), which if present must be small enough to avoid conflict with present experimental data.

The most general 2HDM adds a significant number of new parameters to the Standard Model.   Recall that the Standard Model (where neutrino masses are zero and
are not counted as separate parameters) is governed by 19 parameters, which include three gauge couplings, $\Theta_{\rm QCD}$, nine quark and lepton masses,
three Cabibbo-Kobayashi-Maskawa (CKM) angles, one CKM phase, and two parameters of the Higgs sector that can be taken to be the Higgs vacuum expectation value, $v\simeq 246$~GeV and the Higgs mass (e.g., see Ref.~\cite{Langacker:2017uah}).
In the most general 2HDM, the two Higgs sector parameters of the SM are expanded to eleven, and new Higgs-fermion Yukawa matrix couplings arise that are in principle independent of
the quark and lepton masses.   In light of \eq{YUK2}, these new Yukawa matrix couplings correspond to the six $3\times 3$ hermitian matrices $\rho^F_R$ and $\rho^F_I$ (where $F=U,D,E$ refers to the couplings to up-type quarks, down-type quarks and charged leptons, respectively) and yield 54 new parameters.
Thus, the most general two-Higgs-doublet extension of the SM is governed by 82 parameters!

It is not practical to devise search strategies that scan over all 82 parameters of the general 2HDM.  Moreover, a generic point in this 82-dimensional space would be immediately
ruled out due to scalar-mediated FCNCs that can already be experimentally ruled out.   In the literature, the standard practice is to eliminate tree-level Higgs-mediated FCNCs by
imposing an appropriate discrete symmetry~\cite{Glashow:1976nt,Paschos:1976ay}.
By considering all possible symmetries of this type, one finds four classes of 2HDM Yukawa couplings, which are called Types I, II, X and Y in the literature~\cite{Hall:1981bc,Barger:1989fj,Aoki:2009ha}.  From a purely phenomenological point of view, this assumption is too strong as it reduces the size of the 2HDM parameter space more strictly than necessary.   Indeed, it is sufficient to simply require that the hermitian matrices $\rho^F_R$ and $\rho^F_I$ are diagonal.   One can reduce the number of 2HDM parameters even further by
assuming that $\rho^F_R$ and $\rho^F_I$ are each proportional to the $3\times 3$ identity matrix (with coefficients, called flavor-alignment parameters, that depend on $F$), which yields the flavor-aligned 2HDM (A2HDM)~\cite{Pich:2009sp}.\footnote{Flavor-aligned extended Higgs sectors can naturally arise from symmetries of ultraviolet completions of low-energy effective theories of flavor as shown in Refs.~\cite{Serodio:2011hg,Knapen:2015hia,Egana-Ugrinovic:2018znw,Egana-Ugrinovic:2019dqu}.   In such models, departures from exact flavor alignment due to renormalization group running down to the electroweak scale are typically small enough~\cite{Braeuninger:2010td,Gori:2017qwg} to be consistent with all known experimental FCNC bounds. \label{fnalign}}
Although there are theoretical arguments for favoring the stricter Types I, II, X and Y structures (which are renormalization group stable~\cite{Ferreira:2010xe} and hence can be realized without an artificial fine-tuning of parameters in the 82-dimensional 2HDM parameter space), ultimately it will be experiment that will determine the structure of the Yukawa interactions.   Indeed, if potential signals of an extended Higgs sector arise, we believe that it is prudent to employ the less restrictive A2HDM framework in order to test the validity of the 2HDM interpretation of the data.

Likewise, one must decide whether to include new sources of CP-violation in the 2HDM when confronting potential signals of extended Higgs sector phenomenology.
Experimental constraints exist due to the absence of evidence for an electric dipole moment of the electron~\cite{Roussy:2022cmp,Altmannshofer:2020shb}.  In this work, we choose to assume a CP-conserving scalar sector
for simplicity to reduce the number of parameters of the model and simplify the subsequent analysis.   This is accomplished by taking the flavor-alignment parameters to be real and demanding the existence of a Higgs basis in which all the scalar potential parameters are real.
It is quite likely that any initial discovery of new scalars at the LHC will be insensitive to assumptions regarding possible CP-violating parameters associated with the extended Higgs sector.
However, it is certainly worth considering the phenomenological implications
of scalar sector CP violation, which we will leave for a future work.

In Section~\ref{sec:theory}, we review the theoretical structure of the 2HDM.   We explicitly specify the parameters that govern the A2HDM parameter space, specializing to the case where no CP-violating parameters (beyond the CKM phase) are present.  In particular, the flavor-alignment parameters are defined such that they are manifestly basis-independent quantities and hence directly related to physical observables.
In Section~\ref{sec:anomalies}, we survey the search for non-SM-like Higgs bosons at the LHC and discuss a few excesses in different search channels that have been reported by the ATLAS and CMS Collaborations~\cite{ATLAS:2020gxx,ATLAS:2020zms,ATLAStautau,CMS:2019pzc} .
We propose to analyze two different A2HDM scenarios that could yield excesses of events that would be compatible with the reported LHC data.  In Scenario~1, we take $m_A=610$~GeV and $m_H=290$~GeV, motivated by an ATLAS excess of events with a local (global) significance of  3.1(1.3)$\sigma$~\cite{ATLAS:2020gxx}, which is compatible with the interpretation of $gg\to A\to ZH$, where $H\to b\bar{b}$ and $Z\to\ell^+\ell^-$ (where $\ell=e$, $\mu$).
Scenario 2 is based on an ATLAS excess of $\tau^+\tau^-$ events with an invariant mass of around 400 GeV, with local significances of 2.2$\sigma$ in the $gg$ fusion production channel and
2.7$\sigma$ in the $b$-associated production channel~\cite{ATLAS:2020zms,ATLAStautau,CMS:2019pzc}.
The CMS Collaboration sees no excess in the $\tau^+\tau^-$ channel~\cite{CMS:2022goy}, but still leaves some room for a possible signal.  However, the CMS Collaboration observes an excess of $t\bar{t}$ events with an invariant mass of around 400 GeV, with a local (global) significance of 3.5(1.9)$\sigma$, that favors identifying the excess with $A$ production~\cite{CMS:2019pzc}.

We propose to examine whether Scenarios 1 or 2 is compatible with an A2HDM interpretation.
In Section~\ref{sec:scans}, we discuss the details of our A2HDM parameter scan and the relevant
theoretical and experimental constraints that must be applied to the A2HDM parameter space.
Some of the indirect experimental constraints that we have applied (such as the oblique $T$-parameter and the constraints due to the observed $b\to s\gamma$ and $B_s$--$\overline{B}_s$ mixing) are based on theoretical calculations that are reviewed in three appendices.
In Section~\ref{sec:scen1}, we exhibit the regions of the A2HDM parameter scans that are consistent with Scenario 1.  We show that these regions include a subregion that coincides with Type-I Yukawa couplings.
In order to facilitate future experimental heavy Higgs boson searches at the LHC, we provide two specific A2HDM benchmarks for Scenario 1.  One benchmark is specifically targeted to a Type-I 2HDM, whereas a second benchmark corresponds to a more general A2HDM parameter point that  is incompatible with a Type I, II, X or Y structure.

In Section~\ref{sec:scen2}, we exhibit the regions of the A2HDM parameter scans that are consistent with Scenario 2.   We demonstrate that none of A2HDM parameter regions favored by Scenario 2 are consistent with Type I, II, X or Y Yukawa couplings.  We construct one A2HDM benchmark for Scenario 2 that can be used to suggest additional heavy scalar channels that might be probed in future runs at the LHC.

In Section~\ref{sec:conclude}, we summarize our results and suggest other heavy scalar signatures that could be used to confirm the presence of new scalars beyond the SM-like Higgs boson, should any of the small excesses that have been previously reported by the ATLAS and CMS collaborations turn out to be something more than a statistical fluctuation.


\section{Theoretical background}
\label{sec:theory}

In this section, we briefly review the theoretical structure of the 2HDM.
The scalar fields of the 2HDM consist of two
identical complex hypercharge-one, SU(2) doublets
$\Phi_i(x)\equiv (\Phi^+_i(x)\,,\,\Phi^0_i(x))$,
labeled by the ``Higgs flavor'' index $i\in\{1,2\}$.
The most general renormalizable SU(2)$_L\times$U(1)$_Y$ invariant scalar potential is given in the so-called $\Phi$-basis
by
\beqa  \label{pot}
\mathcal{V}&=& m_{11}^2\Phi_1^\dagger\Phi_1+m_{22}^2\Phi_2^\dagger\Phi_2
-[m_{12}^2\Phi_1^\dagger\Phi_2+{\rm h.c.}]+\half\lambda_1(\Phi_1^\dagger\Phi_1)^2
+\half\lambda_2(\Phi_2^\dagger\Phi_2)^2
+\lambda_3(\Phi_1^\dagger\Phi_1)(\Phi_2^\dagger\Phi_2)\nonumber\\
&&\quad
+\lambda_4(\Phi_1^\dagger\Phi_2)(\Phi_2^\dagger\Phi_1)
+\left\{\half\lambda_5(\Phi_1^\dagger\Phi_2)^2
+\big[\lambda_6(\Phi_1^\dagger\Phi_1)
+\lambda_7(\Phi_2^\dagger\Phi_2)\big]
\Phi_1^\dagger\Phi_2+{\rm h.c.}\right\}\,,
\eeqa
where $m_{11}^2$, $m_{22}^2$, and $\lambda_1,\cdots,\lambda_4$ are real parameters
and $m_{12}^2$, $\lambda_5$, $\lambda_6$ and $\lambda_7$ are
potentially complex parameters.  We assume that the
parameters of the scalar potential are chosen such that
the minimum of the scalar potential respects the
U(1)$\ls{\rm EM}$ gauge symmetry.  Then, the scalar field
vacuum expectations values (vevs) are of the form\footnote{Without loss of generality $\widehat{v}_1$ can be chosen to be nonnegative by applying a suitable U(1)$_Y$ transformation that has no effect on the scalar potential parameters.}
\beq \label{vhat}
\langle\Phi_i\rangle=
\frac{v}{\sqrt{2}}\begin{pmatrix} 0\\ \widehat{v}_i\end{pmatrix}\,,
\eeq
where $ \widehat{v}$ is a complex vector of unit norm,
\beq \label{veedef}
\widehat{v}=(\widehat{v}_1,\widehat{v}_2) =(c_\beta\,,\, s_\beta e^{i\xi}),
\eeq
$c_\beta\equiv \cos\beta$ and $s_\beta\equiv\sin\beta$ (such that $0\leq\beta\leq\half\pi$, $0\leq \xi< 2\pi$), and $v$ is determined by the Fermi constant,
\beq
v\equiv \frac{2m_W}{g}=(\sqrt{2}G_F)^{-1/2}\simeq 246~{\rm GeV}\,.\label{v246}
\eeq

\subsection{The Higgs basis}

One is always free to redefine the scalar field basis by a unitary transformation, $\Phi_i\to U_{ij}\Phi_j$.    The $\Phi$-basis that was used to define the scalar potential given in \eq{pot} has no physical significance in a generic 2HDM.  However, starting from an arbitrary $\Phi$-basis, one can always transform to the so-called Higgs basis~\cite{Georgi:1978ri,Lavoura:1994yu,Lavoura:1994fv,Botella:1994cs,Branco:1999fs,Davidson:2005cw}, where the Higgs basis fields, denoted by
$\mathcal{H}_1$ and $\mathcal{H}_2$ are defined
by the linear combinations of $\Phi_1$ and $\Phi_2$ such that $\langle \mathcal{H}_1^0\rangle=v/\sqrt{2}$ and $\langle \mathcal{H}_2^0\rangle=0$.
In particular~\cite{Haber:2006ue,Boto:2020wyf},
\beq \label{hbasisdef}
\mathcal{H}_1=(\mathcal{H}_1^+\,,\,\mathcal{H}_1^0)\equiv \widehat v_i^{\,\ast}\Phi_i\,,\qquad\qquad
\mathcal{H}_2=(\mathcal{H}_2^+\,,\,\mathcal{H}_2^0)\equiv e^{i\eta} \epsilon_{ij}\widehat{v}_i\Phi_j\,,
\eeq
where $\epsilon_{12}=-\epsilon_{21}=1$ and $\epsilon_{11}=\epsilon_{22}=0$, and there is an implicit sum over repeated indices.   The phase factor $e^{i\eta}$ indicates that the Higgs basis is not unique, since one can always rephase the Higgs basis field $\mathcal{H}_2$ while maintaining $\langle \mathcal{H}_2^0\rangle=0$.

In terms of the Higgs basis fields defined in \eq{hbasisdef}, the scalar potential is given by,
 \beqa
 \mathcal{V}&=& Y_1 \mathcal{H}_1^\dagger \mathcal{H}_1+ Y_2 \mathcal{H}_2^\dagger \mathcal{H}_2 +[Y_3 e^{-i\eta}
\mathcal{H}_1^\dagger \mathcal{H}_2+{\rm h.c.}]
\nn\\
&&\quad
+\half Z_1(\mathcal{H}_1^\dagger \mathcal{H}_1)^2+\half Z_2(\mathcal{H}_2^\dagger \mathcal{H}_2)^2
+Z_3(\mathcal{H}_1^\dagger \mathcal{H}_1)(\mathcal{H}_2^\dagger \mathcal{H}_2)
+Z_4(\mathcal{H}_1^\dagger \mathcal{H}_2)(\mathcal{H}_2^\dagger \mathcal{H}_1) \nn \\
&&\quad
+\left\{\half Z_5 e^{-2i\eta}(\mathcal{H}_1^\dagger \mathcal{H}_2)^2 +\big[Z_6 e^{-i\eta} (\mathcal{H}_1^\dagger
\mathcal{H}_1) +Z_7 e^{-i\eta} (\mathcal{H}_2^\dagger \mathcal{H}_2)\big] \mathcal{H}_1^\dagger \mathcal{H}_2+{\rm
h.c.}\right\}\,.\label{higgspot}
\eeqa
The minimization of the scalar potential in the Higgs basis yields
\beq \label{minconds}
Y_1=-\half Z_1 v^2\,,\qquad\quad Y_3=-\half Z_6 v^2\,.
\eeq

Note that under a change of scalar field basis, $\Phi_i\to U_{ij}\Phi_j$ (where $U$ is a unitary $2\times 2$ matrix), the parameters $Y_1$, $Y_2$ and $Z_1,\ldots Z_4$ are invariant whereas
\beq \label{rephasing}
 [Y_3, Z_6, Z_7, e^{i\eta}]\to (\det~U)^{-1}[Y_3, Z_6, Z_7, e^{i\eta}] \quad{\rm and}\quad
Z_5\to  (\det~U)^{-2} Z_5\,.
\eeq
It follows that the Higgs basis fields $\mathcal{H}_1$ and $\mathcal{H}_2$ and the coefficients appearing in \eq{higgspot} [as well as the scalar potential $\mathcal{V}$] are invariant quantities
with respect to U(2) transformations.

\subsection{Scalar mass eigenstates}
\label{masseigenstates}

Given the scalar potential and its minimization conditions, one can determine the masses of the neutral scalars.
After removing the massless Goldstone boson, $G^0=\sqrt{2}\,\Im~\!\mathcal{H}_1^0$ from the $4\times 4$ neutral scalar squared-mass matrix,
the physical neutral scalar mass-eigenstate fields are obtained by diagonalizing the resulting $3\times 3$ neutral scalar squared-mass matrix,
\beq  \label{matrix33}
\mathcal{M}^2=v^2\left( \begin{array}{ccc}
Z_1&\quad \Re(Z_6 e^{-i\eta}) &\quad -\Im(Z_6 e^{-i\eta})\\
\Re(Z_6 e^{-i\eta})  &\quad \half\bigl[Z_{34}+\Re(Z_5 e^{-2i\eta})\bigr]+Y_2/v^2 & \quad
- \half \Im(Z_5  e^{-2i\eta})\\ -\Im(Z_6  e^{-i\eta}) &\quad - \half \Im(Z_5  e^{-2i\eta}) &\quad
\half\bigl[Z_{34}-\Re(Z_5 e^{-2i\eta})\bigr]+Y_2/v^2\end{array}\right),
\eeq
which is expressed with respect to the $\{\sqrt{2}\,\Re \mathcal{H}_1^0-v,\sqrt{2}\,\Re \mathcal{H}_2^0,\sqrt{2}\,\Im \mathcal{H}_2^0\}$ basis, where
\beq \label{zeethreefour}
Z_{34}\equiv Z_3+Z_4\,.
\eeq
The squared masses of the physical neutral scalars, denoted by $m_k^2$ ($k=1,2,3$) with no implied mass ordering, are the eigenvalues of $\mathcal{M}^2$, which are independent of the choice of~$\eta$.

The real symmetric squared-mass matrix $\mathcal{M}^2$ can be diagonalized by
a real orthogonal transformation of unit determinant,
\beq \label{rmrt}
R\mathcal{M}^2 R^{\T}= {\rm diag}~(m_1^2\,,\,m_2^2\,,\,m_3^2)\,,
\eeq
where $R\equiv R_{12}R_{13}R_{23}$   is the product of three rotation matrices parametrized by $\theta_{12}$, $\theta_{13}$ and $\theta_{23}$, respectively~\cite{Haber:2006ue}.
Since the matrix elements of $\mathcal{M}^2$ are independent of the scalar field basis [in light of \eq{rephasing}], it follows that the mixing angles $\theta_{ij}$ are basis-invariant parameters.
The physical neutral mass-eigenstate scalar fields are
\beq \label{hsubkay}
h_k=q_{k1}\bigl(\sqrt{2}\,\Re \mathcal{H}_1^0-v\bigr)+\frac{1}{\sqrt{2}}\bigl(q_{k2}^*\mathcal{H}_2^0 e^{i\theta_{23}}+{\rm h.c.}\bigr)\,,
\eeq
where $q_{k1}$ and $q_{k2}$ are exhibited in Table~\ref{tabinv}.  The charged scalar mass eigenstates are defined by,
\beq
G^\pm=\mathcal{H}_1^\pm\,,\qquad\quad H^\pm\equiv e^{\pm i\theta_{23}}\mathcal{H}_2^\pm\,,
\eeq
where $G^\pm$ are the massless charged Goldstone fields.  Note that we have rephased the charged Higgs fields as a matter of convenience.
The mass of the charged Higgs scalar is given by,
\beq \label{plusmass}
m_{H^\pm}^2=Y_2+\half Z_3 v^2\,.
\eeq

\begin{table}[t!]
\centering
\begin{tabular}{|c||c|c|}\hline
$\phaa k\phaa $ &\phaa $q_{k1}\phaa $ & \phaa $q_{k2} \phaa $ \\ \hline
$1$ & $c_{12} c_{13}$ & $-s_{12}-ic_{12}s_{13}$ \\
$2$ & $s_{12} c_{13}$ & $c_{12}-is_{12}s_{13}$ \\
$3$ & $s_{13}$ & $ic_{13}$ \\
\hline
\end{tabular}
\caption{\small The basis-invariant quantities $q_{k\ell}$ are functions of
the neutral Higgs mixing angles $\theta_{12}$ and $\theta_{13}$, where
$c_{ij}\equiv\cos\theta_{ij}$ and $s_{ij}\equiv\sin\theta_{ij}$.
\label{tabinv}}
\end{table}

Inverting \eq{hsubkay}, one can now express the Higgs basis fields in terms of the mass eigenstate fields,
\beq \label{Hbasismassbasis}
\mathcal{H}_1=\begin{pmatrix} G^+ \\[3pt] \displaystyle\frac{1}{\sqrt{2}}\left(v+iG+\sum_{k=1}^3 q_{k1}h_k\right)\end{pmatrix},
\qquad\quad
e^{i\theta_{23}}\mathcal{H}_2=\begin{pmatrix} H^+ \\[3pt] \displaystyle\frac{1}{\sqrt{2}}\sum_{k=1}^3 q_{k2}h_k\end{pmatrix}.
\eeq
Although ${\theta}_{23}$ is an invariant parameter, it has no physical significance since
it can be eliminated by rephasing $\mathcal{H}_2\to e^{-i\theta_{23}}\mathcal{H}_2$.
Thus, without loss of generality, we henceforth set $\theta_{23}=0$.

\subsection{2HDM Yukawa couplings}
\label{2HDMYUKS}

Given the most general Yukawa Higgs-quark Lagrangian involving the scalar fields of the 2HDM and the interaction eigenstate quark fields, one can derive expressions for the
$3\times 3$ complex up-type and down-type quark mass matrices by setting the neutral Higgs fields to their vacuum expectation values.
Each of the two quark mass matrices can then be diagonalized via singular value decomposition, which yields a pair of unitary matrices that are then employed in defining the left-handed and right-handed quark mass-eigenstate fields, respectively.

After determining the quark mass eigenstate fields and the Higgs mass eigenstate fields, the
resulting 2HDM Yukawa couplings in their most general form are (cf.~eq.~(58) of Ref.~\cite{Boto:2020wyf}),
\beqa
 && \hspace{-0.2in} -\mathscr{L}_Y =\frac{1}{\sqrt{2}}\overline U \biggl\{ q_{k1}\kappa^U +
q^*_{k2}\,\rho^U P_R+
q_{k2}\,\rho^{U\dagger} P_L\biggr\}U h_k
 \nonumber \\
&&\qquad \,\, + \frac{1}{\sqrt{2}}\overline D
\biggl\{ q_{k1} \kappa^{D\dagger} +
q_{k2}\,\rho^{D\dagger} P_R+
q^*_{k2}\,\rho^D P_L\biggr\}Dh_k
\nonumber \\
&&\qquad \,\, +\biggl\{\overline U\left[K\rho^{D\dagger}
P_R-\rho^{U\dagger} KP_L\right] D\mathcal{H}^+
+\overline
U\left[K\kappa^{D\dagger} P_R-\kappa^U KP_L\right] DG^+
+{\rm
h.c.}\biggr\}\,, \label{Yukawas}
\eeqa
where there is an implicit sum over $k\in\{1,2,3\}$, $P_{R,L}\equiv \half(1\pm\gamma\ls{5})$,
$K$ is the CKM matrix, the mass-eigenstate down-type and up-type quark fields are
$D=(d,s,b)^{\T}$ and $U\equiv (u,c,t)^{\T}$, respectively,
and the $3\times 3$ Yukawa coupling matrices $\kappa^U$ and $\kappa^D$ are related
to the corresponding up-type and down-type diagonal quark mass matrices,
\beq \label{MQ}
M_U=\frac{v}{\sqrt{2}}\kappa^U={\rm diag}(m_u\,,\,m_c\,,\,m_t)\,,\qquad
M_D=\frac{v}{\sqrt{2}}\kappa^{D\,\dagger}={\rm
diag}(m_d\,,\,m_s\,,\,m_b) \,.
\eeq
The Yukawa coupling matrices $\rho^U$ and $\rho^D$ are independent basis-invariant complex
$3\times 3$ matrices.

It is convenient to rewrite the Higgs-quark Yukawa couplings in terms of
the following two $3\times 3$ hermitian matrices that are invariant
with respect to the rephasing of the Higgs basis field $\mathcal{H}_2$,
\beq \label{rhoRI}
\rho^F_R \equiv \frac{v}{2\sqrt{2}}\,M^{-1/2}_F
(\rho^F +
\rho^{F\,^\dagger})M^{-1/2}_F\,,
\qquad\quad
\rho^F_I \equiv \frac{v}{2\sqrt{2}\,i}M^{-1/2}_F
(\rho^F -
\rho^{F\,\dagger})M^{-1/2}_F\,,
\eeq
for $F=U,D$,
where the $M_F$ are the diagonal
fermion mass matrices [cf.~\eq{MQ}].
Then, the Yukawa couplings take the following form:\footnote{\Eq{YUK2} is easily extended to include the Higgs boson couplings to leptons.   Since neutrinos are massless in the two-Higgs doublet extension of the Standard Model, one simply
replaces $D\to E=(e,\mu,\tau)^{\T}$ and $U\to N=(\nu_e,\nu_\mu,\nu_\tau)^{\T}$, with $M_E={\rm diag}(m_e,m_\mu,m_\tau)$ and $M_N=0$  in \eqs{Yukawas}{YUK2}.}
\beqa
\!\!\!\!\!\!\!\!\!\!
-\mathscr{L}_Y &=& \frac{1}{v}\,\overline U \sum_{k=1}^3 M_U^{1/2}\biggl\{q_{k1}\mathds{1}
+ \Re(q_{k2})\bigl[\rho^U_R+i\gamma\ls{5}\rho^U_I\bigr]+\Im(q_{k2})\bigl[\rho^U_I-i\gamma\ls{5}\rho^U_R\bigr]\biggr\} M_U^{1/2}Uh_k \nonumber \\
&&  +\frac{1}{v}\,\overline D\sum_{k=1}^3 M_D^{1/2}
\biggl\{q_{k1}\mathds{1}  +\Re(q_{k2})\bigl[\rho^D_R-i\gamma\ls{5}\rho^D_I\bigr]+\Im(q_{k2})\bigl[\rho^D_I+i\gamma\ls{5}\rho^D_R\bigr]\biggr\} M_D^{1/2}Dh_k  \nonumber \\
&& +\frac{\sqrt{2}}{v}\biggl\{\overline{U}\bigl[KM_D^{1/2}(\rho^D_R-i\rho^D_I)
M_D^{1/2}P_R-M_U^{1/2}(\rho^U_R-i\rho^U_I) M_U^{1/2}KP_L\bigr] DH^+ +{\rm
h.c.}\biggr\}, \label{YUK2}
\eeqa
where $\mathds{1}$ is the $3\times 3$ identity matrix.
The appearance of unconstrained hermitian $3\times 3$ Yukawa matrices
$\rho^F_{R,I}$ in \eq{YUK2} indicates the presence of potential flavor-changing neutral Higgs--quark and lepton interactions.
If the off-diagonal elements of $\rho^F_{R,I}$ are unsuppressed, they will generate tree-level Higgs-mediated flavor changing neutral currents (FCNCs)
that are incompatible with the strong suppression of FCNCs observed in Nature.

The flavor-aligned 2HDM (often denoted by A2HDM) posits that the Yukawa matrices $\kappa^F$ and $\rho^F$ [cf.~\eq{Yukawas}] are proportional.  In light of \eq{MQ},
$\kappa^F=\sqrt{2}M_F/v$ is diagonal.  Thus in the A2HDM, the $\rho^F$ are likewise diagonal,
which implies that tree-level Higgs-mediated FCNCs are absent.
We define the basis-invariant \textit{flavor-alignment parameters} $a^F$ via,
\beq \label{aligned}
\rho^F=a^F \kappa^F\,,\qquad\quad \text{for $F=U,D,E$},
\eeq
where the (potentially) complex numbers $a^F$ are invariant under the rephasing of the Higgs basis field $\mathcal{H}_2\to e^{i\chi}\mathcal{H}_2$. It follows from \eq{rhoRI} that
\beq
\rho_R^F=(\Re a^F)\mathds{1}\,,\qquad\quad \rho_I^F=(\Im a^F)\mathds{1}\,.
\eeq
Inserting the above results into \eq{YUK2}, the Yukawa couplings take the following form:
\beqa
-\mathscr{L}_Y &=& \frac{1}{v}\,\overline U M_U\sum_{k=1}^3 \biggl\{q_{k1}
+ q_{k2}^* a^U P_R+q_{k2} a^{U*}P_L\biggr\} Uh_k   \nonumber \\
&& +\frac{1}{v}\sum_{F=D,E} \biggl\{\overline F  M_F\sum_{k=1}^3 \bigl(q_{k1}
+ q_{k2} a^{F*} P_R+q^*_{k2}a^{F}P_L\bigr) Fh_k\biggr\}
 \nonumber \\
&& +\frac{\sqrt{2}}{v}\biggl\{\overline U\bigl[a^{D*}KM_DP_R-a^{U*}M_U KP_L\bigr] DH^+ + a^{E*}\overline NM_E P_R EH^+
+{\rm
h.c.}\biggr\}. \label{YUK4}
\eeqa
\begin{table}[t!]
\centering
{\addtolength\tabcolsep{10pt}
\begin{tabular}{|c||c|c|c|c|c|c|}
\hline & $\Phi_1$ & $\Phi_2$ & $U_R$ & $D_R$ & $E_R$ &
 $U_L$, $D_L$, $N_L$, $E_L$ \\  \hline
Type I  & $+$ & $-$ & $-$ & $-$ & $-$ & $+$ \\
Type II & $+$ & $-$ & $-$ & $+$ & $+$ & $+$ \\
Type X   & $+$ & $-$ & $-$ & $-$ & $+$ & $+$ \\
Type Y  & $+$ & $-$ & $-$ & $+$ & $-$ & $+$ \\
\hline
\end{tabular}}
\caption{\small Four possible $\mathbb{Z}_2$ charge assignments for scalar and fermion fields~\cite{Aoki:2009ha}.
 The $\mathbb{Z}_2$ symmetry is employed to constrain the Higgs-fermion Yukawa couplings, thereby implementing the conditions
 for the natural absence of tree-level Higgs-mediated FCNCs.
\label{Tab:type}}
\end{table}

Special cases of the A2HDM arise if the flavor alignment is the consequence of a symmetry.  For Yukawa couplings of
Type-I, II, X and Y, one imposes a $\mathbb{Z}_2$ symmetry on the dimension-4 terms of the Higgs Lagrangian in the $\Phi$-basis, where the
$\mathbb{Z}_2$ charges are exhibited in Table~\ref{Tab:type}~\cite{Aoki:2009ha}.  It then follows that $\lambda_6=\lambda_7=0$.  One can show that there exists a scalar field basis where
$\lambda_6=\lambda_7=0$ if and only if the following two conditions are satisfied~\cite{Boto:2020wyf}:
\beqa
&& (Z_1-Z_2)\bigl[(Z_3+Z_4) |Z_{67}|^2-Z_2 |Z_6|^2-Z_1 |Z_7|^2-(Z_1+Z_2)\Re(Z_6^* Z_7)+\Re(Z_5^* Z^2_{67})\bigr]\nonumber \\
&& \qquad\qquad\qquad\qquad\qquad -2|Z_{67}|^2\bigl(|Z_6|^2-|Z_7|^2\bigr)=0\,,\label{cond5}\\
&& (Z_1-Z_2)\Im(Z_6^* Z_7)+\Im\bigl(Z_5^* Z_{67}^2\bigr)=0\,,\label{cond6}
\eeqa
where $Z_{67}\equiv Z_6+Z_7$.
In models with Type I and II Yukawa couplings,
the matrices $\rho^U$ and $\rho^D$ are diagonal and fixed as follows,
\beqa
&& \text{Type I:}~~~\rho^U=\frac{e^{i(\xi+\eta)}\sqrt{2}M_U\cot\beta}{v}\,,\qquad\quad  \rho^D=\frac{e^{i(\xi+\eta)}\sqrt{2}M_D\cot\beta}{v}\,, \label{rhoType1a}\\
&& \text{Type II:}~~\rho^U=\frac{e^{i(\xi+\eta)}\sqrt{2}M_U\cot\beta}{v}\,,\qquad\quad  \rho^D=-\,\frac{e^{i(\xi+\eta)}\sqrt{2}M_D\tan\beta}{v}\,,\label{rhoType2a}
\eeqa
where $\tan\beta=|\widehat{v}_2/\widehat{v}_1|$ [cf.~\eq{veedef}].
In Type X models, the quarks possess Type-I Yukawa couplings whereas the leptons possess Type-II Yukawa couplings.
In Type Y models, the quarks possess Type-II Yukawa couplings whereas the leptons possess Type-I Yukawa couplings.

The Type-I, II, X and Y 2HDMs are indeed special cases of the A2HDM, where we can identify the corresponding complex flavor-alignment parameters as follows,
\begin{enumerate}
\item
Type-I: $a^U=a^D=a^E=e^{i(\xi+\eta)}\cot\beta$.
\item
Type-II: $a^U=e^{i(\xi+\eta)}\cot\beta$ and $a^D=a^E=-e^{i(\xi+\eta)}\tan\beta$.
\item
Type-Y: $a^U=a^E=e^{i(\xi+\eta)}\cot\beta$ and $a^D=-e^{i(\xi+\eta)}\tan\beta$.
\item
Type-X $a^U=a^D=e^{i(\xi+\eta)}\cot\beta$ and $a^E=-e^{i(\xi+\eta)}\tan\beta$.
\end{enumerate}
Note that in the generic A2HDM, $\tan\beta$ is not a physical parameter, since the $\Phi$-basis has no physical significance.  However, after imposing a $\mathbb{Z}_2$ symmetry on the dimension-4 terms of the Higgs Lagrangian, the $\Phi$-basis where the $\mathbb{Z}_2$ symmetry is manifestly realized becomes meaningful, in which case $\tan\beta$ is promoted to a physical parameter of the model~\cite{Haber:2006ue,Boto:2020wyf}.

\subsection{The CP-conserving 2HDM}
\label{CP2HDM}

In the case of a CP-conserving Higgs scalar potential and vacuum, the results of Sections~\ref{masseigenstates} and \ref{2HDMYUKS} simplify significantly.
In particular, we can fix the Higgs basis up to a potential sign ambiguity by rephasing the Higgs basis field $\mathcal{H}_2$ such that $Y_3$, $Z_5$, $Z_6$ and $Z_7$ are all simultaneously real, which yields the so-called real Higgs basis.
Following Refs.~\cite{Haber:2006ue,Boto:2020wyf}, we set $s_{13}=0$, $c_{13}=1$ and $e^{i\eta}=\pm 1$.
 The remaining ambiguity in defining the real Higgs basis is  due to the possibility of transforming $\mathcal{H}_2\to -\mathcal{H}_2$, in which case $Y_3$, $Z_6$ and $Z_7$ change sign (whereas all other scalar potential parameters in the real Higgs basis, including $Z_5$ are unchanged).   In particular,
$e^{i\eta}$ changes sign under $\mathcal{H}_2\to -\mathcal{H}_2$.
One can identify\footnote{If $Z_6=Z_7=0$, then the sign of $Z_5$ is no longer invariant with respect to transformations that preserve the real Higgs basis (since the sign of $Z_5$ changes under $\mathcal{H}_2\to \pm i\mathcal{H}_2$).   In this case, it would be more appropriate to define $\varepsilon\equiv e^{2i\eta}=\sgn Z_5$.}
 \beq \label{twosigns}
\varepsilon\equiv e^{i\eta}=\begin{cases} \sgn Z_6\,, & \quad \text{if $Z_6\neq 0$},\\   \sgn Z_7\,,& \quad \text{if $Z_6=0$ and $Z_7\neq 0$}.\end{cases}
\eeq
The corresponding $q_{kj}$ given in Table~\ref{tabinv}  simplify to the results given in Table~\ref{cptabinv}.
\begin{table}[t!]
\centering
\begin{tabular}{|c|| c ||c|c|}\hline
$\phaa k\phaa $ & $\phaa h_k$\phaa\ & \phaa $q_{k1}\phaa $ & \phaa $q_{k2} \phaa $ \\
\hline
$1$ & $h$ & $\,\,\,s_{\beta-\alpha}$ & $\varepsilon c_{\beta-\alpha}$ \\
$2$ & $\!\!\!-\varepsilon H$ & $-\varepsilon c_{\beta-\alpha}$ & $s_{\beta-\alpha}$ \\
$3$ & $\varepsilon A$ & $0$ & $i$ \\ \hline
\end{tabular}
\caption{\small Basis-invariant combinations $q_{kj}$ defined in Table~\ref{tabinv} in the CP-conserving limit, corresponding to a real Higgs basis where $\varepsilon=\pm 1$ with the choice of sign defined by \eq{twosigns}.         \label{cptabinv}
}
\end{table}

To make contact with the standard notation of the CP-conserving 2HDM, under the assumption that the lighter of the two neutral CP-even Higgs bosons is SM-like, we make the following identifications,\footnote{This means that the signs of the fields $H$ and $A$ and the sign of $\cbma$ all flip under the redefinition of the Higgs basis field $\mathcal{H}_2\to -\mathcal{H}_2$.   In the CP-conserving 2HDM literature, in models in which the choice of the $\Phi$-basis is physically meaningful (e.g., due to the presence of a discrete $\mathbb{Z}_2$ symmetry of the scalar potential), it is traditional to impose one further restriction that $\tan\beta$ is real and positive.  This removes the final sign ambiguity in defining the real Higgs basis.}

\beq\label{eq:epsilon6}
h=h_1\,,\qquad H=-\varepsilon h_2\,,\qquad A=\varepsilon h_3\,,\qquad H^\pm\to \varepsilon H^\pm\,,
\eeq
where the neutral CP-odd Higgs mass eigenstate is related to the Higgs basis fields by $A=\varepsilon\sqrt{2}\,\Im \mathcal{H}_2^0$.
In a real $\Phi$-basis of scalar fields, the mixing angle that diagonalizes the CP-even Higgs squared-mass matrix
is denoted by~$\alpha$.   However, in the generic CP-conserving 2HDM, the $\Phi$-basis has no physical meaning, which implies that the angles $\alpha$ and $\beta$ are not (separate) physical quantities.
It is therefore more convenient to analyze the CP-even Higgs squared-mass matrix in the real Higgs basis [cf.~\eq{matrix33}],
\beq
\mathcal{M}^2=\begin{pmatrix} Z_1 v^2 & \,\,\, \varepsilon Z_6 v^2 \\ \varepsilon Z_6 v^2 & \,\,\, m_A^2+Z_5 v^2\end{pmatrix},
\eeq
where
\beq
m_A^2=Y_2+\half(Z_3+Z_4-Z_5)v^2\,.
\eeq
The CP-even Higgs mass eigenstates, $h$ and $H$ (with $m_{h}\leq m_{H}$), are then related to the neutral fields of the Higgs basis via
\beq \label{hH}
\begin{pmatrix} H\\ h\end{pmatrix}=\begin{pmatrix} \cbma & \,\,\, -\sbma \\
\sbma & \,\,\,\phantom{-}\cbma\end{pmatrix}\,\begin{pmatrix} \sqrt{2}\,\,{\rm Re}~\mathcal{H}_1^0-v \\
\varepsilon \sqrt{2}\,{\rm Re}~\mathcal{H}_2^0
\end{pmatrix}\,.
\eeq
Comparing with \eq{hsubkay} after setting $s_{13}=\theta_{23}=0$, we can then identify,
\beq \label{anglelimit}
c_{12}=\sbma\,,\qquad\quad s_{12}=-\varepsilon\,\cbma\,.
\eeq
The angle $\beta-\alpha$ is defined modulo $\pi$.  It is conventional to take $0\leq\beta-\alpha\leq\pi$, in which case $0\leq\sbma\leq 1$.  Note that the signs of $\cbma$ and $\varepsilon$ are
not physical as they change when redefining the Higgs basis field $\mathcal{H}_2\to -\mathcal{H}_2$.
However, the product $\varepsilon\,\cbma$ is invariant with respect to this sign change and hence is a physical quantity.  Moreover, if $\sbma\cbma\neq 0$ then
\eq{zee6cs} implies that $Z_6\neq 0$ and $\varepsilon \cbma<0$ in the convention for $\beta-\alpha$ adopted above.\footnote{Although it is possible to tune the parameters of the 2HDM such that $m_h=m_H$, this parameter choice does not yield a phenomenologically viable scenario, and is thus excluded from further consideration.}
In this case, \eq{twosigns} yields $\varepsilon=\sgn Z_6$, and it follows that
\beq \label{cbmasign}
\varepsilon\,\cbma=-|\cbma|\,.
\eeq

Given the values of $\beta-\alpha$ and the masses of $h$, $H$, $A$ and $H^\pm$, four of the seven
real Higgs basis parameters $Z_i$ are determined:\footnote{Note that \eq{zee6cs} is consistent with the condition $\varepsilon \cbma<0$ in the convention for $\beta-\alpha$ adopted above where $0\leq\sbma\leq 1$.}
\beqa
Z_1 v^2&=& m_h^2 s^2_{\beta-\alpha}+m_H^2 c^2_{\beta-\alpha}\,, \label{zee1cs}\\
Z_4 v^2 &=& m_h^2c_{\beta-\alpha}^2+ m_H^2s_{\beta-\alpha}^2+m_A^2-2m^2_{H^\pm}\,,\\
Z_5 v^2 &=& m_h^2  c^2_{\beta-\alpha} + m_H^2 s^2_{\beta-\alpha} -m_A^2\,, \\
Z_6 v^2 &=& -(m_H^2-m_h^2)\sbma\cbma\,. \label{zee6cs}
\eeqa

Formally, the CP-conserving 2HDM is defined such that the only source of CP violation enters via the nontrivial phase of the CKM matrix $K$ that appears in the respective interactions of the $W^\pm$ and the $H^\pm$ with fermion pairs.  In particular, in
the CP-conserving 2HDM,
the flavor-alignment parameters are real.
The Yukawa couplings given in \eq{YUK4} then take the following form,

\beqa
\!\!\!\!\!\!\!\!\!\!
-\mathscr{L}_Y &=& \frac{1}{v}\sum_{F=U,D,E} \overline F  M_F\bigl[\sbma
-a^F|\cbma|  \bigr] Fh
-\frac{1}{v}\sum_{F=U,D,E} \varepsilon\,\overline F  M_F\bigl[|\cbma|
+a^F\sbma\bigr]FH
 \nonumber \\
  &&
-\frac{i}{v}\sum_{F=U,D,E}\varepsilon_F\,\varepsilon\,a^F\, \overline F  M_F
 \gamma\ls{5} FA
 \nonumber \\
&& +\frac{\sqrt{2}}{v}\,\varepsilon\, \biggl\{\overline U\bigl[a^{D}KM_DP_R-a^{U}M_U KP_L\bigr] DH^+
+a^{E} \overline{N} M_EP_R EH^+ +{\rm
h.c.}\biggr\},
\label{YUK5}
\eeqa
where $\varepsilon$ is defined in \eq{twosigns} and we have introduced the notation,
\beq \label{ef}
\varepsilon_F=\begin{cases} +1 & \quad \text{for $F=U$}\,,\\
-1 & \quad \text{for $F=D, E$}\,.\end{cases}
\eeq
In this paper, we shall interpret LHC searches for new scalar states in terms of the CP-conserving A2HDM.  Thus, we shall employ the Yukawa couplings of \eq{YUK5} in which the real parameters $a^U$ and $a^D$ can take either sign and only the absolute value of $\cbma$ is physical, in light of \eq{cbmasign}.\footnote{Of course, the explicit factors of $\varepsilon$ are not physical as previously noted. Indeed, $\varepsilon$
can always be absorbed into the definitions of the $H$, $A$ and $H^\pm$ fields.}

If the absence of neutral Higgs mediated FCNCs is enforced naturally via a symmetry (which may be softly broken by dimension-2 squared-mass terms), then one should impose a
 $\mathbb{Z}_2$ symmetry on the dimension-4 terms of the Higgs Lagrangian in the $\Phi$-basis as specified in Table~\ref{Tab:type}, which implies that $\lambda_6=\lambda_7=0$.   It is convenient to define the quantity,
 \beq \label{eq:TZ2}
 T_{Z_2}\equiv \bigl|(Z_1-Z_2)[Z_1 Z_7+Z_2 Z_6-(Z_3+Z_4+Z_5)(Z_6+Z_7)]+2(Z_6+Z_7)^2(Z_6-Z_7)\bigr|\,.
 \eeq
Applying \eq{cond6} to the real Higgs basis of a CP-conserving 2HDM, it follows that the real Higgs basis parameters satisfy $T_{Z_2}=0$ if and only if a $\mathbb{Z}_2$ symmetry is present in some scalar field basis.

Moreover,
it is conventional to rephase the $\Phi$-basis scalar fields such that $\xi=0$ (i.e., the vevs are real and nonnegative), in which case one can identify $e^{i(\xi+\eta)}=\varepsilon$
and  $\tan\beta\equiv \langle\Phi_2^0\rangle/\langle\Phi_1^0\rangle$ [cf.~\eq{veedef}].
In particular, the CP-conserving Type-I, II, X and~Y 2HDMs are special cases of the A2HDM, where we can identify the corresponding real flavor-alignment parameters as follows,
\beqa
&&
\text{Type-I:~~~\,$a^U=a^D=a^E=\varepsilon \cot\beta$.} \label{typeone} \\
&&
\text{Type-II:~~~\!$a^U=\varepsilon\cot\beta$ and $a^D=a^E=-\varepsilon\tan\beta$.}\label{typetwo} \\
&&
\text{Type-Y:~~~$a^U=a^E=\varepsilon\cot\beta$ and $a^D=-\varepsilon\tan\beta$.}\label{typewhy} \\
&&
\text{Type-X:~~~$a^U=a^D=\varepsilon\cot\beta$ and $a^E=-\varepsilon\tan\beta$.\phantom{xxxxxxxxxxxx}}\label{typeex}
\eeqa
Inserting the CP-conserving Type~I or Type II values of the flavor-alignment parameters in \eq{YUK5}
and writing the Yukawa couplings of $h$ and $H$ in terms of $\cbma$ rather than its absolute value [using \eq{cbmasign}],
we see that the factors of $\varepsilon$ now cancel exactly, as they must since there is no remaining two-fold ambiguity in defining the real Higgs basis once the ratio of vevs has been chosen to be non-negative.  In particular, in the conventions of the Type-I, II, X and~Y 2HDMs adopted above (where $\xi=0$), the sign of $\cbma$ is now a physical parameter.  Moreover, the sign $\varepsilon$ is now fixed as determined by \eq{cbmasign}.  We shall adopt this approach in the Type-I 2HDM benchmark presented in Section~\ref{bench1}.

In contrast, a less common approach for examining the Type-I, II, X and Y limits of the CP-conserving A2HDM is to allow for both values of $e^{i\xi}=\pm 1$.  In this case, one can extend the definition of $\beta$ such that $-\half\pi\leq\beta\leq\half\pi$, in which case the parameter $\tan\beta$ can be of either sign.
\Eqst{typeone}{typeex} remain valid, but now we see that neither $\tan\beta$ nor $\cbma$ is physical (since both change sign when redefining the Higgs basis field $\mathcal{H}_2\to -\mathcal{H}_2$), although
the product  $\cbma\tan\beta$ is physical.\footnote{In this case, one could adopt a convention where $\cbma$ is always negative while allowing for both signs of $\tan\beta$, which is equivalent to employing \eq{YUK5} with $\varepsilon=+1$ [in light of \eq{cbmasign}].}

\section{Hints for heavy neutral scalars}
\label{sec:anomalies}

The ATLAS and CMS Collaborations have initiated dedicated searches for new elementary scalar bosons, which if discovered would signal the existence of an extended Higgs sector beyond the SM.
If evidence for new scalar states were discovered, then it would be tempting to provide a theoretical framework in interpreting the discoveries.   For example, if both neutral and (singly) charged scalars were observed, a compelling framework for the extended Higgs sector would be to posit the existence of multiple generations of hypercharge one electroweak doublets.  In this paper, we shall employ the 2HDM in the analysis of evidence for the presence of new scalar states.

However, despite its relative simplicity, the 2HDM in its most general form has too many new parameters.  Two simplifications can be invoked to reduce the number of parameters to a more practical number.   First, we shall impose flavor-aligned Yukawa couplings as discussed in Section~\ref{2HDMYUKS}.    The requirement of approximate flavor-aligned Yukawa couplings is
a consequence of the experimentally observed suppression of FCNCs.  Note that we shall not assume a priori that the flavor alignment is a consequence of a symmetry, which is often imposed as a theoretical requirement to avoid unnatural fine tuning in the Yukawa sector.  In our view, it is better to allow this question to be addressed by future experimental studies if the existence of an extended Higgs sector is confirmed.
Second, we shall impose a CP symmetry on the scalar potential, as discussed in Section~\ref{CP2HDM}, and on the Yukawa interactions of the neutral scalars.

Given the wide ranging search for new scalars at the LHC, one expects from
the statistical fluctuations inherent in the data that a few 3$\sigma$ ``signals'' should appear that {\em na\"{\i}vely} could be interpreted as evidence for new scalar states.  Any such fluctuation can be interpreted as a local excess above the expected SM backgrounds.
But, due to the look elsewhere effect~\cite{Gross:2010qma},
the interpretation of the events above background as a global excess is much less significant (typically in the range of
$1\sigma$~to~$2\sigma$).

Nevertheless, if new scalar states exist in a mass range accessible to the LHC, then they should show up initially as small excesses in the ATLAS and CMS searches.   Of course, if any such excesses truly represent physics beyond the Standard Model, then the significance of the initially observed small excesses will grow with the accumulation of more data.   In our view, focusing on any one particular excess is most likely to reveal a statistical fluctuation rather than a true signal.   It is likely to be more fruitful to focus on multiple excesses that can be interpreted within a single theoretical framework.

Following this strategy, we have examined a variety of excesses that have been reported in searches for new scalars by ATLAS and CMS.   Based on a number of excesses (with local significances around 3$\sigma$), we have constructed two scenarios that have a natural interpretation within the CP conserving flavor-aligned A2HDM framework.   We view this exercise as instructive rather than making a claim for an actual discovery.   One particular benefit of our approach is that, by imposing the more generic flavor aligned Yukawa couplings (as opposed to the more constrained Types I, II, X or Y scenarios), we can let the data decide whether the more constrained versions of the parameter space are viable.

One noteworthy excess has been reported by the ATLAS Collaboration (based on 139$~{\rm fb}^{-1}$ of data), which is interpreted as the simultaneous observation of two heavy scalars~\cite{ATLAS:2020gxx}.  First, a heavy CP-odd scalar $A$ is produced by gluon-gluon fusion.  It then decays to a $Z$ boson
 and a heavy CP-even scalar $H$ (i.e., $A\to ZH$), where the $Z$ is detected via its leptonic decay  ($e^+e^-$ and $\mu^+\mu^-$) and the
 $H$ decays directly via $H\to b\bar{b}$.  The ATLAS Collaboration reports that ``the most significant excess for the gluon-gluon fusion production signal assumption is at the ($m_A$, $m_H$) = (610, 290) GeV signal point, for which the local (global) significance is 3.1 (1.3) standard deviations.''   The ATLAS heat plots provided in Fig.~9 of Ref.~\cite{ATLAS:2020gxx} yield a 95\% CL upper limit of $\sigma\times{\rm BR}(A\to ZH)\times {\rm BR}(H\to b\bar{b})\lsim 0.08$ [$0.03$]~pb observed [expected] for ($m_A$, $m_H$) = (610, 290) GeV.\footnote{The ATLAS $\sigma\times{\rm BR}$ limits quoted here were obtained with the help of the Digital Color Meter application that resides on all Mac computers.}
This result suggests a scenario of interest, which we denote as Scenario~1.  In this scenario, we shall interpret the ATLAS excess in the $\ell^+\ell^- b\bar{b}$ channel as having arisen in the A2HDM where $\sigma\times{\rm BR}(A\to ZH)\times {\rm BR}(H\to b\bar{b})\simeq 0.06\pm 0.02$~pb for $m_A=610$~GeV and $m_H=290$~GeV.
In this case, all neutral Higgs boson masses are fixed, whereas the charged Higgs mass will be taken as a free parameter to be scanned.

The ATLAS Collaboration has also searched for evidence of $A\to ZH$ via the  $\ell^+\ell^- b\bar{b}$ channel, where the $A$ is produced in association with $b\bar{b}$.  In contrast to the gluon fusion channel, only a small excess is seen for ($m_A$, $m_H$) = (610, 290)~GeV~\cite{ATLAS:2020gxx} in the $b$-associated production channel, corresponding to a 95\% CL upper limit of $\sigma\times{\rm BR}(A\to ZH)\times {\rm BR}(H\to b\bar{b})\lsim 0.05$ [$0.03$]~pb observed [expected].
Indeed, we shall impose this 95\% CL upper limit constraint on our A2HDM interpretation of
Scenario~1.
On the other hand,
the ATLAS Collaboration reports that ``for $b$-associated production, the most significant excess is at the ($m_A$, $m_H$) = (440, 220) GeV signal point, for which the local (global) significance is 3.1 (1.3) standard deviations.''
The heat plots provided in Fig.~9 of Ref.~\cite{ATLAS:2020gxx} yield a 95\% CL upper limit of $\sigma\times{\rm BR}(A\to ZH)\times {\rm BR}(H\to b\bar{b})\lsim 0.15$ [$0.07$]~pb observed [expected] for ($m_A$, $m_H$) = (440, 220) GeV.
However, if one were to interpret this latter excess as an A2HDM  signal
where $m_A=440$~GeV and $m_H=220$~GeV with a cross section of order 0.1~pb, then
in the A2HDM parameter regime of interest
one would predict a production cross section for $gg$ fusion production of~$A$ that is excluded by Fig.~9(b) of Ref.~\cite{ATLAS:2020gxx}.  Thus, we shall henceforth assume that
any ATLAS excess in the $b$-associated production channel reported in Ref.~\cite{ATLAS:2020gxx} is a statistical fluctuation.

Finally, in a recent paper, the ATLAS Collaborations reports on a search for $gg\to A\to ZH$  where $H\to hh$ and both final state Higgs bosons decay to $b\bar{b}$~\cite{ATLAS:2022fpx}.
No excess is seen in this channel for ($m_A$, $m_H$) = (610, 290) GeV, corresponding to
 a 95\% CL upper limit of $\sigma\times{\rm BR}(A\to ZH\to Zhh\to Zb\bar{b}b\bar{b})\lsim 10$ [$8$] fb observed [expected].\footnote{These results are obtained from
the heat plots provided in Fig.~9 of Ref.~\cite{ATLAS:2022fpx}, under the assumption
that the width of $A$ is narrow compared to the experimental mass resolution.}  Again, we shall assume that the most significant excess observed for a somewhat higher value of $m_A$ in this search channel is a statistical fluctuation.

Other interesting excesses have also been observed in the search for a scalar mass around 400~GeV.
For example, the ATLAS Collaboration conducted a search for a heavy scalar that decays into $\tau^+\tau^-$~\cite{ATLAS:2020zms}.   Two possible production mechanisms were considered: gluon-gluon fusion (ggF) into a heavy scalar (singly produced) and the production of a heavy scalar in association with a $b\bar{b}$ pair ($bb\phi$).   Employing $139~{\rm fb}^{-1}$ of data and scanning over possible scalar masses, the ATLAS Collaboration concluded that
``for ggF, the lowest local $p_0$, the probability that the background can produce a fluctuation greater than the excess observed in data, is 0.014 ($2.2\sigma$) at $m_\phi=400$~GeV, while for $bb\phi$ production it is 0.003 ($2.7\sigma$) at $m_\phi=400$~GeV.''  The best fit point at  $m_\phi=400$~GeV corresponds roughly to~\cite{ATLAStautau}:
\beqa
&&\sigma_{\rm ggF}\times {\rm BR}(\phi\to\tau^+\tau^-)\simeq 20^{+37}_{-20}~{\rm fb}\,,\label{ATLAStautau1}\\
&&\sigma_{bb\phi}\times {\rm BR}(\phi\to\tau^+\tau^-)\simeq 38^{+30}_{-29}~{\rm fb}\,,\label{ATLAStautau2}
\eeqa
where the error bars correspond to a region of 68\% CL.\footnote{These results are more correctly represented (as shown in Fig.~08 of Ref.~\cite{ATLAStautau}) as ellipses in a two dimensional plane of $\sigma_{\rm ggF}\times {\rm BR}(\phi\to\tau^+\tau^-)$ vs.~$\sigma_{bb\phi}\times {\rm BR}(\phi\to\tau^+\tau^-)$.}
The ATLAS analysis does not distinguish between a CP-even scalar ($H$) and a CP-odd scalar ($A$).

The CMS Collaboration has also searched for a heavy scalar that decays into $\tau^+\tau^-$ and does not see any excess~\cite{CMS:2022goy}.   Employing $138~{\rm fb}^{-1}$ of data,
maximum likelihood estimates and 95\% CL contours obtained from scans of the signal likelihood are provided in Fig.~11 of Ref.~\cite{CMS:2022goy} for selected values of the scalar mass.   Although the 95\% CL contours for $m_\phi=400$~GeV are not given, we can interpolate between two mass values provided ($m_\phi=250$ and 500~GeV) by making use of the observed 95\% CL upper limits on the product of the cross sections (ggF and $bb\phi$) and branching fraction for $\phi\to\tau^+\tau^-$ as a function of $m_\phi$ shown in Fig.~10 of Ref.~\cite{CMS:2022goy}.
We then obtain an approximate 95\%~CL exclusion contour in the plane of
$\sigma_{\rm ggF}{\rm BR}(\phi\to\tau^+\tau^-)$ vs.~$\sigma_{bb\phi}{\rm BR}(\phi\to\tau^+\tau^-)$ for $m_\phi=400$~GeV, which is exhibited by the dashed contour in Fig.~\ref{fig:scen2_Atau} (see Section~\ref{sec:scen2}).   In particular, the $95\%$~CL exclusion contours intersect the
$x$ and $y$ axis at the following values:\footnote{To obtain \eqs{limitgg}{limitbb} from Fig.~10 of Ref.~\cite{CMS:2022goy}, we have multiplied the latter results by $(5.99/3.84)^{1/2}$ (see Table 40.2 of Ref.~\cite{ParticleDataGroup:2022pth}),
which yields the 95\%~CL values for the $x$ and $y$ axis intercepts of the two-dimensional contour.}
\beqa
&&\sigma_{\rm ggF}\times {\rm BR}(\phi\to\tau^+\tau^-)\lesssim 40~{\rm fb}\quad \text{at 95\%~CL}\,,\label{limitgg}\\
&&\sigma_{bb\phi}\times {\rm BR}(\phi\to\tau^+\tau^-)\lesssim 40~{\rm fb}\quad \text{at 95\%~CL}\,.
\label{limitbb}
\eeqa
These results still leave room for the possibility that the ATLAS excess in the $\tau^+\tau^-$ channel could correspond to a real signal.

\begin{table}[t!]
\centering
\begin{tabular}{|c|c|c||c|}
\hline
& $m_H$ & $m_A$ & A2HDM interpretation of the excess of events \TBstrut \\
\hline
{\bf Scenario 1} & 290~GeV & 610~GeV & $\phm gg\rightarrow A\rightarrow ZH$, where $H \rightarrow b\bar{b}$ and $Z\to \ell^+\ell^-$~~\cite{ATLAS:2020gxx} $\phm$ \TBstrut  \\
\hline
\hline
&&& $gg\to A\to \tau^+\tau^-$~~\cite{ATLAStautau}\Tstrut \\[4pt]
{\bf Scenario 2} & $>450$~GeV & 400~GeV & $gg\to b\bar{b}A$, where $A\to \tau^+\tau^-$~~\cite{ATLAStautau} \\[4pt]
&&& $gg\to A\to t\bar{t}$~~\cite{CMS:2019pzc}\Bstrut  \\
\hline
\end{tabular}
\caption{\small  Scenario 1 is based on an ATLAS excess of events with a local (global) significance of 3.1(1.3)$\sigma$, where the observed lepton is $\ell=e$, $\mu$.
Scenario 2 is based on an ATLAS excess of $\tau^+\tau^-$ events with an invariant mass of around 400 GeV, with local significances of 2.7$\sigma$ in the $gg$ fusion production channel and
2.2$\sigma$ in the $b$-associated production channel.  The CMS Collaboration sees no excess, but still leaves some room for a possible signal.   The ATLAS data does not distinguish between $H$ and $A$ production.  However, the CMS excess of $t\bar{t}$ events with an invariant mass of around 400 GeV, with a local (global) significance of 3.5(1.9)$\sigma$, favors identifying the excess at 400 GeV with $A$ production.
\label{scenarios}}
\centering
\end{table}

Meanwhile, the CMS Collaboration has also searched for a heavy CP-even scalar and/or a heavy CP-odd scalar decaying into $t\bar{t}$~\cite{CMS:2019pzc}.  Based on 35.9$~{\rm fb}^{-1}$ of data and scanning over a range of masses, the CMS Collaboration concludes that ``a moderate signal-like deviation compatible with an $A$ boson with a mass of 400 GeV is observed'' with a local significance of $3.5\pm 0.3$ standard deviations.
Taking the look-elsewhere effect into consideration, the CMS Collaboration quotes a global significance of 1.9 standard deviations, which corresponds to a $p$-value of 0.028.

Taken together, the ATLAS and CMS results quoted above suggest a second scenario of interest, which we denote as Scenario~2.  In this scenario, we interpret the ATLAS excess in the $\tau^+\tau^-$ channel and the CMS excess in the $t\bar{t}$ channel as having arisen in the A2HDM where $m_A=400$~GeV.  Note that in Scenario 2, we take $m_H>450$~GeV to avoid the possibility of both $H$ and $A$ contributing to the $\tau^+\tau^-$ and $t\bar{t}$ excesses.  Smaller values of $m_H$ are excluded based on the ATLAS search for $gg\to A\to ZH$~\cite{ATLAS:2020gxx}.
 The two A2HDM scenarios introduced above are summarized in Table~\ref{scenarios}.

Other extended Higgs model interpretations of the $\tau^+\tau^-$ and $t\bar{t}$ excesses described above have been previously considered in the literature.
 Both the $\tau^+\tau^-$ and $t\bar{t}$ excesses were suggested as possible evidence for a CP-odd Higgs boson of mass 400 GeV in
 Refs.~\cite{Richard:2020cav,Arganda:2021yms,Biekotter:2021qbc}. In particular, the analysis of Ref.~\cite{Richard:2020cav} interprets the $\tau^+\tau^-$ and $t\bar{t}$ excesses
 without specifying the extended Higgs sector model (along with a comment that these excesses cannot be accommodated by the Higgs sector of the MSSM).\footnote{See also Ref.~\cite{Kundu:2022bpy} for a follow-up to the results initially reported in Ref.~\cite{Richard:2020cav}, which invokes an extended Higgs sector with SU(2)$_L$ triplets~\cite{Georgi:1985nv}.}   The authors of Ref.~\cite{Arganda:2021yms} employ an enlarged composite Higgs sector where the observed Higgs boson is identified as a pseudo-Goldstone boson state.   Finally, the CP-odd scalar of mass 400 GeV is interpreted in the context of an extended Higgs sector consisting of two SU(2)$_L$ doublets and an SU(2)$_L$ singlet.
 More recently, the authors of Ref.~\cite{Botella:2023tiw} assert that it is possible to accommodate the ATLAS $\tau^+\tau^-$ excess in a Type I 2HDM while at the same time explaining the muon $g-2$ anomaly~\cite{Muong-2:2021ojo}.


\section{Parameter Scans}
\label{sec:scans}

We performed parameter scans to probe the CP-conserving A2HDM parameter space for regions which predict the excess of events for Scenarios 1 and 2 and are otherwise consistent with other relevant, measured Higgs observables. A generic point in the CP-conserving A2HDM parameter space is uniquely specified by the following set of real parameters,
\beq
\upsilon, m_{h}, m_{H}, m_{A}, m_{H^{\pm}},  |c_{\beta-\alpha}|, Z_2, Z_3, Z_7, a^U, a^D, a^E.
\eeq
As indicated in \eqst{zee1cs}{zee6cs}, the masses of $h$, $H$, and $A$ can then be used to obtain $Z_1$, $Z_4$, $Z_5$ and $|Z_6|$.
In Scenario 1, after fixing $\upsilon$, $m_h$, $m_H$ and $m_A$, we scan over the remaining parameters.
In Scenario 2, $m_H$ is not fixed so we must scan over this parameter as well.

Given the dimensionality of the parameter space, we can perform a more efficient scan by first
imposing a variety of preemptive conditions
before calculating observables of interest.  The following theoretical and experimental constraints are applied:
\vskip 0.1in
1.~The scalar potential is bounded from below.

\vskip 0.1in
\noindent
By imposing the conditions given by Ivanov in Refs.~\cite{Ivanov:2006yq,Ivanov:2007de}, one obtains
 inequalities involving combinations of the quartic couplings $Z_i$ so that the potential cannot be negative when the scalar fields assume arbitrarily large values.
 \vskip 0.1in
 2.~Tree-level unitarity and perturbativity
  \vskip 0.1in

 \noindent
 Imposing tree-level unitarity on scattering processes involving the scalar bosons and the gauge bosons yields upper bounds on various combinations of the $Z_i$~\cite{Ginzburg:2005dt,Kanemura:2015ska,Bahl:2022lio}.
These conditions are best imposed numerically, as the corresponding algebraic constraints are quite complicated.   Perturbativity is a less well-defined constraint.   In the interest of performing a more efficient scan (without removing a significant number of viable parameter points), we shall first restrict the values of the quartic couplings such that $|Z_i|\leq 4\pi$ before applying the tree-level unitarity conditions.
\vskip 0.1in
3. The lightest Higgs scalar is SM-like
 \vskip 0.1in

\noindent
In Scenarios 1 and 2, the lightest scalar state of the A2HDM is identified with the \pagebreak
observed Higgs boson with mass 125 GeV.   The LHC Higgs data imply that the experimental properties of the 125 GeV mass scalar closely approximate those of the SM Higgs boson~\cite{ATLAS:2022vkf,CMS:2022dwd}.   That is, the observed Higgs signal strengths $\mu^X_{i}$, each of which is defined as 
the ratio of the measured value of the Higgs cross section $\sigma_i(h)$ times the branching ratio into particular final states $X\widebar{X}$
and the product of the corresponding SM predicted values, are close to unity.  
In the A2HDM, the Higgs signal strengths are given by
\beq \label{muvalues}
\mu^X_i \equiv \frac{\sigma_i(h)_{\text{A2HDM}}\times\text{BR}(h\rightarrow X\widebar{X})_{\text{A2HDM}}}{\sigma_i(h)_{\text{SM}}\times\text{BR}(h\rightarrow X\widebar{X})_{\text{SM}}} \sim 1.
\eeq
We have employed the values of the $\mu^X_i$ defined by \eq{muvalues} to place bounds on the magnitudes of the A2HDM coupling modifiers $|f_{\phi,F}|$ [cf.~\eqst{modify1}{modify3}] by demanding that the observed Higgs signal strengths are within the 1$\sigma$ experimental errors for $X=\gamma$ and $Z$, and within two times
the experimental errors for $X=W$, $\tau$, and $b$.
As a secondary check, we have independently verified that the bounds obtained on magnitudes of the coupling modifiers
are sensible in light of the results obtained from the public code
\texttt{HiggsSignals}~\cite{Bahl:2022igd}.\footnote{The public code \texttt{HiggsSignals} makes use of 
experimental data consisting of $n=131$ Higgs boson observables and 
obtains the $\chi^2$ for a particular theoretical model.  For example, a comparison of the experimental data with the SM yields $\chi^2/n\simeq 0.94$.   We have checked that for Scenarios 1 and 2, \texttt{HiggsSignals} yields $0.93\lsim\chi^2/n\lsim 1.02$ for the A2HDM parameter points that are consistent with the constraints imposed in this section.}

\vskip 0.1in
4. Precision electroweak constraints on the oblique parameters
 \vskip 0.1in

\noindent
The experimentally measured oblique parameter $T$ provides the most stringent constraint on the A2HDM parameters~\cite{Haber:2010bw}.  In particular, we impose the requirement that $T$ is within $2\sigma$ of
its central value as reported in Ref.~\cite{ParticleDataGroup:2022pth}.  Imposing this constraint in our parameter scans restricts the possible values of the charged Higgs mass and forces $m_{H^\pm}$ to be within about $\pm60$ GeV of either $m_A$ or $m_H$.  See Appendix~\ref{sec:Tparm} for further details.

\vskip 0.1in
5. Heavy flavor constraints
 \vskip 0.1in

\noindent
A comprehensive treatment of flavor constraints on the A2HDM parameters has been provided in
Ref.~\cite{Enomoto:2015wbn}.   We ensure that these constraints are satisfied in our scans.  In a large
 fraction of the parameter space, the most significant constraint derives from the observed $\bar{B}\to
 X_s\gamma$ decay, which is derived from a computation of the ratio of rates for $b\to s\gamma$ in the A2HDM
 and the SM, respectively.   The results of the latter are summarized in Appendix~\ref{sec:bsg}, and the
 implications for the A2HDM parameter space are discussed there.

 Among the other flavor observables discussed in Ref.~\cite{Enomoto:2015wbn}, we found that the $B_s$--$\overline{B}_s$ mixing data provided an additional constraint that limited the value of $|a^U|$ as a function of the charged Higgs mass.\footnote{In Ref.\cite{UTfit:2022hsi}, the observed value of $\Delta M_{B_s}=17.241(20)~{\rm ps}^{-1}$ obtained from $B_s$--$\overline{B}_s$ oscillation data is compared with the Standard Model prediction, $17.94(69)~{\rm ps}^{-1}$ based on a global fit of flavor observables.  Since the error in the Standard Model prediction is still considerably larger than the precision of the measured value, we chose to identify the 2$\sigma$ error in the theoretical prediction as the upper limit to the contribution to $|\Delta M_{B_s}|$ of new physics beyond the Standard Model.}
 In particular, within the parameter intervals exhibited in Table~\ref{parms},
a rough upper bound of $|a^U|\lesssim 1$ was obtained [although this value can be lower for charged Higgs masses below 500 GeV as shown in Fig.~\ref{fig:aU_mch}(a)], whereas no constraint on $a^D$ was obtained in the scanning region of interest.  As a result, the excluded region of the A2HDM parameter space due to the constraint on $\Delta M_{B_s}$ roughly coincides with the corresponding excluded region of the Type-I 2HDM
 in the $m_{H^\pm}$ vs.~$\tan\beta$ plane exhibited in Fig.~8 of Ref.~\cite{Arbey:2017gmh} after identifying $\tan\beta = 1/|a^U|$.
In particular, imposing the $\Delta M_{B_s}$ constraint can eliminate regions of the 2AHDM parameter space with values of $|a^U|\gsim 1$ that otherwise would be allowed by the $b\to s\gamma$ constraints shown in Fig.~\ref{fig:adau}.  Further details can be found in 
Appendix~\ref{sec:MsubS}.

\vskip 0.1in
6. Searches for new elementary scalar states at the LHC
 \vskip 0.1in

\noindent
Numerous other searches for new elementary scalar states at the LHC have been performed by the ATLAS and CMS Collaborations.   We have checked that the 95\% CL upper limits obtained in all these searches are consistent with the data excesses that we have identified in Scenarios 1 and 2, respectively.\footnote{After completing this work, we used the public code \texttt{HiggsBounds}~\cite{Bahl:2022igd} to ensure that we had not missed any heavy scalar production channels at the LHC that would eliminate the scan points analyzed in Sections~\ref{sec:scen1}
and~\ref{sec:scen2}.}

\vskip 0.1in
Having imposed all the conditions listed above,
we scan values of $|c_{\beta-\alpha}|$ from 0 to 0.45 to keep the Yukawa couplings roughly within 20\% of their SM model values (in light of the precision LHC Higgs data).
The parameter $Z_7$ only enters $Hhh$ and $HH^+H^-$ couplings multiplied by $c^2_{\beta-\alpha}$ which is extremely small for most of the valid points in the scan.  Hence, our results are quite insensitive to the choice of $Z_7$.
As noted above, the charged Higgs mass must be nearly degenerate in mass with $m_A$ or $m_H$ due to the $T$ parameter constraint.
We also require that
$m_{H^\pm}>m_t+m_b$ (thereby avoiding the possibility of an on-shell $t\to H^+b$ decay, which is not observed at the LHC).
Hence, we initially allow for values of $m_{H^{\pm}}$ in the range of $[200,1000]~{\rm GeV}$. We scan over up-type Yukawa coupling parameters $a^U$ between $-1.5$ and $1.5$ and down- and lepton-type Yukawa coupling parameters $a^D$ and $a^E$ between $-50$ and 50, which ensures the absence of Landau poles significantly below the Planck scale~\cite{Gori:2017qwg}.  A summary of the parameter intervals employed in our scans is presented in Table~\ref{parms}.

\begin{table}[h!]  \centering
\begin{tabular}{|c|c|}
\multicolumn{2}{c}{\bf Parameter Intervals Scanned}\\[4pt]
\hline
$|c_{\beta-\alpha}|$ & $0\,,\,0.45$ \TBstrut\\
$Z_2$ & $0\,,\,4.5$\TBstrut \\
$Z_3$ & $-2\,,\,12$ \TBstrut\\
$Z_7$ & $-10\,,\,10$\TBstrut \\
$m_{H^\pm}$ & $200\,,\,1000$ GeV\TBstrut \\
$a^U$ & $-1.5\,,\,1.5$ \TBstrut\\
$a^D$ & $-50\,,\,50$ \TBstrut\\
$a^E$ & $-50\,,\,50$ \TBstrut\\
\hline
\end{tabular}
\caption{\small Parameter intervals scanned in the analysis of Scenarios 1 and 2.
\label{parms}}
\centering
\end{table}

We compute cross sections and branching ratios in the CP-conserving A2HDM for a large number of randomly generated parameter combinations. Cross sections
for the CP-even scalars $h$ (125 GeV) and $H$ (290 GeV) and for the CP-odd scalar $A$ (610 or 400 GeV) are calculated utilizing the public code
\texttt{SusHi} \cite{Harlander:2012pb}. We employ the SM mode of \texttt{SusHi} to compute gluon-gluon fusion and $b$-associated production cross sections
as in the SM but with $m_h$ replaced with a specified Higgs mass $m_{\phi}$ (where $\phi=h,H,A$), and with the top or bottom quark couplings switched
on or off.
 We also make use of the \texttt{SusHi} ``pseudoscalar" setting for computing the $A$ production cross sections -- these processes insert an
 $i\gamma\ls{5}$ into the SM calculation.

For the gluon-gluon fusion production mechanism we neglect the first and second generations of quarks due to their insignificant couplings to the Higgs bosons and separately compute the contributions from top loops, bottom loops, and the interference term. The interference term $\sigma_{\text{int}}$ is then extracted by subtracting the contributions from top loops $\sigma_t$ and bottom loops $\sigma_b$ from the total cross section $\sigma_{\text{tot}}$ for each mass eigenstate $\phi$,
\beq
\sigma(gg\phi)_{\text{int}} = \sigma(gg\phi)_{\text{tot}} - \sigma(gg\phi)_t - \sigma(gg\phi)_b.
\eeq

The gluon-gluon fusion production cross sections for Higgs bosons in the CP-conserving A2HDM are then obtained by inserting the appropriate vertex modifications to the top, bottom, and interference terms,
\beq
\sigma(gg\phi)_{\text{A2HDM}}
=\sigma(gg\phi)_{t}(f_{\phi,U})^2
+\sigma(gg\phi)_{b}(f_{\phi,D})^2
+\sigma(gg\phi)_{\text{int}}(f_{\phi,U}) (f_{\phi,D}),
\eeq
where the A2HDM coupling modifiers $f_{\phi,F}$ (for $F=U,D,E$) are obtained from \eq{YUK5},
\beqa
f_{h,F} &=& s_{\beta-\alpha} - |c_{\beta-\alpha}|  a^F,  \label{modify1}\\
f_{H,F}&=& -|c_{\beta-\alpha}|+ s_{\beta-\alpha}  a^F, \label{modify2} \\
f_{A,F}&=& -\epsilon_F\, a^F, \label{modify3}
\eeqa
after absorbing the factors of $\varepsilon$ into the definitions of the corresponding scalar fields.

The $b$-associated production cross sections are also computed using
the SM mode of \texttt{SusHi}
(which implements the five-flavor scheme) and adjusted to the A2HDM by inserting the coupling modifier,
\beq
\sigma(gg\rightarrow bb\phi)_{\text{A2HDM}} = \sigma(gg\rightarrow bb\phi)(f_{\phi,D})^2.
\eeq

We then compute branching ratios and check them with our modified version of the public code \texttt{2HDMC} \cite{Eriksson:2009ws}, which we have extended to handle the A2HDM Yukawa coupling matrices. Finally, for each parameter point, we use the above result to compute the number of events expected for the two scenarios exhibited in Table~\ref{scenarios}.

\section{Scenario 1: $m_A = 610$ GeV, $m_H = 290$ GeV}
\label{sec:scen1}

Consider the CP-conserving 2HDM with a CP-odd scalar mass of $m_A=610$~GeV and two
CP-even scalars with masses $m_h=125$~GeV and $m_H = 290$ GeV (where $h$ is identified as the Higgs boson observed in LHC data).
This scenario is motivated by a slight
excess of events over the expected backgrounds in the ATLAS search for gluon fusion production of a CP-odd scalar $A$ followed by its decay to $ZH$, with the subsequent
decay of $H\to b\bar{b}$ and $Z\to \ell^+\ell^-$ ($\ell=e$, $\mu$).  Using the results of Ref.~\cite{ATLAS:2020gxx}, we propose that the excess of events noted above lies in the range of
\beq
0.04 \leq \sigma(gg\rightarrow A)\,{\rm BR}(A\rightarrow ZH) \,{\rm BR}(H\rightarrow b\bar{b})  \leq   0.08\,\,\mbox{pb}\,, \label{A1}
\eeq
consistent with the 95\% CL upper limit reported in Ref.~\cite{ATLAS:2020gxx}.
In contrast, no significant excess was seen by the ATLAS Collaboration
for $(m_A,m_H)=(610, 290)$~GeV in the
$Zb\bar{b}b\bar{b}$ final state that can arise either via the $b$-associated production process
$gg\to b\bar{b}A$ where $A\to ZH\to Zb\bar{b}$,
or via $gg\to A\to ZH\to Zhh\to Zb\bar{b}b\bar{b}$,
as discussed in Section~\ref{sec:anomalies}.

A separate search for the gluon fusion production of a CP-even scalar $H$ also did not yield any significant excesses.  For example, the ATLAS Collaboration reported in
Ref.~\cite{ATLAS052}
a 95\%~CL upper limit for $\sigma(gg\to H\to hh)<0.241$~pb for $m_H\sim 290$~GeV.\footnote{The corresponding limit obtained by the CMS Collaboration in Ref.~\cite{CMS-21-011} is nearly identical to the quoted ATLAS result.}
In this section, we shall probe the A2HDM parameter space that is consistent with the Scenario 1 interpretation of the ATLAS excess specified in \eq{A1}, subject to the following conditions,
\beqa
\sigma(gg\rightarrow Ab\bar{b})\,{\rm BR}(A\rightarrow ZH) \,{\rm BR}(H\rightarrow b\bar{b}) &\leq  & 0.05\,\,\mbox{pb}\,,\label{A2} \\
\sigma(gg\to A){\rm BR}(A\to ZH){\rm BR}(H\to hh\to b\bar{b}b\bar{b}) &\leq  &0.01~\text{pb}
\,,\label{A2p} \\
    \sigma(gg\rightarrow H)\,{\rm BR}(H\rightarrow hh) & \leq  & 0.241\,\,\mbox{pb}\,, \label{A3}
\eeqa
derived from the 95\% CL upper limits reported in Refs.~\cite{ATLAS:2020gxx,ATLAS:2022fpx,ATLAS052}, respectively.

\subsection{A2HDM Interpretation of Scenario 1}

\begin{figure}[ht!]
  \centering
  \includegraphics[height=8cm,angle=0]{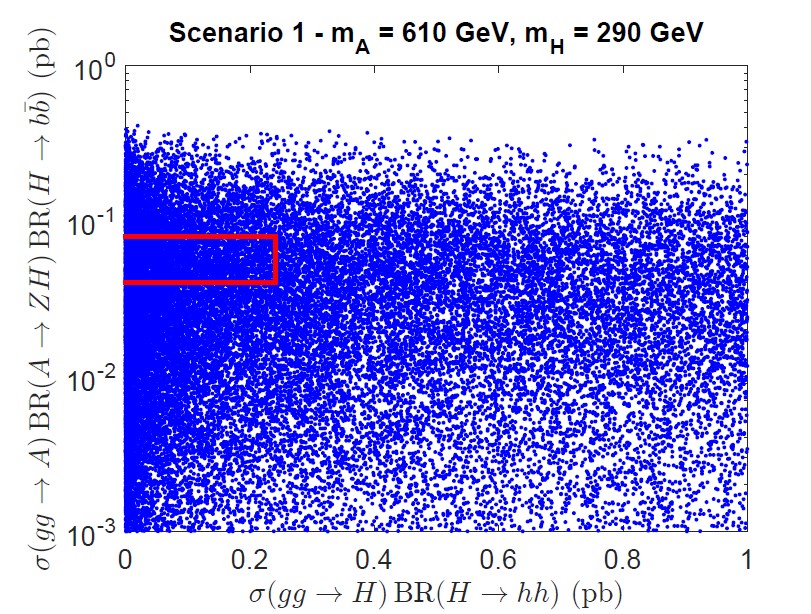}
  \caption{\small Signal rates for the production of $A$ and $H$ in Scenario 1.  The A2HDM parameter generation complies
    with all theoretical and experimental constraints elucidated in Section~\ref{sec:scans}.   Events inside the rectangular box with red boundaries are consistent with the conditions specified in \eqst{A1}{A3},
    which correspond to a  small excess of events
   reported in Ref.~\cite{ATLAS:2020gxx} and  interpreted as $gg\to A\to ZH\to b\bar{b}\ell^+\ell^-$ (with no significant excess in the corresponding $b$-associated production of $A$),
   and with the nonobservation of $gg\to H\to hh$ derived from the 95\%~CL upper limits obtained in Refs.~\cite{ATLAS052,CMS-21-011}.
  \label{fig:scen1}}
\end{figure}

In Fig.~\ref{fig:scen1}, we exhibit the Scenario 1 cross sections for $gg\to A\to ZH\to Zb\bar{b}$ and $gg\to H\to hh$ obtained by a scan of the A2HDM parameter space, while respecting
the theoretical and experimental constraints elucidated in Section~\ref{sec:scans}.  By imposing the conditions specified in \eqst{A1}{A3}, the points of the A2HDM parameter scan are restricted to lie inside the rectangular box with red boundaries.

To provide some understanding of the expected magnitudes of the various relevant cross sections, it is straightforward to apply the program \texttt{SusHi}~\cite{Harlander:2012pb}
to the A2HDM cross sections for $m_A=610$~GeV and $m_H= 290$~GeV as a function of the flavor-alignment parameters $a^U$ and
$a^D$.  By rescaling the cross sections obtained by \texttt{SusHi} for a SM-like Higgs boson with the masses indicated above (and employing the
\texttt{SusHi} switch for a pseudoscalar $Af\bar{f}$ coupling) with the appropriate squared flavor-alignment parameter, and neglecting the interference of the
$t$ and $b$ quark loops in $gg\to A$, $H$ (which is less than a $5\%$ effect), we obtain
\begin{align}
\sigma(gg\rightarrow A) &\simeq\,2.85\,|a^U|^2\; \mbox{pb},\qquad\qquad
\sigma(gg\rightarrow H) \simeq\,10.28\,|a^U|^2\;\mbox{pb}, \label{eq:sig1}\\
\sigma(gg\rightarrow b\bar{b}A) &\simeq\,1.11\,|a^D|^2\;\mbox{{\rm fb}}, \qquad\quad
\,\,\sigma(gg \rightarrow b\bar{b}H) \simeq\,26.98\,|a^D|^2\;\mbox{{\rm fb}}\,.
\label{eq:sig2}
\end{align}

\begin{figure}[t!]
\begin{tabular}{cc}
\includegraphics[height=6cm,angle=0]{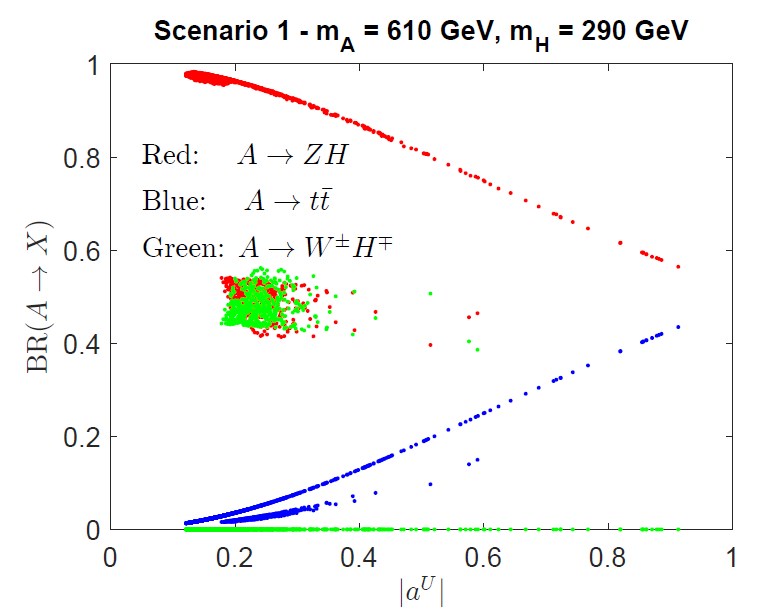}&
\includegraphics[height=6cm,angle=0]{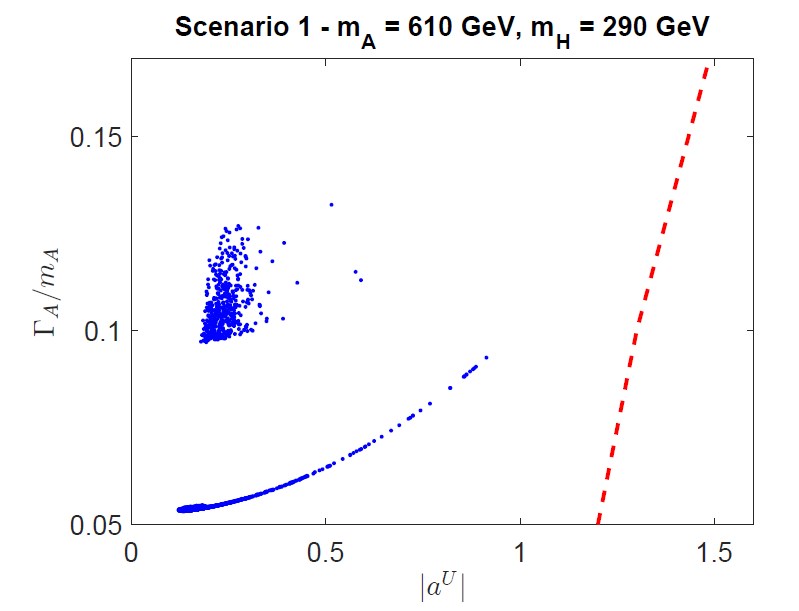}\\
 (a) & (b)
\end{tabular}
\caption{\small Results of a scan over A2HDM parameter points in Scenario 1 that
satisfy the theoretical and experimental constraints elucidated in Section~\ref{sec:scans}
and the constraints of \eqst{A1}{A3}.  Panel
(a) shows the branching ratios for the decay of $A$ into its dominant final state channels, and panel (b) shows the total $A$ width divided
 by its mass, as functions of $|a^U|$.
Applying the 95\% CL upper limit for $\sigma(gg \to A\to t\bar{t})$ reported in Ref.~\cite{CMS:2019pzc} eliminates points to the right of the
 dashed line shown in panel~(b) from further consideration.
The two distinct branches of a given color of points correspond to the cases where the decay $A\to W^\pm H^\mp$ is either kinematically
allowed or disallowed.
}
\label{fig:BRA}
\end{figure}

\begin{figure}[t!]
\begin{tabular}{cc}
\includegraphics[height=6cm,angle=0]{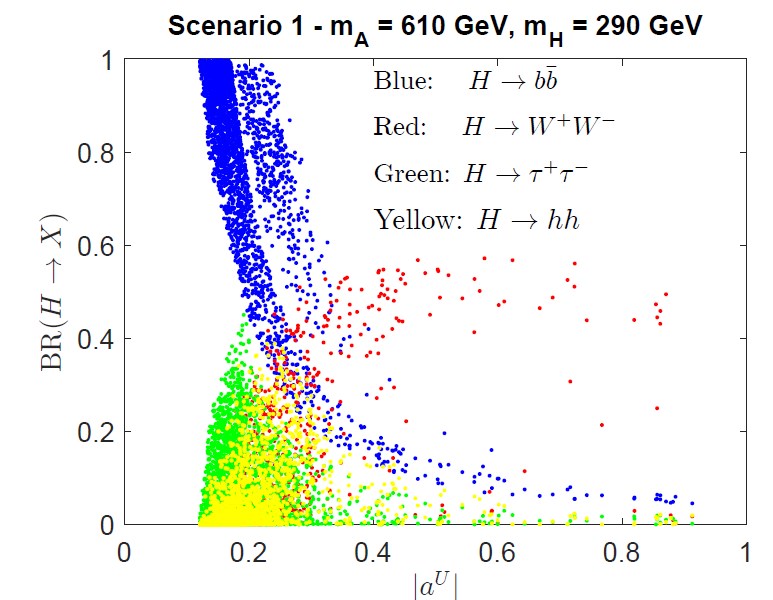}&
\includegraphics[height=6cm,angle=0]{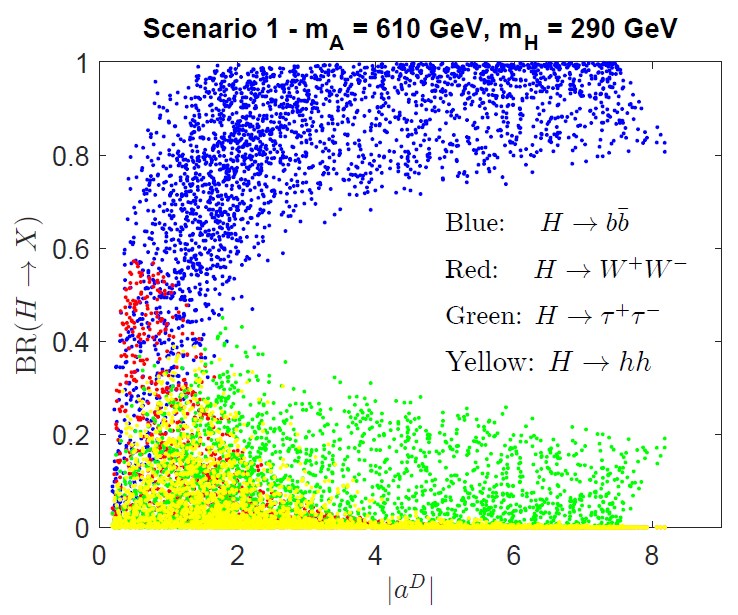}\\
(a) & (b)
\end{tabular}
\caption{\small
Results of a scan over A2HDM parameter points in Scenario 1 that
satisfy the theoretical and experimental constraints elucidated in Section~\ref{sec:scans}
and the constraints of \eqst{A1}{A3}.
Panels (a) and (b) exhibit the branching ratios of the four main $H$ decay channels
as functions of $|a^U|$ (allowing~$a^D$ to vary) and $|a^D|$ (allowing $a^U$ to vary), respectively.
\\[-20pt]}
\label{fig:scen1_BRH}
\end{figure}

In Fig.~\ref{fig:BRA}(a), we exhibit the three main decay channels of $A$ in Scenario 1.
One of the experimental
 constraints that has been applied derives from the search for $A\to t\bar{t}$ reported by the CMS Collaboration in Ref.~\cite{CMS:2019pzc}.   Indeed, the latter constraint rules out any points to the right of the dashed line shown in Fig.~\ref{fig:BRA}(b).
In light of Fig.~\ref{fig:BRA}(a), $A\rightarrow ZH$ is a dominant decay channel
for the CP-odd scalar in Scenario 1 if $|a^U|\lsim1$.
In Fig.~\ref{fig:scen1_BRH}, we exhibit four dominant $H$ decay channels: $H\rightarrow b\bar{b}$ (blue), $H\rightarrow W^+W^-$ (red), $H\rightarrow \tau^+\tau^-$ (green) and $H\rightarrow hh$ (yellow).\footnote{The $H\rightarrow ZZ$ channel would also have a significant decay fraction, but it suffices to show results for $W^+W^-$, since both branching ratios are related via constant phase space factors so that ${\rm BR}(H\rightarrow WW)
\simeq 2.23\, {\rm BR}(H\rightarrow ZZ)$.}
For example, using \eq{eq:sig1} it follows that if $|a^U|\sim 1$ then ${\rm BR}(H\rightarrow hh)$ cannot be larger than about 2\% in light of \eq{A3}, as shown in Fig.~\ref{fig:scen1_BRH}(a).
For example, \eq{eq:sig1} implies that for values of
$|a^U|\sim1$, a signal of about 0.06 pb in the channel
$gg\rightarrow A\rightarrow ZH\rightarrow Zb\bar{b}$ can be achieved with ${\rm BR}(H\to b\bar{b})\sim 5$\%.

\begin{figure}[h!]
\begin{tabular}{cc}
\includegraphics[height=6cm,angle=0]{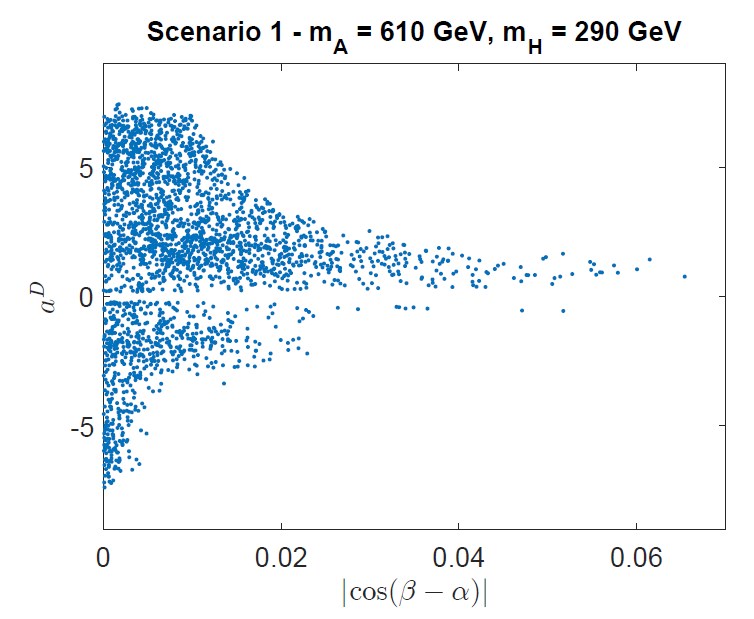}&
\includegraphics[height=6cm,angle=0]{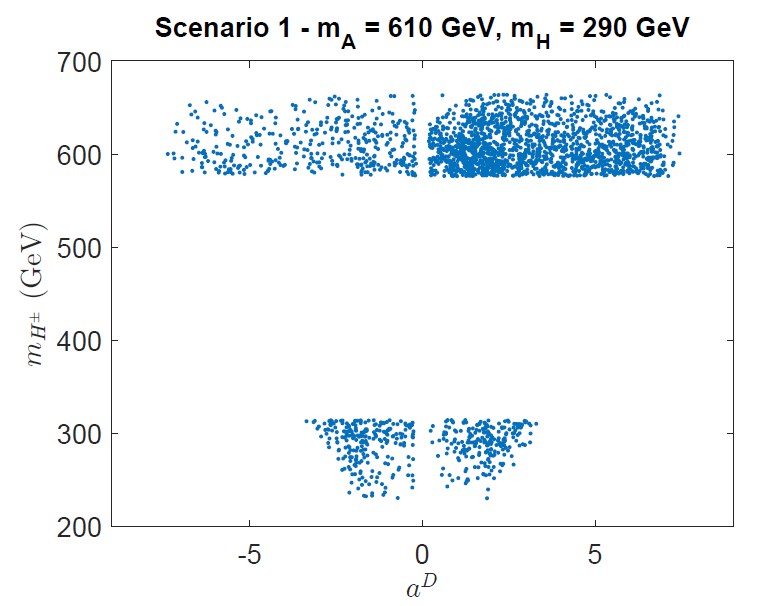}\\
 (a) & (b)
\end{tabular}
\caption{\small Results of a scan over A2HDM parameter points in Scenario 1 that
satisfy the theoretical and experimental constraints elucidated in Section~\ref{sec:scans}
and the constraints of \eqst{A1}{A3}.  Panel
(a) show the allowed values of the flavor-alignment parameter $a^D$ as a function of $|\hspace{-0.125em}\cos(\beta-\alpha)|$ and panel
(b) shows the charged Higgs mass as a function of $a^D$.
}
\label{fig:param}
\end{figure}

We now proceed to analyze more closely the A2HDM parameter space in Scenario 1.
The scatter plot exhibited in
Fig.~\ref{fig:param}(a) indicates that $|\cbma|\lsim 0.07$ (i.e., close to the Higgs alignment limit) as a consequence of the precision Higgs data as discussed in Section~\ref{sec:scans}.
Moreover, there is no preference in Scenario 1 for a specific sign of the alignment
parameter $a^D$.   The region of $a^D$ consistent with the ATLAS excess in the production of $A$ via gluon fusion at $(m_A, m_H)= (610, 290)$~GeV is roughly
$0.18 \lesssim |a^D| \lesssim 8.2$.
In particular, both very small and very large values of $|a^D|$ are excluded.
Both regimes can be understood as follows.
From \eq{YUK5}, we see that
the down-quark coupling modifier for $H$ is
\beq
f_{H,d} \,=-\varepsilon(|c_{\beta-\alpha}|+s_{\beta-\alpha}  a^D) \,.
\eeq
This means that in the approximate Higgs alignment limit of the A2HDM,
where $|\cbma|\ll 1$, the
coupling modifier $f_{H,d}$ will grow with $a^D$.  Consequently, $a^D \sim 0$ is excluded, since
the $A$ signal excess depends on a branching ratio of the decay $H\rightarrow b\bar{b}$ that is not too small.  On the other hand, for very large absolute values of $|a^D|$, both
${\rm BR}(H\rightarrow b\bar{b})$ and $\sigma(gg\to Ab\bar{b})$ become too large to be consistent with
the constraints exhibited in \eqst{A1}{A3}.

The allowed regions of the charged Higgs mass are exhibited
Fig.~\ref{fig:param}(b).  The charged Higgs mass is constrained to lie either in the
interval $575 \lesssim m_{H^\pm} \lesssim 665$ GeV or $220 \lesssim m_{H^\pm} \lesssim 330$ GeV,
as anticipated in the discussion of precision electroweak constraints in Section~\ref{sec:scans}.
In particular, since there is a sizable mass splitting
between $H$ and $A$ in Scenario 1, in order for the oblique $T$ parameter to be suitably small in the approximate Higgs alignment limit, it is necessary for the charged Higgs
mass to be roughly degenerate in mass with either $A$ or $H$, as explained in Appendix~\ref{sec:Tparm}.

The parameter points in Scenario 1 must also be consistent with the non-observation of $H$ and $A$ production and subsequent decay in other search channels examined
by the ATLAS collaboration in their analysis of the Run 2 LHC data.
In particular, for $(m_A, m_H)=(610, 290)$ GeV, there is no experimental evidence (yet) for $H$ and $A$ production followed by the subsequent decay of the scalar into
other final states, such as $ZZ$, $W^+W^-$, $\tau^+\tau^-$ and $\gamma\gamma$.

\begin{figure}[b!]
\begin{tabular}{cc}
\includegraphics[height=6cm,angle=0]{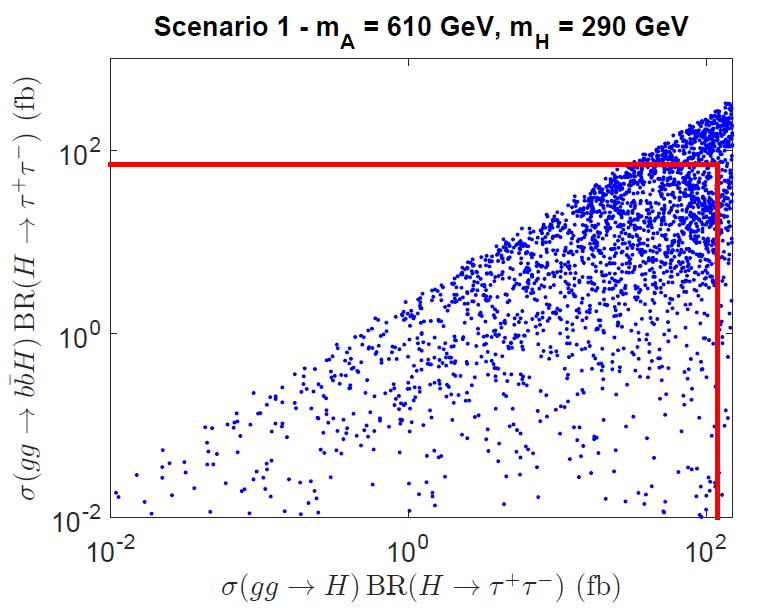}&
\includegraphics[height=6cm,angle=0]{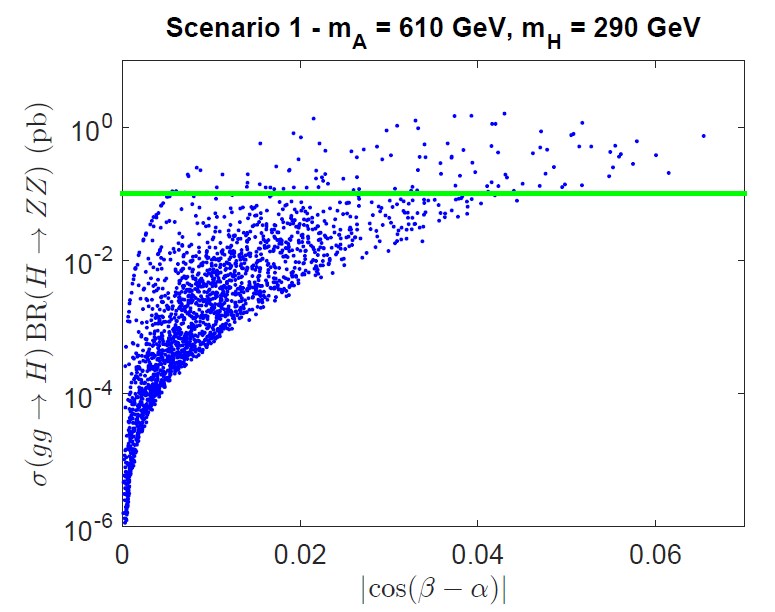}\\
 (a) & (b)
\end{tabular}
\caption{\small Results of a scan over A2HDM parameter points in Scenario 1 that
satisfy the theoretical and experimental constraints elucidated in Section~\ref{sec:scans}
and the constraints of \eqst{A1}{A3}.
Panel (a) shows the cross sections for gluon fusion production and $b$-associated production of $H$
multiplied by ${\rm BR}(H\to\tau^+\tau^-)$, and panel
(b) shows $\sigma(gg\to H)\times{\rm BR}(H\to ZZ)$
as a function of $|\hspace{-0.125em}\cos(\beta-\alpha)|$.
The vertical and horizontal red lines shown in panel (a) correspond to the
$1\sigma$ exclusion limit (for $m_H=300$~GeV) exhibited in \eqs{tau1}{tau2}, respectively~\cite{ATLAStautau}. The blue points in panel (b)
correspond to the points in panel (a) that lie below the red horizontal line and to the left of the red vertical line,
and the horizontal green line shown in panel (b), which corresponds to \eq{zeelim}, reflects the
95\% CL upper bound reported in Ref.~\cite{ATLAS:2020tlo}.
}
\label{fig:tau}
\end{figure}
The ditau
channel is one of the main search channels for new scalars, and there are stringent constraints from LHC Run 2 data.
In particular, we can use the results of the ATLAS Collaboration reported in Fig.~08 of Ref.~\cite{ATLAStautau} to determine whether $m_H\sim 290$~GeV and/or $m_A\sim 610$~GeV might not have already been
excluded by existing ditau data.

We begin by examining the implications of the ditau data for the production of $H$.
In Fig.~\ref{fig:tau} (a), we show the results of a scan over A2HDM parameter points in Scenario 1 that
satisfy the theoretical and experimental constraints elucidated in Section~\ref{sec:scans}
and the constraints of \eqst{A1}{A3}.
Although most of these points survive, some of them are excluded by the
ATLAS search for $H\to \tau^+\tau^-$.  In particular, applying the limits reported by
the ATLAS Collaboration in Ref.~\cite{ATLAStautau} for $m_H=300$~GeV, we find that
at the 1$\sigma$ exclusion level,
\beqa
\sigma(gg\rightarrow H)\,{\rm BR}(H\rightarrow\tau^+\tau^-) &\leq& 120 \,\,\mbox{{\rm fb}} \,, \label{tau1}\\
\sigma(gg\to b\bar{b}H)\,{\rm BR}(H\rightarrow\tau^+\tau^-) &\leq& 70 \,\,\mbox{{\rm fb}}\,.
\label{tau2}
\eeqa
The exclusion limits of \eqs{tau1}{tau2} correspond to the vertical and the horizontal red lines, respectively, of Fig.~\ref{fig:tau} (a).  Thus, all points exhibited in Fig.~\ref{fig:tau} (a) that lie below the horizontal red line and to the left of the vertical red line are consistent with the exclusion limits obtained in Ref.~\cite{ATLAStautau}.
After excluding the points that lie above the horizontal red line and/or to the right of the vertical red line in Fig.~\ref{fig:tau}(a), we increased the scanning statistics to
obtain the plot of $\sigma(gg\to H)\times{\rm BR}(H\to ZZ)$ as a function of $\cos(\beta-\alpha)$
shown in Fig.~\ref{fig:tau}(b).
The horizontal green line shown in Fig.~\ref{fig:tau}(b) corresponds to
the 95\%~CL exclusion limit,
\beq \label{zeelim}
\sigma(gg\rightarrow H)\,{\rm BR}(H\rightarrow ZZ)\leq 0.1~\text{pb}\,,
\eeq
reported by the ATLAS Collaboration in Ref.~\cite{ATLAS:2020tlo}.
Note that the points that lie below the horizontal green line of Fig.~\ref{fig:tau}(b), which take into account the constraint of \eq{zeelim}, satisfy $\cos(\beta - \alpha)\sim 0.045$.
Imposing
similar constraints from the exclusion limit for $\sigma(gg\to H\to W^+ W^-)$ obtained in Ref.~\cite{ATLAS:2017uhp} does not eliminate additional points from the A2HDM scan exhibited in  Fig.~\ref{fig:tau}.

\begin{figure}
  \begin{tabular}{cc}
\includegraphics[height=6cm,angle=0]{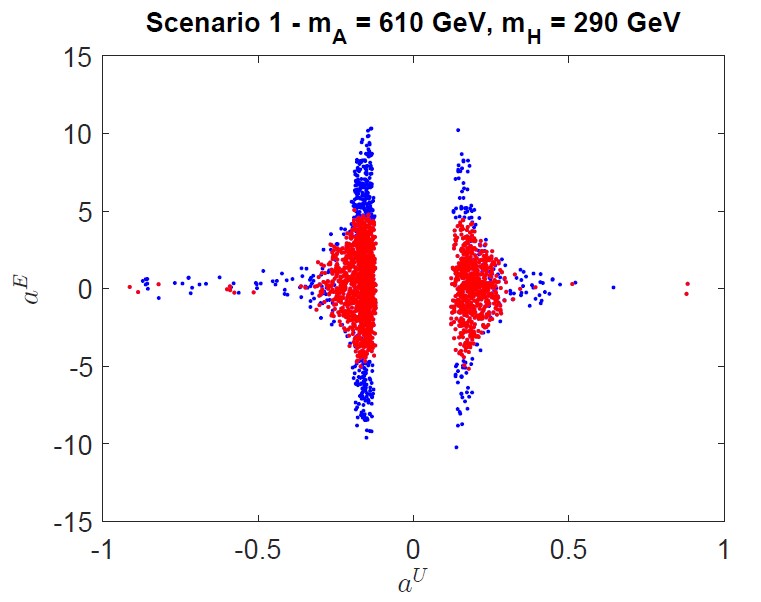} &
 \includegraphics[height=6cm,angle=0]{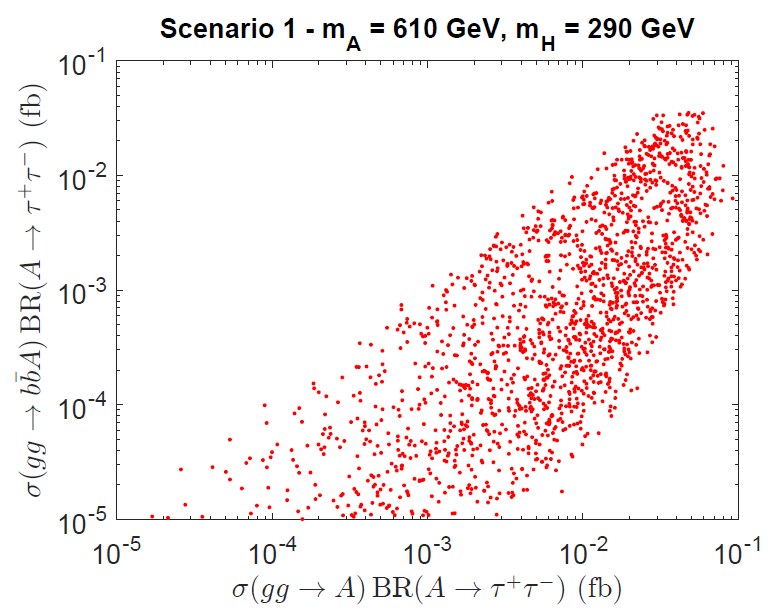} \\
(a) & (b)
\end{tabular}
  \caption{\small Results of a scan over A2HDM parameter points in Scenario 1 that
satisfy the theoretical and experimental constraints elucidated in Section~\ref{sec:scans}
and the constraints of \eqst{A1}{A3}.   The values of the
 flavor-alignment parameters $a^E$ and $a^U$ are then plotted as blue points in panel (a).
 The red points of panel (a) correspond to those that survive the additional constraints of \eqst{tau1}{zeelim}.  The values of the cross sections for gluon fusion production and $b$-associated production of $A$ multiplied by ${\rm BR}(H\to\tau^+\tau^-)$, subject to the same constraints as the red points of panel (a), are plotted as red points in panel (b).}
   \label{fig:aEaU}
\end{figure}

The ditau exclusion limits of \eqs{tau1}{tau2}
have implications for
the flavor-alignment parameter $a^E$, as exhibited in
Fig.~\ref{fig:aEaU}(a).
In particular, after applying all the relevant constraints,
$|a^E| \lesssim 5$, where the precise upper bound can be even more restrictive
depending on the value of $|a^U|$.   Note that the A2HDM parameter space of interest is insensitive to the sign of $a^U$.  This is a consequence of the fact that precision Higgs data is compatible with either sign of $a^U$.   In light of the top quark Yukawa modifier,
\beq
f_{h,t} \,=\, s_{\beta-\alpha} - |c_{\beta-\alpha}| a^U \,,
\label{eq:fht}
\eeq
we see that due to the constraint on $c_{\beta-\alpha}$ exhibited in Fig.~\ref{fig:tau}(a) and the upper
bound of $|a^U|\lsim 1.0$,\footnote{We scanned over values of $a^U$ such that $|a^U|\lsim 1.5$ in order to avoid a Landau pole significantly below the Planck scale.  As noted in Section~\ref{parms}, the constraint of $\Delta M_{B_s}$ reduces this upper bound to roughly $|a^U|\lsim 1$ (with a weak dependence on the charged Higgs mass).}
 it follows that
$f_{h,t}\simeq 1$, regardless of
the sign of $a^U$.
A second feature that is evident from Fig.~\ref{fig:aEaU}(a) is the existence of a lower bound,
$|a^U| \gtrsim 0.11$.  This arises from the fact that the gluon fusion production cross section of $A$ relies on the $At\bar{t}$ coupling that is proportional to $a^U$ [cf.~\eq{eq:sig1}].   Thus, to be consistent with \eq{A1}, $|a^U|$ cannot be arbitrarily small.

\begin{figure}[b!]
  \begin{tabular}{cc}
\includegraphics[height=6cm,angle=0]{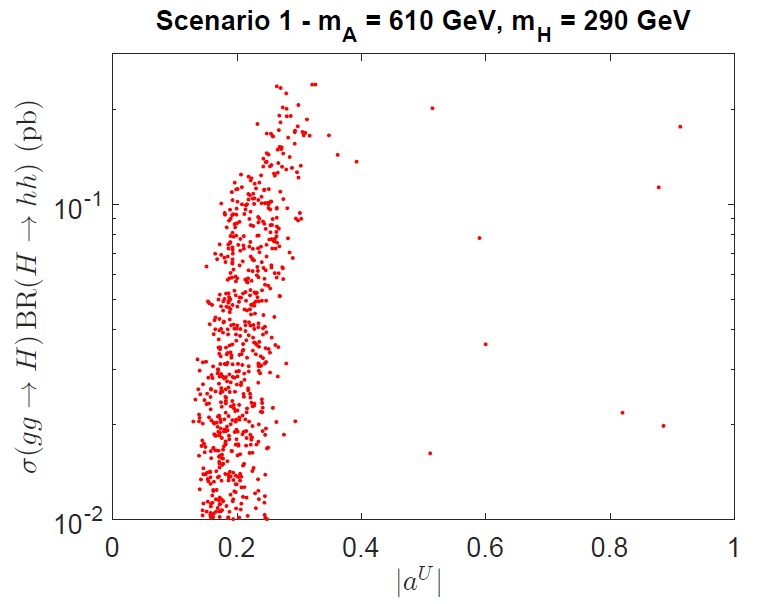} &
\includegraphics[height=6cm,angle=0]{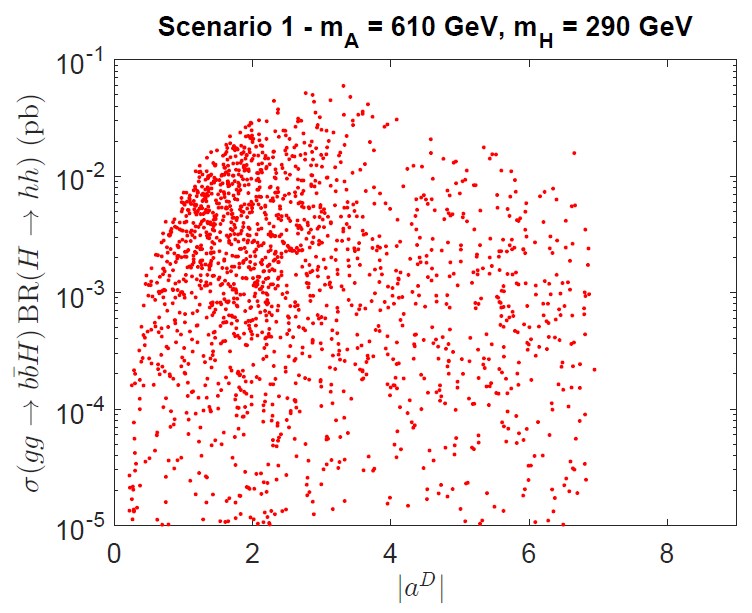} \\
(a) & (b)
\end{tabular}
 \caption{\small
 Results of a scan over A2HDM parameter points in Scenario 1 that
satisfy the theoretical and experimental constraints elucidated in Section~\ref{sec:scans}
and the constraints of \eqst{A1}{A3} and \eqst{tau1}{zeelim}.  In panel (a) the values of the cross section for gluon fusion production
of $H$ multiplied by ${\rm BR}(H\to hh)$ are plotted as a function of $|a^U|$. In panel (b), the
values of the cross section for $b$-associated production of $H$ multiplied by ${\rm BR}(H\to hh)$ are plotted as a function of $|a^D|$.
  \label{fig:bbHhh}}
\end{figure}

Next, consider the implications of the ditau data for the production of $A$.   In particular, applying the limits reported by
the ATLAS Collaboration in Ref.~\cite{ATLAStautau} for $m_A=600$~GeV, we find that
at the 2$\sigma$ exclusion level,
\beqa
\sigma(gg\rightarrow A)\,{\rm BR}(A\rightarrow\tau^+\tau^-) &\leq& 12.5 \,\,{\rm fb} \,, \label{tau3} \\
\sigma(gg\rightarrow b\bar{b}A)\,{\rm BR}(A\rightarrow\tau^+\tau^-) &\leq& 4.5 \,\,{\rm fb}\,.  \label{tau4}
\eeqa
No further reduction of the A2HDM parameter space follows after imposing the $A\to \tau^+\tau^-$ exclusion limits of \eqs{tau3}{tau4}.  In particular, the red parameter points of Fig.~\ref{fig:aEaU}(b) lie considerably below the ATLAS exclusion limits quoted above, which is not surprising given that the $\tau^+\tau^-$ decay mode is subdominant.  Indeed, in light of Fig.~\ref{fig:BRA}(a), if $m_{H^\pm}\sim m_A$ then the dominant $A$ decay channels are $A\rightarrow ZH$ and $A\rightarrow t\bar{t}$, which account for more than 99\% of all $A$ decays.  As expected, the branching ratio for $A\rightarrow t\bar{t}$ increases with $|a^U|$, since the $At\bar{t}$ coupling is proportional to $a^U$.  Moreover, as illustrated in Fig.~\ref{fig:BRA}(b), the total width of $A$ also increases with $|a^U|$.   In Scenario 1, an experimental upper limit on $|a^U|$ can be deduced based solely on the absence of evidence for $A\to t\bar{t}$ for $m_A=610$~GeV, as reported in Ref.~\cite{CMS:2019pzc} by the CMS Collaboration, where the region of $|a^U|\gsim 1.2$ is excluded. However, after applying the full constraints of the Scenario 1 parameter scan, the results of Fig.~\ref{fig:aEaU}(a) imply a somewhat stronger constraint of $|a^U|\lsim 0.9$.   Nevertheless there is still room for a large enough value of $|a^U|$ to yield
a sizable $A\to t\bar{t}$ signal rate, which therefore
remains a tantalizing possibility to be probed in the near future at the LHC.

Having successfully accommodated the proposed signal of Scenario 1 [cf.~\eq{A1}] in the framework of the CP-conserving 2HDM, one can now propose additional Higgs signals that could be discovered in future LHC searches.  As an example,
we can provide the possible values of the cross section that are consistent with the ATLAS data excess given in \eq{A1}
for the gluon fusion production and the $b$-associated production of $H$, followed by its decay into a pair of Higgs bosons, which are exhibited in the plots shown in Fig.~\ref{fig:bbHhh}.
In particular, given that ${\rm BR}(h\to b\bar{b})\simeq 58\%$, the $b$-associated production process would yield spectacular events with six $b$-quarks in the final state.

Although the proposed signal of $A\to ZH$ specified in \eq{A1} can be successfully accommodated in the framework of the A2HDM, one can now ask whether
this signal is viable in two Higgs doublet models with natural flavor conservation.
In light of \eqst{typeone}{typeex}, the following four special cases of the A2HDM are of interest:
\beqa
&&
\text{Type-I:~~~\,$a^U=a^D=a^E$,} \qquad\qquad
\text{Type-X:~~~$a^U=a^D=-\frac{1}{a^E}$,}  \label{TypeoneX} \\
&&
\text{Type-II:~~~\!$-\frac{1}{a^U}=a^D=a^E$,} \qquad\quad
\text{Type-Y:~~~$a^U=-\frac{1}{a^D}=a^E$.}
\label{TypetwoY}
\eeqa
In all four cases above, the corresponding conditions are a consequence of a $\mathbb{Z}_2$ symmetry that is preserved by all dimension-4 terms of the Higgs Lagrangian.  Consequently, the condition $T_{Z_2}=0$ must also be satisfied [cf.~\eq{eq:TZ2}].

To investigate whether the surviving points of the A2HDM parameter scans presented above
are consistent with any of the relations given in \eqs{TypeoneX}{TypetwoY}, we introduce the following four quantities,
\beqa
T_I &\equiv& \left|1 - a^D/a^U\right| + \left|1 - a^E/a^U \right|\,+\,T_{Z_2}\,,\qquad\quad
T_X \equiv \left|1 - a^D/a^U\right| + \left|1 + a^E a^U \right|\,+\,T_{Z_2}\,, \label{TsubIX} \\
T_{I\!I} &\equiv& \left|1 + a^D a^U\right| + \left|1 - a^E/a^U \right| \,+\,T_{Z_2}\,,\qquad\quad\,\,\,
T_Y \equiv \left|1 + a^D a^U\right| + \left|1 - a^E/a^U \right| \,+\,T_{Z_2}\,.\label{TsubIIY}
\eeqa
By design, $T_I=0$ only for the Type-I 2HDM,  $T_{I\!I}=0$ only for the Type-II 2HDM,
 $T_Y=0$ only for the Type-Y 2HDM,  and $T_X=0$ only for the Type-X 2HDM.

 In light of Fig.~\ref{fig:param}(b), $m_{H^\pm}\lsim 670$~GeV for the A2HDM scan points that satisfy all of the Scenario~1 constraints.  However, as shown in Ref.~\cite{Misiak:2020vlo}, the constraint imposed by the $b\to s\gamma$ measurement yields $m_{H^\pm}\gsim 800$~GeV in the Type-II 2HDM, and the same rough upper bound applies to the Type-Y 2HDM~\cite{Arbey:2017gmh}.  Hence, Scenario 1 is incompatible with the Type-II and Type-Y 2HDM.  In contrast, the range of charged Higgs masses
 shown in Fig.~\ref{fig:param}(b) are not excluded by the $b\to s\gamma$ measurement in the Type I and  X 2HDM~\cite{Arbey:2017gmh}.  Thus, we now explore whether or not Scenario 1 is compatible with some range of parameters within the Type-I or X 2HDM.

\begin{figure}[t!]
\centering
\includegraphics[height=8cm,angle=0]{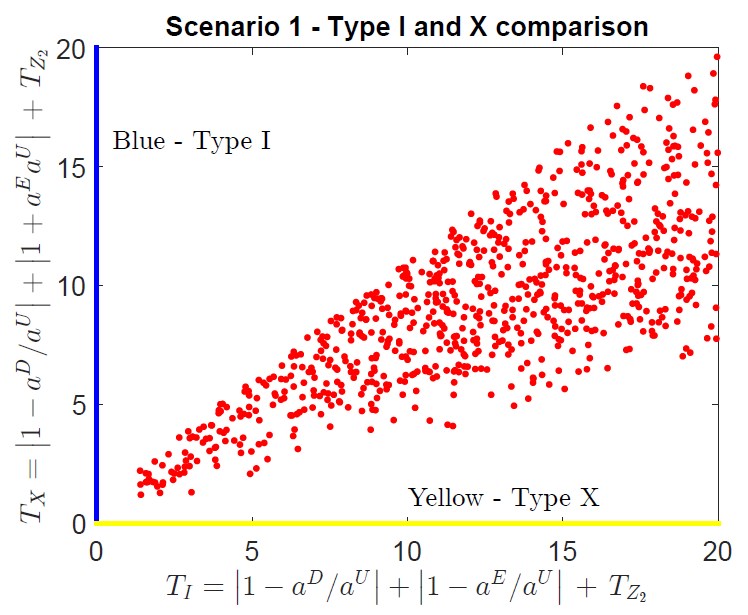}
  \caption{\small
  Results of a scan over A2HDM parameter points in Scenario 1 that
satisfy the theoretical and experimental constraints elucidated in Section~\ref{sec:scans}
and the constraints of \eqst{A1}{A3} and \eqst{tau1}{zeelim}.   We plot the values
of $T_I$ vs.~$T_X$ for all surviving scan points.   Points that lie along the horizontal (yellow) axis
would be consistent with a Type-X 2HDM.  Points that lie along the vertical (blue) axis would be consistent with a Type-I 2HDM.   }
  \label{fig:types}
\end{figure}

In Fig.~\ref{fig:types}, we have plotted the values of $T_I$ and $T_X$
for those A2HDM scan points that survive the theoretical and experimental constraints elucidated in Section~\ref{sec:scans} and the constraints of \eqst{A1}{A3} and \eqst{tau1}{zeelim}.
Given that the scan points of Fig.~\ref{fig:types} appear to be approaching the horizontal (Type-X) or the vertical (Type I) axes, it seems plausible (with a higher statistics scan) that Scenario~1 could be compatible within the Type-I and/or X 2HDM frameworks.   To investigate this possibility in more detail, we have performed
dedicated Type-I and Type-X scans for Scenario 1 and confirmed that it is indeed possible to fit all Scenario 1 constraints in the Type-I 2HDM but \textit{not} in the Type-X 2HDM.

The failure to find solutions within the Type-X 2HDM framework can be attributed to the constraints of the $H\to\tau^+\tau^-$ decays.  For all points that satisfied the constraints of
 eqs.~\eqref{A1}--\eqref{A3}, we obtained values of $\sigma(gg\rightarrow H)\,
 {\rm BR}(H\rightarrow\tau^+\tau^-)$ between 0.9 and 1.3 pb, in violation of the bound of
 eq.~\eqref{tau1}.  In particular, imposing the bounds of eqs.~\eqref{A1}--\eqref{A3} on
the Type-X (and Type-I) scans require values of $1/a^U = \tan\beta\gsim 2$, which implies that in the Type-X 2HDM we also have $a^E = \tan\beta\gsim 2$.  That is, the decay rate for
$H\rightarrow\tau^+\tau^-$ is enhanced by a factor of $\tan^2\beta$, thereby producing values of
 $\sigma(gg\rightarrow H \rightarrow\tau^+\tau^-)$ that exceed the observed upper bound given in \eq{tau1}.

\renewcommand{\arraystretch}{1.5}
\setlength{\tabcolsep}{10pt}
\begin{table}[t!]
\centering
\begin{tabular}{|c||c|c|}
\hline
& Type-I 2HDM & A2HDM  \TBstrut  \\
\hline\hline
 $\sigma(gg\rightarrow H)\,{\rm BR}(H\rightarrow\tau^+\tau^-)$ &  $\lsim 90$~fb & $\lsim 120$~fb \TBstrut   \\
 \hline
$\sigma(gg\rightarrow b\bar{b}H)\,{\rm BR}(H\rightarrow\tau^+\tau^-)$ &
 $\lsim 0.26$~fb &  $\lsim 70$~fb  \TBstrut  \\
 \hline
$\sigma(gg\rightarrow A)\,{\rm BR}(A\rightarrow\tau^+\tau^-)$ &
$\lsim 1.6\times 10^{-3}$~fb& $\lsim 0.1$~fb \TBstrut   \\
\hline
$\sigma(gg\rightarrow b\bar{b}A)\,{\rm BR}(A\rightarrow\tau^+\tau^-)$ &
 $\lsim 6.1\times 10^{-7}$~fb  & $\lsim 0.05$~fb  \TBstrut  \\
\hline
\end{tabular}
\caption{\small In the parameter scans subjected to all Scenario 1 constraints, the maximal values of $\sigma\times{\rm BR}$ is shown for four different production processes of neutral heavy scalars that decay to $\tau^+\tau^-$.   Results in the case of the A2HDM are taken from Figs.~\ref{fig:tau}(a)
and \ref{fig:aEaU}(b).  Results in the case of the Type-I 2HDM are obtained from a dedicated scan.
\label{tab:typeonecomp}}
\centering
\end{table}

Although it is possible to accommodate Scenario 1 in a Type-I 2HDM, more flexibility is achieved by employing the generic A2HDM framework.   For example, in Table~\ref{tab:typeonecomp}, we exhibit the maximal $\sigma\times{\rm BR}$ for the production of $H$ and $A$ (either via gluon fusion or via $b$-associated production) followed by the decay into $\tau^+\tau^-$.   If the Scenario 1 data excesses persist, the detection of the $\tau^+\tau^-$ decay mode in multiple channels would be inconsistent with a Type-I interpretation but could be compatible in the more general A2HDM framework.  This exercise illustrates the advantage of employing the A2HDM in analyzing evidence for the production of new scalar states.   In particular, the larger parameter space (relative to the special cases of the A2HDM corresponding to Type I, II, X, and Y Yukawa couplings) provides the freedom to
independently vary the $a^U$, $a^D$ and $a^E$ flavor-alignment parameters, thereby providing the A2HDM with greater flexibility in interpreting different signals of heavy scalar production and decay.

If the proposed signal of \eq{A1} is confirmed, then one should expect to discover a charged Higgs boson either in the mass range of  $[220\,,\,320]$~GeV or in the range of $[570\,,\,670]$ GeV,
as indicated in Fig.~\ref{fig:param}.
There is an extensive set of experimental results on searches for charged scalars in the literature.
The most recent results for charged Higgs boson searches at the LHC yield upper bounds on
$\sigma(pp\rightarrow H^\pm \rightarrow tb)$~\cite{CMS:2020imj}
and $\sigma(pp\rightarrow H^\pm \rightarrow \tau\nu)$~\cite{CMS:2019bfg}.\footnote{Here we use the notation $H^\pm\to tb$ to mean either $H^+\to t\bar{b}$ or $H^-\to \bar{t}b$. Likewise, $H^\pm \to\tau\nu$ denotes $H^+\to \tau^+\nu_\tau$ or $H^-\to \tau^-\bar{\nu}_\tau$.}
We have computed the decay
branching ratios for the charged scalar in the A2HDM, and we have used the results of the LHC Cross Section
Working Group~\cite{LHCXSWG} to obtain the LHC production cross section for $H^\pm$ as a function of the A2HDM flavor-alignment parameters.
The A2HDM parameter space obtained in our analysis of Scenario 1 is such that constraints derived from the non-observation of the charged Higgs boson
decaying to $\tau\nu$ are satisfied.  However,
scans with higher values of $|a^U|$ can
produce values of $\sigma(pp\rightarrow H^\pm \rightarrow tb)$ that lie above the observed bound
obtained by the CMS Collaboration using 35.9 ${\rm fb}^{-1}$ of data~\cite{CMS:2020imj}.
As a result, we obtain an upper bound on $|a^U|$ that
depends on the range of charged Higgs masses: $|a^U| \lesssim 0.5$ for $m_{H^\pm} \in [220\,,\,320]$~GeV and
$|a^U| \lesssim 1$ for $m_{H^\pm} \in [570\,,\,670]$~GeV.
These bounds on $|a^U|$ will improve when the full Run 2 dataset and future Run 3 data are analyzed,
with the real possibility of a discovery of the charged Higgs boson via its $tb$ decay mode.

\subsection{A2HDM Benchmarks for Scenario 1}
\label{bench1}

We present two benchmarks for Scenario 1 in Tables~\ref{tab:BP1A_param} and \ref{tab:BP1B_param}, chosen to illustrate
A2HDM parameter sets that would yield other heavy scalar channels that could
be probed in future runs at the LHC.

The parameters of the first benchmark (denoted B1a) corresponds to a CP-conserving Type-I 2HDM, which is a special case of the A2HDM where the flavor-alignment parameters are related according to \eq{typeone} and satisfy $a^U = a^D = a^E = \varepsilon/\tan\beta$, which defines $\tan\beta$ of the Type-I Yukawa sector.
Adopting the convention where $\tan\beta$ is positive then fixes $\varepsilon$ to be the (common) sign of the flavor-alignment parameters.
 The parameters of benchmark B1a are displayed in Table~\ref{tab:BP1A_param}.

\renewcommand{\arraystretch}{1.5}
\setlength{\tabcolsep}{10pt}
\begin{table}[ht!]
\centering
\begin{tabular}{|c|c|c|c|c|}
\hline
\multicolumn{5}{|c|}
{Benchmark B1a -- Type-I 2HDM} \\
\hline
$m_{H^\pm}$ (GeV) & $\cos(\beta-\alpha)$ & $Z_2$ & $Z_7$ & $\tan\beta$  \\
\hline\hline
650 & $-0.0013$ & 2.27 & 0.58 & 4.0 \\
\hline
\end{tabular}
\caption{\small Parameters characterizing Benchmark B1a, for which $m_h = 125$, $m_A = 610$ and $m_H = 290$ GeV.  The corresponding A2HDM flavor-alignment parameters satisfy
$a^U = a^D = a^E = 1/\tan\beta\simeq 0.25$.  Note that $\cos(\beta-\alpha)<0$ in light of \eqs{cbmasign}{typeone}.
The parameter $Z_3= 11.89$ is obtained by imposing the
condition for a softly-broken $\mathbb{Z}_2$ symmetric scalar potential by setting $T_{Z_2}=0$ [cf.~\eq{eq:TZ2}].
\label{tab:BP1A_param}}
\centering
\end{table}

\renewcommand{\arraystretch}{1.5}
\setlength{\tabcolsep}{10pt}
\begin{table}[h!]
\begin{subtable}[c]{0.5\textwidth}
\centering
\begin{tabular}{|c|c|}
\hline
 $\sigma(gg\rightarrow H)$ (pb) & 0.65 \\
\hline
$\sigma(gg\rightarrow b\bar{b}H)$ (pb) & 1.9$\times$10$^{-3}$ \\
\hline\hline
${\rm BR}(H\rightarrow ZZ)$ & 0.0053  \\
\hline
${\rm BR}(H\rightarrow b\bar{b})$ & 0.47 \\
\hline
${\rm BR}(H\rightarrow\tau^+\tau^-)$ & 0.053 \\
\hline
${\rm BR}(H\rightarrow hh)$ & 0.023 \\
\hline
${\rm BR}(H\rightarrow gg)$ & 0.41 \\
\hline
$\Gamma_H$ (GeV) & 7$\times$10$^{-4}$ \\
\hline
\end{tabular}
\end{subtable}
\begin{subtable}[c]{0.5\textwidth}
\begin{tabular}{|c|c|}
\hline
 $\sigma(gg\rightarrow A)$ (pb) & 0.18 \\
\hline
$\sigma(gg\rightarrow b\bar{b}A)$ (pb) & 6.9$\times$10$^{-5}$ \\
\hline\hline
${\rm BR}(A\rightarrow ZH)$ & 0.94 \\
\hline
${\rm BR}(A\rightarrow t\bar{t})$ & 0.057 \\
\hline
${\rm BR}(A\rightarrow b\bar{b})$ & 1.9$\times$10$^{-5}$  \\
\hline
${\rm BR}(A\rightarrow\tau^+\tau^-)$ & 2.0$\times$10$^{-6}$ \\
\hline
${\rm BR}(A\rightarrow H^\pm W^\mp)$ & 0 \\
\hline
$\Gamma_A$ (GeV) & 31.99 \\
\hline
\end{tabular}
\end{subtable}
\caption{\small Production cross sections and relevant decay branching ratios for $H$ and $A$ in benchmark B1a.
\label{tab:BP1A_HA}}
\centering
\end{table}

\renewcommand{\arraystretch}{1.5}
\setlength{\tabcolsep}{10pt}
\begin{table}[h!]
\centering
\begin{tabular}{|c|c|}
\hline
 $\sigma(gg\rightarrow tbH^\pm)$ (pb) & 0.0078 \\
\hline\hline
${\rm BR}(H^\pm \rightarrow tb)$ &  0.040 \\
\hline
${\rm BR}(H^\pm \rightarrow \tau^\pm\nu)$ & 2.0$\times$10$^{-6}$  \\
\hline
${\rm BR}(H^\pm \rightarrow HW^\pm)$ &  0.96 \\
\hline
$\Gamma_{H^\pm}$ (GeV) & 45.39 \\
\hline
\end{tabular}
\caption{\small Production cross sections and relevant decay branching ratios for $H^\pm$ in benchmark B1a.
\label{tab:BP1A_Hch}}
\centering
\end{table}

For the
benchmark parameters shown in Table~\ref{tab:BP1A_param},
the main production cross sections
and some of the relevant branching ratios for $H$, $A$ and $H^\pm$ are
exhibited in Tables~\ref{tab:BP1A_HA} and \ref{tab:BP1A_Hch}.
The results of Tables~\ref{tab:BP1A_HA} and \ref{tab:BP1A_Hch} suggest a number of additional channels that could yield possible discoveries in future LHC runs.   The most promising channel for $H$ would be production either indirectly via $gg\to A\to ZH$ or directly via $gg\to H$, with the subsequent decay of $H\to hh\to b\bar{b}b\bar{b}$.   Upper bounds on the former have been reported by the ATLAS Collaboration in Ref.~\cite{ATLAS:2022fpx}, whereas upper bounds on the latter
have been presented by the ATLAS and CMS Collaborations based on their resonant diHiggs searches~\cite{ATLAS:2021ifb,ATLAS:2021tyg,CMS:2022goy,CMS:2022kdx}.   The most promising alternative channel for $A$ would be production via gluon fusion, with the subsequent decay to $t\bar{t}$.   Finally, the most promising channel for $H^\pm$ would be via $tb$ associated production, with the subsequent decay to $HW^\pm$.
Upper bounds for this process have already been established in Ref.~\cite{CMS:2022jqc}.

To emphasize the difference between the Type-I 2HDM and a generic A2HDM, we present a second benchmark in Table~\ref{tab:BP1B_param}.
For the benchmark parameters shown in Table~\ref{tab:BP1B_param},
the main production cross sections
and some of the relevant branching ratios for $H$, $A$ and $H^\pm$ are exhibited in Tables~\ref{tab:BP2B_HA} and~\ref{tab:BP2B_Hch}.

\renewcommand{\arraystretch}{1.5}
\setlength{\tabcolsep}{10pt}
\begin{table}[h!]
\centering
\begin{tabular}{|c|c|c|c|c|c|c|c|}
\hline
\multicolumn{8}{|c|}
{Benchmark B1b -- generic A2HDM} \\
\hline
$m_{H^\pm}$ (GeV) & $|\hspace{-0.125em}\cos(\beta-\alpha)|$ & $Z_2$ & $Z_3$ & $Z_7$ & $a^U$ & $a^D$ & $a^E$ \\
\hline\hline
600 & 0.013 &  1.51 & 9.79 & $-0.20$ &  0.20 & 1.75 & 3.50 \\
\hline
\end{tabular}
\caption{\small Parameters characterizing Benchmark B1b, for which $m_h = 125$, $m_A = 610$ and $m_H = 290$ GeV.
\label{tab:BP1B_param}}
\centering
\end{table}

\renewcommand{\arraystretch}{1.5}
\setlength{\tabcolsep}{10pt}
\begin{table}[h!]
\begin{subtable}[c]{0.5\textwidth}
\centering
\begin{tabular}{|c|c|}
\hline
 $\sigma(gg\rightarrow H)$ (pb) & 0.40 \\
\hline
$\sigma(gg\rightarrow b\bar{b}H)$ (pb) & 0.096 \\
\hline\hline
${\rm BR}(H\rightarrow ZZ)$ & 0.015 \\
\hline
${\rm BR}(H\rightarrow b\bar{b})$ & 0.61 \\
\hline
${\rm BR}(H\rightarrow\tau^+\tau^-)$ &  0.28 \\
\hline
${\rm BR}(H\rightarrow hh)$ & 0.053 \\
\hline
$\Gamma_H$ (GeV) & 0.027 \\
\hline
\end{tabular}
\end{subtable}
\begin{subtable}[c]{0.5\textwidth}
\centering
\begin{tabular}{|c|c|}
\hline
 $\sigma(gg\rightarrow A)$ (pb) & 0.11 \\
\hline
$\sigma(gg\rightarrow b\bar{b}A)$ (pb) & 0.0034 \\
\hline\hline
${\rm BR}(A\rightarrow ZH)$ & 0.96 \\
\hline
${\rm BR}(A\rightarrow t\bar{t})$ & 0.036 \\
\hline
${\rm BR}(A\rightarrow b\bar{b})$ & 9.47$\times$10$^{-4}$  \\
\hline
${\rm BR}(A\rightarrow\tau^+\tau^-)$ & 4.95$\times$10$^{-4}$ \\
\hline
$\Gamma_A$ (GeV) & 31.31 \\
\hline
\end{tabular}
\end{subtable}
\caption{\small Production cross sections and relevant decay branching ratios for $H$ and $A$ in benchmark B1b.
\label{tab:BP2B_HA}}
\centering
\end{table}

\renewcommand{\arraystretch}{1.5}
\setlength{\tabcolsep}{10pt}
\begin{table}[th]
\centering
\begin{tabular}{|c|c|}
\hline
 $\sigma(gg\rightarrow tbH^\pm)$ (pb) & 0.0069 \\
\hline\hline
${\rm BR}(H^\pm \rightarrow tb)$ &  0.035 \\
\hline
${\rm BR}(H^\pm \rightarrow \tau^\pm\nu)$ & 5.18$\times$10$^{-4}$ \\
\hline
${\rm BR}(H^\pm \rightarrow HW^\pm)$ & 0.96  \\
\hline
$\Gamma_{H^\pm}$ (GeV) & 29.38 \\
\hline
\end{tabular}
\caption{\small Production cross sections and relevant decay branching ratios for $H^\pm$ in benchmark B1b.
\label{tab:BP2B_Hch}}
\centering
\end{table}

The results of Tables~\ref{tab:BP2B_HA} and \ref{tab:BP2B_Hch} suggest a number of additional channels that could yield possible discoveries in future LHC runs.
In addition to $gg\to A\to ZH$ and
$gg\to H$, where $H\to hh\to b\bar{b} b\bar{b}$, as previously mentioned, it may be possible to detect
$gg \to b\bar{b}H$ followed by $H\to b\bar{b}$, which would also yield a $b\bar{b}b\bar{b}$ final state but with different kinematics.
The most promising alternative channel for $A$ would be production via gluon fusion,with the subsequent decay to $t\bar{t}$, since there is no longer kinematic access to $H^\pm W^\mp$.
Finally, the most promising channel for $H^\pm$ would again be via $tb$ associated production, with the subsequent decay to $HW^\pm$.

\section{Scenario 2: $m_A = 400$ GeV}
\label{sec:scen2}

Consider the CP-conserving 2HDM with a CP-odd scalar mass of $m_A=400$~GeV and two
CP-even scalars with masses $m_h=125$~GeV and $m_H>450$ GeV (where $h$ is identified as the Higgs boson observed in LHC data).
This scenario is motivated by a slight excess of events over the expected backgrounds in the ATLAS search for resonant $\tau^+\tau^-$ production and in the CMS search for resonant $t\bar{t}$ production due to the production and the subsequent decay of a heavy neutral scalar.  The ATLAS Collaboration observes an excess of events in both gluon fusion and in $b$-associated production of a heavy scalar (where the CP quantum number is not determined).  Although the CMS Collaboration observes no excess in a similar search, there remains some room for the ATLAS excess that is not yet excluded at the 95\% CL by the CMS Collaboration, as discussed below \eq{ATLAStautau2}.  Meanwhile, the CMS Collaboration reports an excess of $t\bar{t}$ pairs that are interpreted as the gluon fusion production of $A\to t\bar{t}$ (in which an alternative interpretation of $H\to t\bar{t}$ is excluded).   As both data excesses are associated with the production of a scalar of mass 400~GeV, we shall assume that the scalar associated with the ATLAS excess is CP-odd.
The CP-even scalar is taken to be heavier, $m_H> 450$~GeV to avoid the possibility for it to contribute to the ATLAS $\tau^+\tau^-$ excess.\footnote{The possibility of $m_H<350$~GeV is less likely to survive the constraints of our model scans, and hence we discard this option in what follows.}
Note that this scenario is completely orthogonal to Scenario 1 since the decay $A\to ZH$ is kinematically excluded.

\subsection{A2HDM Interpretation of Scenario 2}

\begin{figure}[t!]
  \centering
  \includegraphics[height=7.2cm,angle=0]{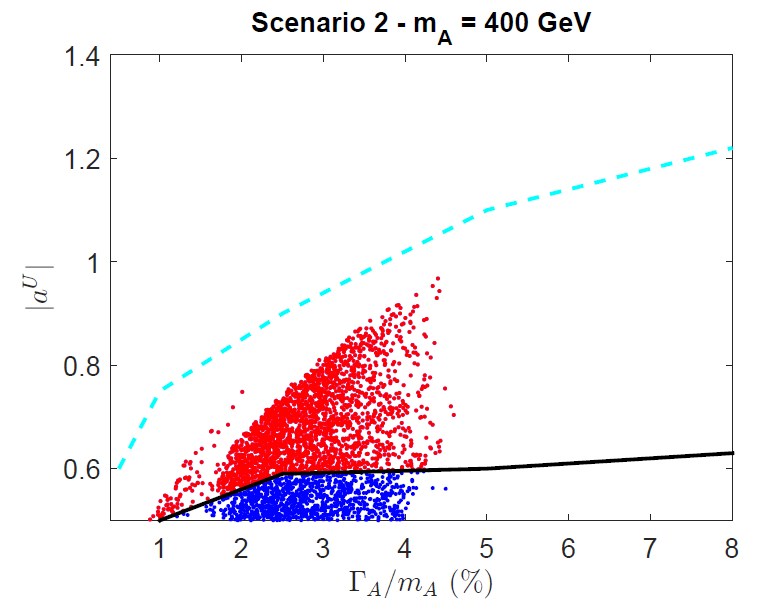}
  \caption{\small The ratio of the $A$ width to its mass (with $m_A=400$~GeV) as a function of the flavor-alignment parameter $a^U$ in Scenario 2 obtained in a scan over A2HDM parameters, subject to
  the theoretical and experimental constraints elucidated in Section~\ref{sec:scans}.
  The dashed cyan (solid black) line shows the observed (expected) 95\% CL upper limit on the gluon fusion cross section for $A\to t\bar{t}$ reported by the CMS Collaboration in Ref.~\cite{CMS:2019pzc},
  translated into an upper limit for $a^U$ as a function of $\Gamma_A/m_A$.  The points of the scan that lie between the dashed cyan and solid black curve are colored red, which constitute the proposed signal of Scenario 2.}
  \label{fig:scen2_gamA}
\end{figure}

We have performed a scan of the A2HDM parameter space for Scenario 2, while respecting
the theoretical and experimental constraints elucidated in Section~\ref{sec:scans}.   To determine the range of interest for the flavor-alignment parameter $a^U$,
we exhibit in
Fig.~\ref{fig:scen2_gamA} the width to mass ratio,
 $\Gamma_A/m_A$,  as a function of $a^U$ and consider the implications of
 the excess of $t\bar{t}$ events reported by the CMS Collaboration in
Ref.~\cite{CMS:2019pzc}, which are interpreted as the production of
$A\to t\bar{t}$ via gluon fusion.
The red and blue points of Fig.~\ref{fig:scen2_gamA} represent the results of our scan prior to imposing
the constraints of the $A\to t\bar{t}$ constraints of Ref.~\cite{CMS:2019pzc}.  The CMS Collaboration expected to obtain a 95\% CL upper limit on the gluon fusion production of $A\to t\bar{t}$ for $m_A=400$~GeV, which when translated into a limit on $a^U$ yields the solid black line shown in
Fig.~\ref{fig:scen2_gamA}.\footnote{We can identify
the coupling modifier $g_{At\bar{t}}$ of Ref.~\cite{CMS:2019pzc} with $a^U$.}
However, due to an excess of events above background (with a local significance of 3.5$\sigma$),
the actual 95\% CL upper limit on the gluon fusion production of $A\to t\bar{t}$ translated into a limit on $a^U$ yields the dashed cyan line shown in Fig.~\ref{fig:scen2_gamA}.   The points in our scan that lie between the black and cyan lines (colored red) will be the parameter points of interest for Scenario~2 going forward.

A more sophisticated approach was undertaken in Ref.~\cite{Biekotter:2021qbc}, in which a chi-squared
analysis was performed to determine the best fit value for the $A$ coupling modifier to top quarks for each value of the $A$ width. The chi-squared analysis of Ref.~\cite{Biekotter:2021qbc} only considered
the CMS data excess in the
$t\bar{t}$ channel~\cite{CMS:2019pzc}, whereas we wish to consider the implications of this data excess
together with the ATLAS data excess in the $\tau^+\tau^-$ channel~\cite{ATLAStautau}.
Nevertheless, we have verified that the red points
chosen in our Fig.~\ref{fig:scen2_gamA} have $\chi^2_{t\bar{t}}\lesssim 5$, for the definition of $\chi^2_{t\bar{t}}$ given in Ref.~\cite{Biekotter:2021qbc}
and the results shown in their Fig. 1.  Moreover, a substantial portion of the red points of our Fig.~\ref{fig:scen2_gamA} have values of $\chi^2_{t\bar{t}}$
well below 1.

\begin{figure}[h!]
  \centering
  \includegraphics[height=7.25cm,angle=0]{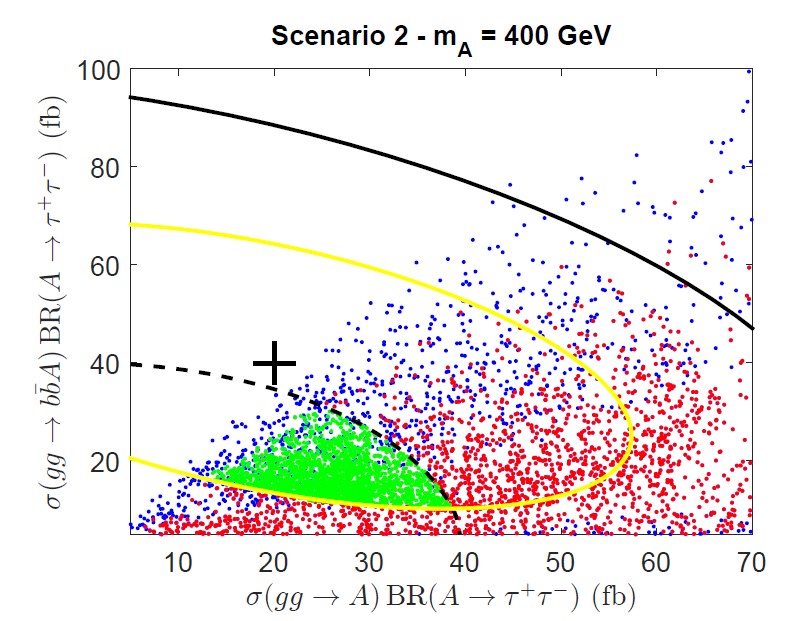}
  \caption{\small
   Results of a scan over A2HDM parameter points in Scenario 2 that
satisfy the theoretical and experimental constraints elucidated in Section~\ref{sec:scans}.
 The blue points exhibit the values of the cross sections for gluon fusion production and $b$-associated production of $A$ multiplied by ${\rm BR}(H\to\tau^+\tau^-)$.  These points are
 colored red if the corresponding values of $a^U$ and $\Gamma/m_A$ lie in the red region of Fig.~\ref{fig:scen2_gamA}.  The $+$ indicates the best fit point for the ATLAS excess of Ref.~\cite{ATLAS:2020zms} interpreted as $A$ production with $m_A=400$~GeV;
 the solid yellow and black curves correspond to the corresponding 1$\sigma$ and 2$\sigma$ contours.
 Red points within the 1$\sigma$ contour are colored green.  Finally, all points outside the boundary of the dashed black ellipse are excluded at the 95\% CL by the ditau search of the CMS Collaboration~\cite{CMS:2022goy}.}
  \label{fig:scen2_Atau}
\end{figure}

Employing the same A2HDM scan described above, we consider the implications of the
ATLAS ditau signal excess interpreted as the production of a CP-odd scalar with $m_A=400$~GeV.
In Fig.~\ref{fig:scen2_Atau}, the best fit point for the ATLAS
 excess of Ref.~\cite{ATLAS:2020zms} is indicated by the $+$ sign;
 the solid yellow and black curves correspond to the corresponding 1$\sigma$ and $2\sigma$ contours.  The nonobservation of the ditau signal by the CMS collaboration~\cite{CMS:2022goy} excludes (at the 95\% CL)
values of the $A$ production cross section multiplied by ${\rm BR}(A\to\tau^+\tau^-)$ that lie
outside the boundary of the dashed black ellipse.\footnote{The dashed black contour exhibited in Fig.~\ref{fig:scen2_Atau} was obtained from the results of Ref.~\cite{CMS:2022goy} by interpolation as explained above \eq{limitgg}.}
The blue points of our initial scan are constrained to lie within the red region of Fig.~\ref{fig:scen2_gamA} as discussed above.   These point are colored red in Fig.~\ref{fig:scen2_Atau}.  In addition, the green points are a subset of the red points that lie within the 1$\sigma$ contour.  Finally, the green points that lie within the dashed black ellipse in the lower left hand corner of Fig.~\ref{fig:scen2_Atau} constitute the A2HDM parameter regime of
interest for Scenario 2.
Note that despite of the more restrictive CMS exclusion limits, there are still a significant number of green scan points for Scenario 2 that lie within the 1$\sigma$ ellipse of the ATLAS ditau data excess.

The sharp lower bound $\sigma(gg\to A)\times{\rm BR}(A \rightarrow \tau^+\tau^-)\gsim 19$~fb exhibited by the green points of  Fig.~\ref{fig:scen2_Atau} is noteworthy.   This lower bound
is a consequence of requiring $|a^U| \gsim
0.5$, so as to be above the solid black line in Fig.~\ref{fig:scen2_gamA} (in order to explain the
observed CMS excess of $A\to t\bar{t}$), while also imposing $|a^E|\gsim 4.9$ (as illustrated in Fig.~\ref{fig:scen2_aEaD} below) in order that our scan points live within the $1\sigma$ ellipse of Fig.~\ref{fig:scen2_Atau}.  In particular, the lower bound on $|a^U|$ forces $\sigma(gg\to A)$ to be roughly above 7.1 pb, whereas the lower bound on $|a^E|$ yields
a minimum branching ratio of ${\rm BR}(A\to\tau^+\tau^-)\gsim 1.2\times 10^{-3}$.  Although these considerations omit the implications of scanning over $a^D$, they do provide a rough understanding of the lower bound on $\sigma(gg\to A\to\tau^+\tau^-)$ observed in Fig.~\ref{fig:scen2_Atau}.

\begin{figure}[t!]
\begin{tabular}{cc}
\includegraphics[height=6cm,angle=0]{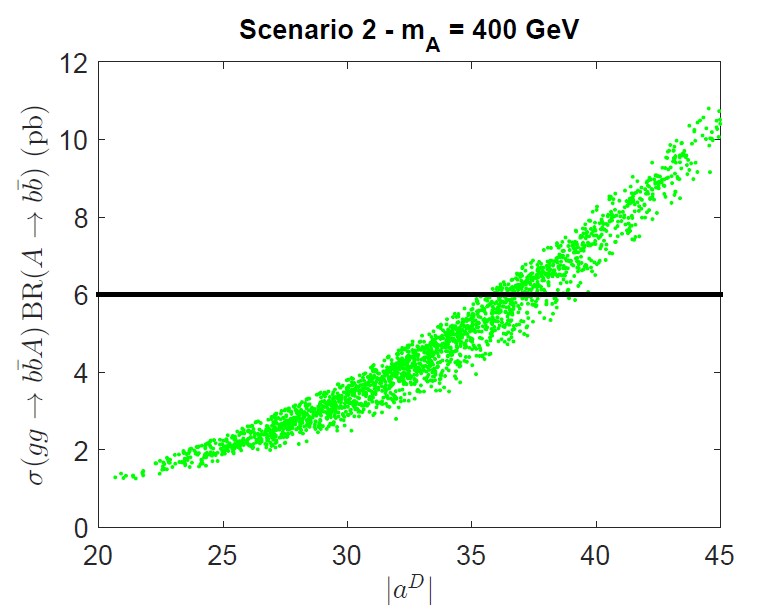}&
\includegraphics[height=6cm,angle=0]{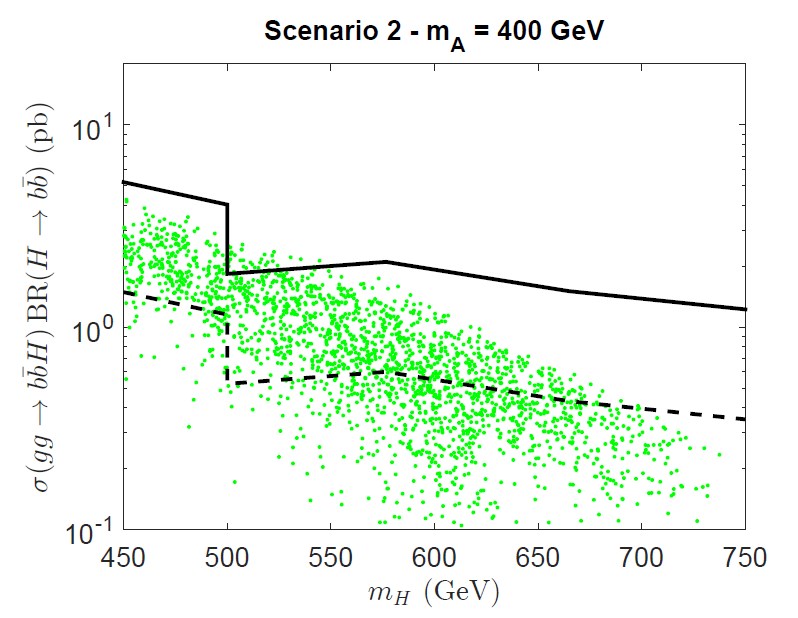}\\
 (a) & (b)
\end{tabular}
  \caption{\small
    Results of a scan over A2HDM parameter points in Scenario 2 that
satisfy the theoretical and experimental constraints elucidated in Section~\ref{sec:scans}.
The points shown are the subset of the green points of
Fig.~\ref{fig:scen2_Atau} that are contained within the dashed ellipse shown there. We exhibit
(a) the predicted values of $\sigma(gg\to b\bar{b}A)\times{\rm BR}(A \rightarrow b\bar{b})$ as a function of $|a^D|$, and
(b) the predicted values of $\sigma(gg\to b\bar{b}H)\times{\rm BR}(H \rightarrow b\bar{b})$ as a function of $m_H$.
Points that lie above the solid line are excluded by the 95\% CL upper limit obtained by the CMS Collaboration in Ref.~\cite{CMS:2018hir}.
The dashed black line is the expected exclusion at Run 3 of the LHC.
  \label{fig:scen2_Abb}}
\end{figure}

In order to achieve a large enough rate in $b$-associated production of $A$ that subsequently decays into $\tau^+\tau^-$, a sufficiently large absolute value of
the flavor-alignment parameter $a^D$ will be required, which tends to enhance the $A\to b\bar{b}$ decay rate. The most restrictive bounds on $\sigma(gg\to b\bar{b}A)\times{\rm BR}(A \rightarrow b\bar{b})$ are provided by the CMS Collaboration~\cite{CMS:2018hir}, with a 95\% CL upper limit of roughly 6 pb for $m_A=400$~GeV.   As shown in Fig.~\ref{fig:scen2_Abb} (a), this latter constraint eliminates
a substantial region of the A2HDM parameter space for Scenario 2 and yields
an upper bound of $|a_D|\lsim$ 40.  
In Fig.~\ref{fig:scen2_Abb} (b) we show the predicted values of
$\sigma(gg\to b\bar{b}H)\times{\rm BR}(H \rightarrow b\bar{b})$ as a function of $m_H$, and we compare these values with the 95\% CL upper limit
from the CMS Collaboration in Ref.~\cite{CMS:2018hir}, obtained with 35.9 fb$^{-1}$ of data and shown as a solid black line in that plot.
As we see, a very small set of our points are already excluded. The dashed line shown corresponds to a {\em na\"{\i}ve} rescaling (by the square root of the corresponding luminosities)
of the current CMS bound based on the full Run 2 dataset and
an anticipated 300 fb$^{-1}$ of data during Run 3 of the LHC.  Indeed, we expect that a substantial portion of the parameter space
for Scenario 2, with $H$ masses up to 700 GeV, could be probed in Run 3.

\begin{figure}[h!]
\begin{tabular}{cc}
\includegraphics[height=6cm,angle=0]{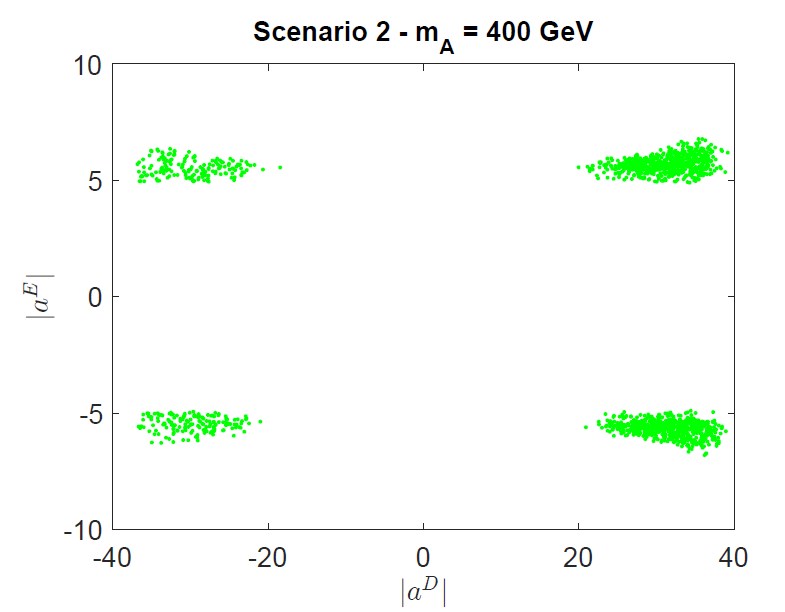}&
\includegraphics[height=6cm,angle=0]{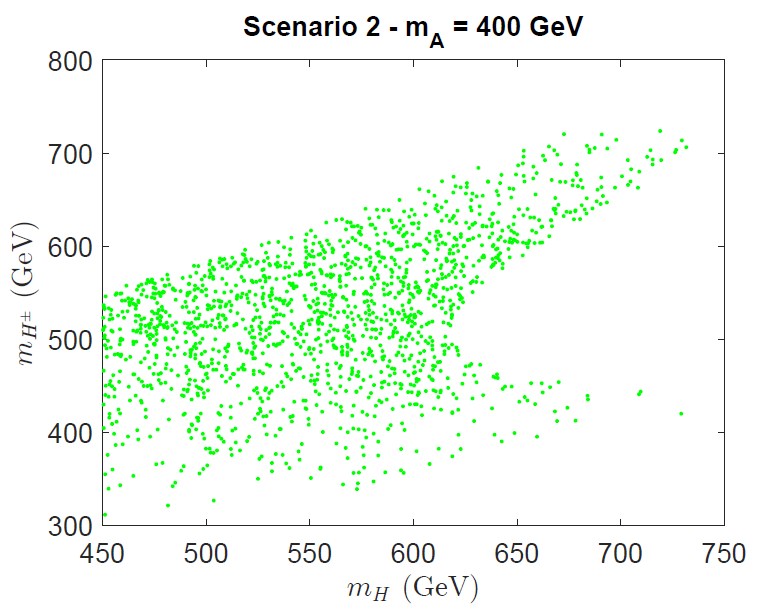}\\
 (a) & (b)
\end{tabular}
  \caption{\small
Results of a scan over A2HDM parameter points in Scenario 2 that lie within the region of interest, corresponding to the area of parameter space occupied by the subset of green points that lie inside the dashed black ellipse in Fig.~\ref{fig:scen2_Atau} and below the dashed line in Fig.~\ref{fig:scen2_Abb}.
Panel (a) exhibits the values of the
flavor-alignment parameters $a^E$ and $a^D$ and panel (b) exhibits the masses of the heavier CP-even scalar $H$ and the charged Higgs boson $H^\pm$, for each scan point that satisfies all the specified constraints.}
  \label{fig:scen2_aEaD}
\end{figure}

We now examine in more detail  the properties of the green A2HDM scan points that lie within the dashed black ellipse of Fig.~\ref{fig:scen2_Atau} and below the solid black lines of Fig.~\ref{fig:scen2_Abb}, which satisfy all known experimental limits while interpreting the ATLAS ditau excess and the CMS $t\bar{t}$ excess as the production of a CP-odd scalar with $m_A\sim 400$~GeV, which from now on we call the ``{\em region of interest}."

In Fig.~\ref{fig:scen2_aEaD}, we show in panel (a) the values of the flavor-alignment parameters
$a^E$ and $a^D$ and in panel~(b) we show
the values of $m_H$ and $m_{H^\pm}$ for the A2HDM scan points that satisfy all the specified constraints.  Note that the values of $|a^D|$ and $|a^E|$ are restricted to lie within a very narrow range of values, $25\lsim |a^D|\lsim 40$ and $5\lsim |a^E|\lsim 7$.   The corresponding lower limiting values are a consequence of the $gg\to b\bar{b}A\to b\bar{b}\tau^+\tau^-$ interpretation of ATLAS excess of Ref.~\cite{ATLAStautau}.  The upper limit on $|a^D|$ is imposed by the solid black lines of Fig~\ref{fig:scen2_Abb} and the upper limit on $|a^E|$ is due in part to our $gg\to A\to t\bar{t}$ interpretation of the CMS excess of Ref.~\cite{CMS:2019pzc}.  In Fig.~\ref{fig:scen2_aEaD}(b), two distinct branches are observed corresponding to $m_H\sim m_A$ and $m_{H^\pm}\sim m_A$,
which arise after imposing the $T$ parameter constraint, as discussed in Section~\ref{sec:scans}.
Although we scan over $h$ and $H^\pm$ masses up to 1~TeV, we find no scan points above about 750~GeV.   This is a consequence of the tree-level unitarity and perturbativity constraints of Section~\ref{sec:scans} that limit the magnitude of the heavy scalar mass splittings.

Although the proposed Scenario 2 signals are viable in the A2HDM framework, one can again ask whether the same signals can be successfully accommodated in two Higgs doublet models with natural flavor conservation. In Section~\ref{sec:scen1} we introduced four quantities, $T_{I,I\!I,X,Y}$ [cf.~\eqs{TsubIX}{TsubIIY}], which if zero would indicate the presence of a (softly-broken) $\mathbb{Z}_2$ symmetry with a Type I, II, X or Y Yukawa coupling pattern.
\begin{figure}[t!]
  \begin{tabular}{cc}
\includegraphics[height=6cm,angle=0]{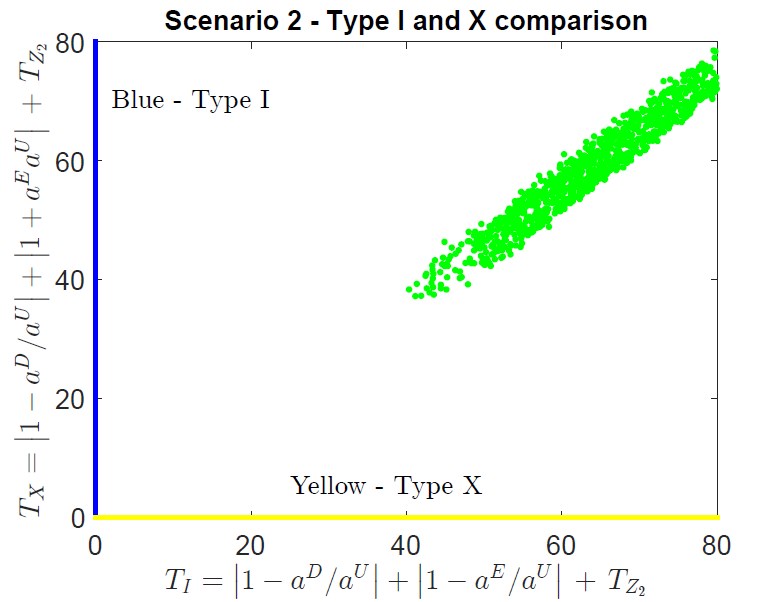}&
\includegraphics[height=6cm,angle=0]{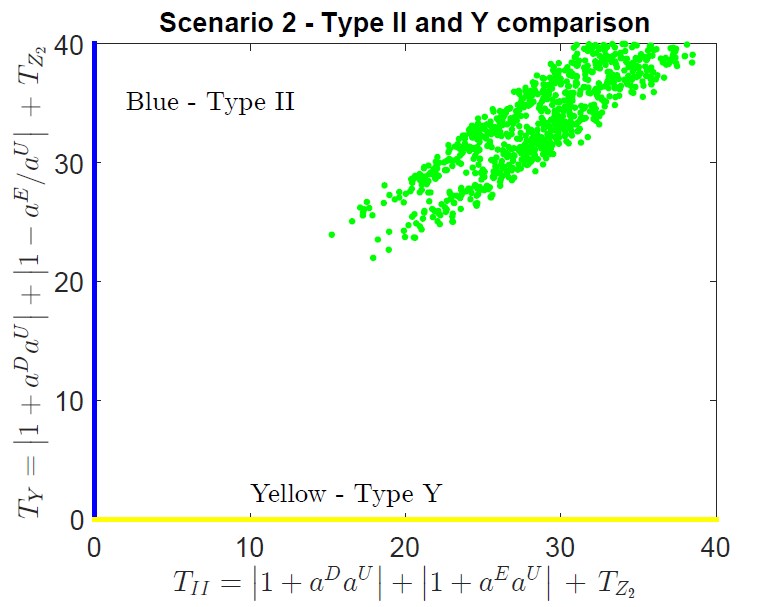}\\
 (a) & (b)
\end{tabular}
  \caption{\small
 Results of a scan over A2HDM parameter points in Scenario 2 that lie within the region of interest, corresponding to the area of parameter space occupied by the subset of green points that lie inside the black dashed ellipse in Fig.~\ref{fig:scen2_Atau} and below the solid black lines in Fig.~\ref{fig:scen2_Abb}.
Panel (a) exhibits the values of $T_I$ vs.~$T_X$.   Points that lie along the horizontal (yellow) axis
would be consistent with a Type-X 2HDM.  Points that lie along the vertical (blue) axis would be consistent with a Type-I 2HDM.   Panel (b) exhibits the values
of $T_{I\!I}$ vs.~$T_Y$.   Points that lie along the horizontal (yellow) axis
would be consistent with a Type-Y 2HDM.  Points that lie along the vertical (blue) axis would be consistent with a Type-II 2HDM. }
  \label{fig:types2}
\end{figure}
\begin{figure}[h!]
  \begin{tabular}{cc}
\includegraphics[height=6cm,angle=0]{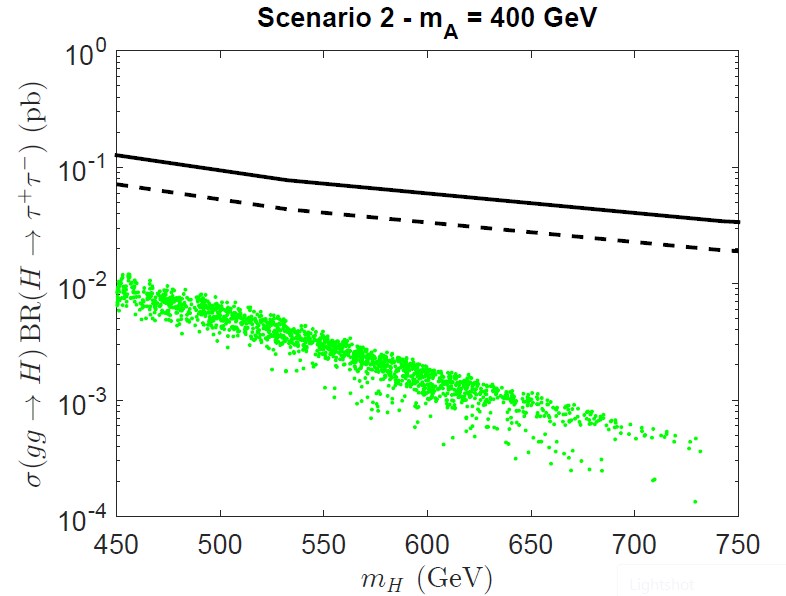}&
\includegraphics[height=6cm,angle=0]{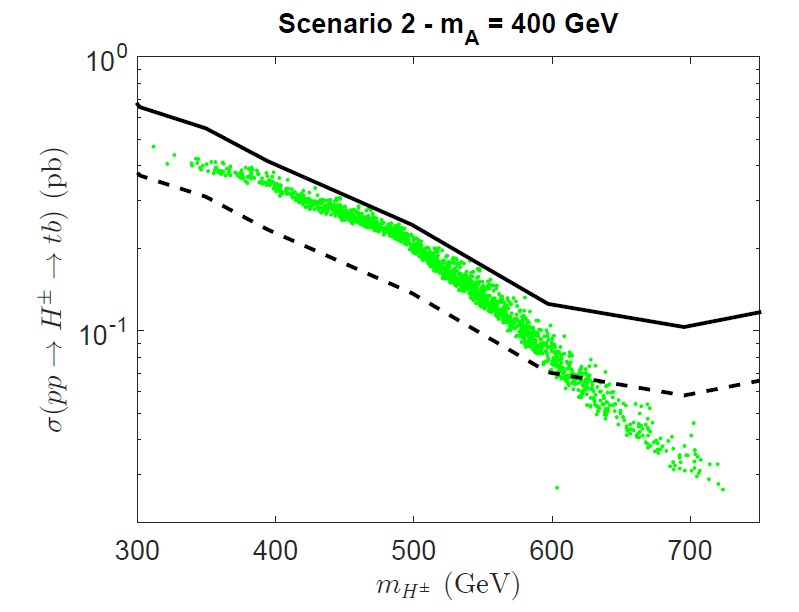}\\
 (a) & (b)
\end{tabular}
  \caption{\small
 The solid black line corresponds to the 95\% CL exclusion limit reported by
the ATLAS Collaboration in Ref.~\cite{ATLAS:2021upq}.  The solid black lines shown in both plots above are based on 139 fb$^{-1}$ of data.   Assuming an additional 300 fb$^{-1}$ of data during Run 3 of the LHC, a {\em na\"{\i}ve} rescaling of the ATLAS exclusion bounds yields the dashed black lines shown in both plots.
}
  \label{fig:scen2_Hch}
\end{figure}
In contrast to Scenario 1, where a Type-I 2HDM provided a viable framework for the interpretation of the ATLAS excess of Ref.~\cite{ATLAS:2020gxx}, the Scenario 2 signals are incompatible with a
Type I, II, X or Y Yukawa coupling pattern, as indicated by the two panels
of Fig.~\ref{fig:types2}.

If the Scenario 2 interpretation of the ATLAS and CMS excesses of Refs.~\cite{ATLAStautau} and \cite{CMS:2019pzc}, respectively, are corroborated by further data, then one would expect to also discover the $H$ and $H^\pm$.  Indeed, some of the scan points of the Scenario 2 region of interest are not too far from the
current exclusion limits derived from ATLAS searches for $H$ production (with subsequent decay to $\tau^+\tau^-$) and $H^\pm$ production (with subsequent decay to $tb$), as exhibited in Fig.~\ref{fig:scen2_Hch}).  In panel (a) of Fig.~\ref{fig:scen2_Hch}, we show that the Scenario 2 scan points in the region of interest reside about a factor of 10 below the 95\% CL exclusion limits for $\sigma(gg\to H){\rm BR}(H\to \tau^+\tau^-)$ obtained by the ATLAS Collaboration~\cite{ATLAStautau}.  In panel~(b),
we show that some of the Scenario 2 scan points in the region of interest lie quite close to the
95\% CL exclusion limits for $\sigma(pp\to H^\pm){\rm BR}(H^\pm\to tb)$ obtained by the ATLAS Collaboration in Ref.~\cite{ATLAS:2021upq}.\footnote{We have also checked that the ATLAS and CMS exclusion limits on $H$ production with subsequent decays into
$ZZ$, $W^+W^-$, $\gamma\gamma$, or $hh$ and $H^\pm$ production with subsequent decays into $\tau\nu$ are less constraining and easily satisfied by all the Scenario 2 scan points in the region of
interest.}
The dashed black lines in both panels of Fig.~\ref{fig:scen2_Hch} show the anticipated exclusion bounds one can expect at the end of Run 3 of the LHC by
performing the appropriate rescaling of the current ATLAS bounds. Although no appreciable changes occur for the
$gg\to H \to \tau^+\tau^-$ channel, it is noteworthy that
Run 3 of the LHC will be able to provide support for the Scenario
2 interpretation if a charged scalar with a mass below (roughly) 650 GeV is
discovered in the tb decay channel, as indicated in
Fig.~\ref{fig:scen2_Hch}(b).

We expect that future LHC searches for $H\to\tau^+\tau^-$ and $H^\pm\to tb$ with larger data sets will begin to probe the Scenario 2 region of interest and thus provide the most likely channels for new discoveries. Moreover, it will eventually be possible to exclude Scenario 2 if a charged scalar with a mass below 750 GeV decaying to $tb$ is not observed.

\subsection{A2HDM Benchmark for Scenario 2}

We present one benchmark for Scenario 2 in Table~\ref{tab:BP2_param}, chosen to illustrate an A2HDM parameter set with significant $H$ and $A$ production cross sections, which would provide additional heavy scalar discovery channels that could
be probed in future runs at the LHC.  In light of the results of Fig.~\ref{fig:types2}, there are no viable A2HDM parameter sets that approximate a Type-I, II, X or Y 2HDM.    For the benchmark parameters
shown in Table~\ref{tab:BP2_param}, the main production cross sections and some of the relevant branching ratios for $H$, $A$ and $H^\pm$ are exhibited in Tables~\ref{tab:BP2_HA} and \ref{tab:BP2_Hch}.\footnote{We do not exhibit the Benchmark B2 values of ${\rm BR}(H\to ZZ)\simeq 4\times 10^{-6}$ and ${\rm BR}(H\to hh)\simeq 10^{-7}$, which are too small to be phenomenologically relevant.}

\renewcommand{\arraystretch}{1.5}
\setlength{\tabcolsep}{10pt}
\begin{table}[h!]
\centering
\begin{tabular}{|c|c|c|c|c|c|c|c|c|}
\hline
\multicolumn{9}{|c|}
{Benchmark B2 -- generic A2HDM} \\
\hline
 $m_H$ & $m_{H^\pm}$ (GeV) & $|\hspace{-0.125em}\cos(\beta-\alpha)|$ & $Z_2$ & $Z_3$ & $Z_7$ & $a^U$ & $a^D$ & $a^E$ \\
\hline\hline
492  & 529 & 0.0018 &  2.42 & 7.58 & -1.39 & 0.60 & 35.07 & 6.32 \\
\hline
\end{tabular}
\caption{\small Parameters characterizing Benchmark B2, for which $m_h = 125$, $m_A = 400$ GeV.
\label{tab:BP2_param}}
\centering
\end{table}

\renewcommand{\arraystretch}{1.5}
\setlength{\tabcolsep}{10pt}
\begin{table}[h!]
\begin{subtable}[c]{0.5\textwidth}
\centering
\begin{tabular}{|c|c|}
\hline
 $\sigma(gg\rightarrow H)$ (pb) & 2.57  \\
\hline
$\sigma(gg\rightarrow b\bar{b}H)$ (pb) & 3.30 \\
\hline\hline
${\rm BR}(H\rightarrow b\bar{b})$ & 0.69  \\
\hline
${\rm BR}(H\rightarrow\tau^+\tau^-)$ & 0.0028 \\
\hline
${\rm BR}(H\rightarrow t\bar{t})$ & 0.31 \\
\hline
$\Gamma_H$ (GeV) & 14.54 \\
\hline
\end{tabular}
\end{subtable}
\begin{subtable}[c]{0.5\textwidth}
\centering
\begin{tabular}{|c|c|}
\hline
 $\sigma(gg\rightarrow A)$ (pb) & 10.87 \\
\hline
$\sigma(gg\rightarrow b\bar{b}A)$ (pb) & 8.97 \\
\hline\hline
${\rm BR}(A\rightarrow t\bar{t})$ & 0.39 \\
\hline
${\rm BR}(A\rightarrow b\bar{b})$ & 0.60  \\
\hline
${\rm BR}(A\rightarrow\tau^+\tau^-)$ & 0.0024 \\
\hline
$\Gamma_A$ (GeV) & 13.94 \\
\hline
\end{tabular}
\end{subtable}
\caption{\small Production cross sections and relevant decay branching ratios for $H$ and $A$ in benchmark B2.
\label{tab:BP2_HA}}
\centering
\end{table}

\renewcommand{\arraystretch}{1.5}
\setlength{\tabcolsep}{10pt}
\begin{table}[h!]
\centering
\begin{tabular}{|c|c|}
\hline
 $\sigma(gg\rightarrow tbH^\pm)$ (pb) & 0.19 \\
\hline\hline
${\rm BR}(H^\pm \rightarrow tb)$ &  0.90 \\
\hline
${\rm BR}(H^\pm \rightarrow \tau^\pm\nu)$ & 0.0023 \\
\hline
${\rm BR}(H^\pm \rightarrow AW^\pm)$ & 0.095 \\
\hline
$\Gamma_{H^\pm}$ (GeV) & 19.05 \\
\hline
\end{tabular}
\caption{\small Production cross sections and relevant decay branching ratios for $H^\pm$ in benchmark B2.
\label{tab:BP2_Hch}\\[-5pt]}
\centering
\end{table}

The results of Tables~\ref{tab:BP2_HA} and \ref{tab:BP2_Hch} suggest a number of additional channels that could yield possible discoveries in future LHC runs.  The most promising channels for $H$ would be via gluon fusion production and/or $b$-associated production followed by $H\to b\bar{b}$ or $t\bar{t}$.   Note that although the $\tau^+\tau^-$ branching ratio is quite small,
$\sigma(gg\to b\bar{b}H)\times{\rm BR}(H\to \tau^+\tau^-)\sim 10$~fb, which will be probed in future runs at the LHC with a sufficiently large data sample (in light of Ref.~\cite{ATLAStautau}).
In addition to $gg\to A\to \tau^+\tau^-$, $gg\to b\bar{b}A\to b\bar{b}\tau^+\tau^-$, and $gg\to A\to t\bar{t}$ (which constitute the current excesses in data that define Scenario~2), it may be possible to detect
 $gg\to b\bar{b}A\to b\bar{b}b\bar{b}$ and $gg\to b\bar{b}A\to b\bar{b}t\bar{t}$.
Finally, the most promising channel for $H^\pm$ would be via $tb$ associated production, with the subsequent decay to either $tb$ or $AW^\pm$.

\section{Conclusions}
\label{sec:conclude}

If new scalars exist with masses below 1 TeV with cross sections governed by the electroweak scale, then searches now being performed at the LHC would be capable of discovering such states given sufficient data.  Indeed, the existence of an extended Higgs sector is primarily a question that must be answered via experimental exploration.   If a convincing signal eventually emerges, then it will be important to provide a model interpretation.

Although there is no concrete argument pointing to a specific extended Higgs sector, there are a number of weak assumptions that can be applied to narrow down the appropriate framework in which to interpret the discovery of a new scalar.  For example, the observed electroweak rho parameter and the absence of significant flavor changing neutral currents places strong constraints on a viable extended Higgs sector.   The former tends to restrict considerations to models of hypercharge-zero singlet and hypercharge-one doublet scalars (e.g., see Ref.~\cite{Gunion:1989we}), although interesting models with Higgs doublets and triplets arranged to have an (admittedly fine-tuned) custodial symmetric scalar potential would also satisfy the electroweak rho parameter constraint~\cite{Georgi:1985nv,Chanowitz:1985ug}.   If a charged scalar were discovered, this would indicate the presence of additional doublet scalars.

For simplicity, we focus on extended Higgs sectors with additional doublets.  Simply adding one doublet already yields a set of new phenomena (charged scalars, new CP-even and CP-odd neutral scalars or new neutral scalars of mixed CP symmetry).  Thus employing the framework of the 2HDM may be sufficient to provide an initial interpretation of evidence that points to the discovery of an extended Higgs sector.  Employing the most general 2HDM is not appropriate given that a significant portion of its parameter space would yield scalar-mediated FCNCs that are too large to be accommodated by the present data and/or an electric dipole moment (edm) for the electron that is inconsistent with the current experimental bounds.   In light of the bounds on edms, one is tempted to restrict the parameters of the 2HDM such that no (significant) new sources of CP violation due to scalar self-interactions or the Higgs-fermion Yukawa couplings are present.  This leads to the framework of the CP-conserving 2HDM.   This is not to say that the LHC experiments should refrain from searching for CP-violating observables.  However, any initial discovery of a new scalar is not likely to be particularly sensitive to the presence of a new scalar-mediated source of CP violation.

This leaves open the question of how to suppress scalar-mediated FCNCs.  Theorists tend to demand that their models should yield suppressed scalar-mediated FCNCs naturally, which is theoretically implemented by a symmetry that is either exact or softly-broken.  Such an assumption restricts the structure of the Higgs-fermion Yukawa couplings to one of four types (called Types I, II, X and Y), which defines a fundamental parameter of the model, called $\tan\beta$.  Experimentalists analyzing their data should ignore such considerations.  After all, determining the structure of the Higgs-fermion Yukawa couplings is an experimental endeavor.  Ideally, the experimental discovery of new scalars will inform theorists on how Nature has chosen to implement the suppression of FCNCs.   With this in mind, we have advocated in this paper that evidence for new scalars at the LHC should be experimentally analyzed within the context of the flavor-aligned 2HDM (or A2HDM), which phenomenologically implements the absence of tree-level scalar-mediated FCNCs by proposing that two initially independent Yukawa coupling matrices are in fact proportional, with the proportionality constant (called the flavor-alignment parameter) an observable to be determined by experiment.   Given that there are three pairs of Yukawa matrices (corresponding to up-type quarks, down-type quarks and charged leptons), there are three flavor-alignment parameters of interest, called $a^U$, $a^D$ and $a^E$.   The Higgs-fermion Yukawa couplings of Types I, II, X and Y are special cases of the A2HDM, so an experimental determination of the flavor-alignment parameters will inform whether one of the Yukawa coupling Types is compatible with the data.

After the completion of Run 2 of the LHC, both ATLAS and CMS have searched a variety of channels for evidence of new scalars.  No statistically significant signal has yet to emerge.  Nevertheless, we believe that it is a useful exercise to exhibit how one could implement the extended Higgs framework described above to probe the properties of any newly discovered scalar(s).   With this in mind, we have reviewed the ATLAS and CMS searches for new scalars and focused on a number of small excesses observed (corresponding typically to local 3$\sigma$ excesses whose significance reduces to $2\sigma$ or below when the look elsewhere effect is taken into account).
We selected two different scenarios: Scenario 1 ($m_A=610$~GeV and $m_H=290$~GeV observed in $gg\to A\to ZH$, where $H\to b\bar{b}$ and $Z\to\ell^+\ell^-$) and Scenario 2 ($m_A=400$~GeV, with $A\to t\bar{t}$ and $A\to\tau^+\tau^-$ decays observed).  Treating the small excesses observed as a potential signal of new scalars, we examined whether such excesses could be produced for some reasonable set of A2HDM parameters, and if so whether it would be possible to deduce whether the underlying fermion-Higgs Yukawa couplings were consistent with the symmetry-based Types I, II, X or Y 2HDMs.

Our analysis of both scenarios resulted in the following conclusions.  Scenario 1 can be realized in a Type I 2HDM, but this solution is viable only for a rather restricted region of the model parameters.  More generally, there is a larger region of the parameter space that is consistent with the A2HDM framework, but inconsistent with all of the symmetry-based Types I, II, X or Y models.   In contrast, the Scenario 2 excesses observed at LHC, while again consistent with the A2HDM framework, exhibits no allowed parameter points consistent with the symmetry-based Types I, II, X or Y models.   This result highlights one of the main points of our study.  If the Scenario 2 excesses had been real (rather than the more likely statistical fluctuation of Standard Model backgrounds), an analysis of these data with a prejudice for the consistency with the symmetry-based Types I, II, X or Y Higgs-fermion Yukawa interactions would have concluded that these data are incompatible with the 2HDM.  Of course, such a conclusion is inappropriate, in light of the compatibility of Scenario 2 with the
more general A2HDM framework.  Indeed, there is more flexibility in fitting a given scenario by employing the A2HDM as compared to each of the symmetry-based Higgs-fermion Yukawa interactions of the 2HDM due to the existence of three independent flavor-alignment parameters, $a^U$, $a^D$ and $a^E$, which allows one to independently fit the constraints on flavor and leptonic observables of the model.  In contrast, in the symmetry-based Higgs-fermion Yukawa interactions of the 2HDM,  the coupling modifiers of the non-SM scalars to fermions (in the approximate Higgs alignment limit)
are governed by a single parameter, $\tan\beta$, which results in strong correlations among the up-quark, down-quark and charged lepton Yukawa couplings.

If the excesses observed by the ATLAS and CMS Collaborations that are the basis for either of the two scenarios analyzed in this paper were confirmed in Run 3 of the LHC, then our analysis also provides predictions for new non-SM Higgs signals that could provide support for the A2HDM framework.
In Scenario 1, the allowed A2HDM parameter space yields a signal for $gg\to H\to hh$ that is close to the current experimental bound, in light of
Fig.~\ref{fig:bbHhh}.  Moreover, the cross section for $b$-associated production of $H$ followed by $H\to hh$ can be as large as 0.07 pb.  Thus, resonant production of $hh$ in subsequent runs at the LHC will provide a critical check of the A2HDM interpretation of Scenario 1.  In addition, the A2HDM parameter space consistent with Scenario 1 yields cross sections for $gg$ fusion and $b$-associated production of $H$ followed by $H\to \tau^+\tau^-$ that can be as large as 120~fb and 70 fb, respectively, which could be detected with the future higher luminosity runs of the LHC.  In contrast,
 $gg$ fusion and $b$-associated production of $A$ followed by $A\to \tau^+\tau^-$ yields smaller cross sections (roughly 0.1 fb and 0.05 fb, respectively), although still significantly larger than one would obtain in a Type-I 2HDM.   Finally, Scenario 1 predicts a rather restricted range of masses for $H^\pm$:
 either $220\lsim m_{H^\pm}\lsim 320$~GeV (i.e., $m_{H^\pm}\sim m_H$) or $570\lsim m_{H^\pm}\lsim 670$~GeV (i.e., $m_{H^\pm}\sim m_A$).
 The predicted signal rates for $pp\rightarrow H^\pm \rightarrow tb$ are close to the current experimental bounds, and would provide another important consistency check of the A2HDM interpretation of Scenario 1 if observed.
In Scenario 2, the allowed A2HDM parameter space exhibited in Fig.~\ref{fig:scen2_Hch} implies potentially significant signal rates for
$\sigma(gg\to H){\rm BR}(H\to \tau^+\tau^-)$ and $\sigma(pp\to H^\pm){\rm BR}(H^\pm\to tb)$: 10 times below the current experimental
sensitivity for the former, and close to the current exclusion bounds for the latter, for $400\lsim m_{H^\pm}\lsim 600$~GeV.
Both of these signals, if present at their expected rates, could be probed at Run 3 of the LHC.  We would also expect a potentially observable signal rate for
$gg$ fusion and $b$-associated production of $H$ followed by $H\to t\bar{t}$.
Finally, in light of Fig.~\ref{fig:scen2_Abb}, there
is a strong possibility of observable signal rates in $b$-associated production of $A$ and $H$, followed by their decays to $b\bar{b}$.
Observations of any of these non-SM Higgs production and decay processes would provide important consistency checks of the A2HDM interpretation of Scenario 2.

Ultimately, one would like to reduce further the specific extended Higgs sector model assumptions employed in analyzing a potential LHC discovery of a new scalar.   Ideally, one should use experimental results to determine how many new doublets make up the extended Standard Model, and whether any singlets are present.   It would be interesting to conceive of a scenario in which a general 2HDM is incompatible with an observed signal but a different extended Higgs structure can successfully explain the observed data.  An obvious example would be the discovery of a doubly charged Higgs boson, which of course is absent in the 2HDM but present in models that contain hypercharge-two Higgs triplets~\cite{Gunion:1989ci}.  However, if the discovery of a new scalar at the LHC indicates the observation of a new colorless neutral or singly-charged scalar, it is not so clear what type of scenario is needed that would require new physics beyond the 2HDM.

Meanwhile, we look forward to new data being taken at Run 3 of the LHC to see whether any of the data excesses that appeared in Run 2 persist or whether any evidence for the production of new scalars emerges.  The discovery of a new scalar would have profound implications for physics at the electroweak scale and perhaps provide a first glimpse of the physics beyond the Standard Model that is necessary to address some of the most pressing problems associated with our current theory of fundamental particles and their interactions.

\section*{Acknowledgments}

We are very grateful for many valuable conversations with Wolfgang Altmannshofer, Henning Bahl, Mike Hance, and Sven Heinemeyer.  H.E.H. is supported in part by the U.S. Department of Energy Grant
No.~\uppercase{DE-SC}0010107.
H.E.H. also acknowledges the Aspen Center for Physics, which is supported by National Science Foundation grant PHY-2210452, where the revised version of this manuscript was completed.
P.M.F. is supported
by \textit{Funda\c c\~ao para a Ci\^encia e a Tecnologia} (FCT)
through contracts
UIDB/00618/2020, UIDP/00618/2020, CERN/FIS-PAR/0004/2019, CERN/FIS-PAR/0014/2019 and CERN/FIS-PAR/ 0025/2021.

\vskip 0.5in

\begin{appendices}

\section{The oblique $T$ parameter of the CP-conserving 2HDM}
\label{sec:Tparm}
\renewcommand{\theequation}{A.\arabic{equation}}
\setcounter{equation}{0}

In the CP-conserving 2HDM, the $T$ parameter is given by~\cite{Haber:2006ue}:
\beqa
\alpha T &=& \frac{3g^{\prime\,2}\cos^2(\beta-\alpha)}
{64\pi^2(m_Z^2-m_W^2)}\Biggl\{
\mathcal{F}(m_Z^2,m_{H}^2)
-\mathcal{F}(m_W^2,m_{H}^2)
-\mathcal{F}(m_Z^2,m_{h}^2) +\mathcal{F}(m_W^2,m_{m_h})\Biggr\}
\nonumber\\
&&\qquad\quad +\frac{g^2}{64\pi m_W^2} \Biggl\{
\mathcal{F}({m^2_{H^\pm}},m_{A}^2)+\sin^2(\beta-\alpha)\left[
\mathcal{F}({m^2_{H^\pm}},m_{H}^2)-\mathcal{F}({m^2_{A}},m_{H}^2)\right]
\nonumber \\[-10pt]
&&\qquad\qquad\qquad\qquad\qquad\quad
+\cos^2(\beta-\alpha)\left[
\mathcal{F}({m^2_{H^\pm}},m_{h}^2)-\mathcal{F}({m^2_{A}},m_{h}^2)\right]
\Biggr\}\,, \label{thdmtcp}
\eeqa
where $\alpha\simeq 1/137$ is the fine structure constant and the function
$\mathcal{F}$ is defined by
\beq
\mathcal{F}(m_1^2,m_2^2) \equiv \half(m_1^2+m_2^2)-\frac{m_1^2m_2^2}{m_1^2-m_2^2}
\ln\left(\frac{m_1^2}{m_2^2}\right)\,.
\eeq
Note that
\beq
\mathcal{F}(m_1^2,m_2^2)=\mathcal{F}(m_2^2,m^2_1)\,,\qquad\qquad
\mathcal{F}(m^2,m^2)=0\,.
\eeq
For a custodial symmetric scalar potential, the term proportional to
$g^2$ in \eq{thdmtcp} must vanish, i.e.
\beqa \label{vanish}
&& \mathcal{F}({m^2_{H^\pm}},m_{A}^2)+\sin^2(\beta-\alpha)\left[
\mathcal{F}({m^2_{H^\pm}},m_{H}^2)-\mathcal{F}({m^2_{A}},m_{H}^2)\right] \nonumber \\[6pt]
&& \qquad \qquad \qquad
+\cos^2(\beta-\alpha)\left[
\mathcal{F}({m^2_{H^\pm}},m_{h}^2)-\mathcal{F}({m^2_{A}},m_{h}^2)\right]
=0\,.
\eeqa
Note that if
$\sin(\beta-\alpha)\cos(\beta-\alpha)\neq 0$, the only
solution to \eq{vanish} is $m_{H^\pm}^2=m_{A}^2$.

Precision electroweak data implies that $T$ is close to zero~\cite{Tparm}.  Thus, we shall require that
the expression on the left hand side of \eq{vanish} must be close to zero.  One way this can be achieved is if $m_{H^\pm}^2\simeq m_{A}^2$.   However, the precision Higgs data implies that
$|\hspace{-0.125em}\cos(\beta-\alpha)|$ is small (under the assumption that the SM-like Higgs boson is the lighter of the two CP-even scalars).   Thus, a second way to achieve a very small value for the left hand side of \eq{vanish} is to demand that
\beq \label{vanish2}
|\mathcal{F}({m^2_{H^\pm}},m_{A}^2)+
\mathcal{F}({m^2_{H^\pm}},m_{H}^2)-\mathcal{F}({m^2_{A}},m_{H}^2)|=\mathcal{O}\bigl((\cos^2(\beta-\alpha)\bigr)\ll 1\,.
\eeq
\Eq{vanish2} is approximately satisfied if either $m_{H^\pm}^2\simeq m_{A}^2$ or $m_{H^\pm}^2\simeq m_{H}^2$.

\bigskip

\section{\texorpdfstring{$b\rightarrow s\gamma$}{b\textrightarrow s\lowercasegamma}
constraints on the CP-conserving A2HDM parameter space}
\label{sec:bsg}
\renewcommand{\theequation}{B.\arabic{equation}}
\setcounter{equation}{0}

In the A2HDM, tree-level Higgs-mediated FCNCs are absent due to the flavor-alignment conditions specified in \eq{aligned}.  However, Higgs-mediated FCNCs can be generated at the one-loop level due to charged Higgs boson exchange.   Indeed, there are numerous flavor observables that can potentially provide constraints on $m_{H^\pm}$  and the flavor-alignment parameters.  In Ref.~\cite{Enomoto:2015wbn}, a comprehensive analysis of the constraints on the A2HDM is provided based on the theoretical predictions and experimental analyses of a variety of processes: $B \rightarrow \tau \nu, D \rightarrow \mu \nu, D_{s} \rightarrow \tau \nu$, $D_{s} \rightarrow \mu \nu, K \rightarrow \mu \nu, \pi \rightarrow \mu \nu, B_{s}^{0} \rightarrow \mu^{+} \mu^{-}, B_{d}^{0} \rightarrow \mu^{+} \mu^{-}, \tau \rightarrow K \nu, \tau \rightarrow \pi \nu, \bar{B} \rightarrow X_{s} \gamma$, $K$--$\bar{K}$ mixing, $B_{d}^{0}$--$\bar{B}_{d}^{0}$ mixing, and $B_{s}^{0}$--$\bar{B}_{s}^{0}$ mixing.  A subset of these flavor observables has also been considered in Ref.~\cite{Penuelas:2017ikk}.

In a significant fraction of the parameter space, the main constraints on the A2HDM parameters can be obtained by comparing the experimental observation with the Standard Model prediction for the rate of inclusive radiative decay, $b\to s\gamma$
(or more precisely, the process $\bar{B} \rightarrow X_{s} \gamma$, where $X_s$ is any hadronic state that contains an $s$ quark).  In practice, one takes the observed photon energy $E_\gamma$ to be larger than some cutoff, $E_0$.   For example, the prediction for the branching ratio of $b\to s\gamma$ in the Standard Model obtained in Ref.~\cite{Misiak:2020vlo} is,
\beq  \label{Eq:bsgamma}
 {\rm BR}(b \to s \gamma)_{E_\gamma > E_0=1.6~{\rm GeV}} = (3.40 \pm 0.17) \times 10^{-4}\,,
 \eeq
 which is to be compared with the current world average of the experimentally measured branching ratio compiled by the HFLAV Collaboration~\cite{HFLAV:2022pwe},
\beq  \label{Eq:bsgammaExp}
 {\rm BR}(b \to s \gamma)_{E_\gamma > E_0=1.6~{\rm GeV}} = (3.49 \pm 0.19) \times 10^{-4}\,.
 \eeq

The SM prediction exhibited in \eq{Eq:bsgamma} is modified in the 2HDM due to charged Higgs boson exchange,
\beq
   {\rm BR}(b \to s \gamma)_{E_\gamma > E_0}= {\rm BR}(b \to s \gamma)^{\rm SM}+\delta {\rm BR}(b \to s \gamma)\,.
   \eeq
In the 2HDM, the dominant contributions to $\delta {\rm BR}$ arise through the effective operators
\beq
\mathcal O_7 = \frac{e}{16\pi^2}m_b(\bar s_L \sigma^{\mu\nu} b_R) F_{\mu\nu}\,,\qquad\quad
\mathcal O_8 = \frac{g_s}{16\pi^2}m_b(\bar s_L \sigma^{\mu\nu} t^a b_R) G_{\mu\nu}^a\,,
\eeq
corresponding to one-loop electroweak and QCD penguin diagrams, respectively, at lowest order (LO).
Next-to-leading order (NLO) corrections have also been obtained in Refs.~\cite{Ciuchini:1997xe,Ciafaloni:1997un,Borzumati:1998tg,Borzumati:1998nx,Bobeth:1999ww}.
A convenient numerical formula  based on Refs.~\cite{Kagan:1998ym,Hurth:2003dk,Lunghi:2006hc} has been provided in Ref.~\cite{Enomoto:2015wbn} in terms of the Wilson coefficients evaluated at the scale $\mu_t=160$~GeV,
\begin{align}
  & \delta {\rm BR}(b\to s\gamma)  = 10^{-4} \times \left( \frac{r_V^{\,}}{0.9626_{\,}} \right) \text{Re}
 \Bigg [
 -8.100\, \mathcal C_7^\text{LO} -2.509\, \mathcal C_8^\text{LO} +2.767\, \mathcal C_7^\text{LO} \mathcal C_8^{\text{LO}*}  \notag \\
 & \hspace{6em}  + 5.348 \left|\mathcal C_7^\text{LO} \right|^2 +0.890\, \left|\mathcal C_8^\text{LO}\right|^2 -0.085\, \mathcal C_7^\text{NLO} -0.025\, \mathcal C_8^\text{NLO}  \\
 & \hspace{6em}  +0.095\, \mathcal C_7^\text{LO} \mathcal C_7^\text{NLO*} +0.008\, \mathcal C_8^\text{LO} \mathcal C_8^\text{NLO*}
    +0.028\, \left( \mathcal C_7^\text{LO} \mathcal C_8^\text{NLO*} +\mathcal C_7^\text{NLO} \mathcal C_8^\text{LO*} \right)
 \Bigg] \,, \notag
 \label{Eq:bsgammaTHDM}
\end{align}
where $\mathcal{C}_i^{\rm LO}$ and $\mathcal{C}_i^{\rm NLO}$ indicate the charged Higgs  contributions from $\mathcal O_i$ for $i=7,8$ at leading and next-to-leading order, respectively, and
$r_V$ is the ratio of the product of CKM matrix elements~\cite{ParticleDataGroup:2022pth},
\begin{align}
 r_V \equiv \left| \frac{V_{ts}^*V_{tb}}{V_{cb}} \right|^2\simeq  0.964\,.
\end{align}

In the A2HDM, the forms of $\mathcal{C}_i^{\rm LO}$ and $\mathcal{C}_i^{\rm NLO}$ for
$i=7, 8$ are given by,
\beqa
 \mathcal C_i^\text{LO} &=& \tfrac13 (a^U)^2  \, G_1^i (y_{H^\pm}^t) - a^U a^D\, G_2^i (y_{H^\pm}^t) \,,
 \label{Eq:bsgLO}  \\
 \mathcal C_i^\text{NLO} &=&  ( a^U )^2\, C_1^i (y_{H^\pm}^t) -a^U a^D \, C_2^i (y_{H^\pm}^t) +\Bigl[ (a^U)^2 D_1^i (y_{H^\pm}^t) - a^U a^D D_2^i (y_{H^\pm}^t) \Bigr] \ln \frac{\mu_t^2}{m_{H^\pm}^2} \,, \phantom{xxxxxx}
 \label{Eq:bsgNLO}
\eeqa
where $y_{H^\pm}^t \equiv m_t^2/m_{H^\pm}^2$ and the explicit expressions for the
loop functions $G_a^i$, $C_a^i$, and $D_a^i$ are given in
Appendix 5.1 of Ref.~\cite{Enomoto:2015wbn}.

In light of \eqs{Eq:bsgamma}{Eq:bsgammaExp}, the contributions to $\delta {\rm BR}(b\to s\gamma)$ from new physics beyond the Standard Model cannot be larger than a few times $10^{-5}$.  However, in order to rule out a particular new physics scenario at roughly 95\% CL, one should consider the 2$\sigma$ error bars on the SM prediction and the experimental observation.  In this paper, we have imposed the requirement that
\beq \label{deltalim}
|\delta {\rm BR}(b\to s\gamma)|\leq 4\times 10^{-5}\,.
\eeq
Using \eq{deltalim}, one can constrain the $\{a^U,a^D,m_{H^\pm}\}$ parameter space of the CP-conserving A2HDM.  Applying these constrains to the special cases of $a^U=a^D$ and $a^U=-1/a^D$ yields the $b\to s\gamma$ constraints of the Type-I and II 2HDM, respectively.

\begin{figure}[t!]
\centering
\includegraphics[height=8cm,angle=0]{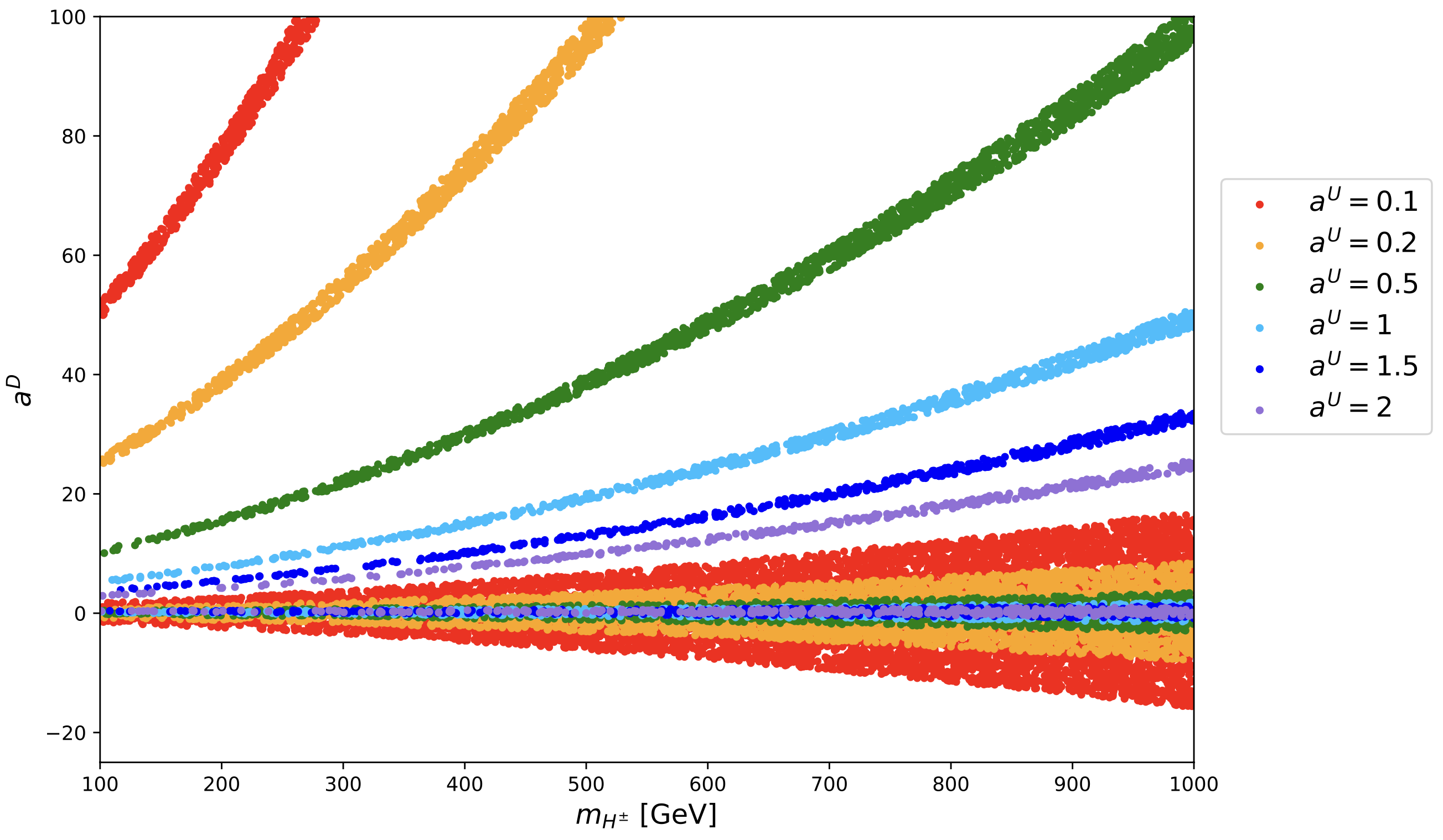}
  \caption{\small Regions of the A2HDM parameter space that satisfy $|\delta {\rm BR}(b\to s\gamma)|\leq 4\times 10^{-5}$.
  \label{fig:adautotal}}
\end{figure}

The most recent $b\to s\gamma$ constraints of the Type-I and II 2HDM can be found in Refs.~\cite{Arbey:2017gmh,Misiak:2020vlo}.  More general A2HDM constraints
can be found in Refs.~\cite{Enomoto:2015wbn,Eberhardt:2020dat,Karan:2023kyj}.  In this appendix, we shall provide more detailed plots of the $b\to s\gamma$ constraints on the CP-conserving A2HDM.   These constraints are then used in our survey of 2HDM models that are consistent with the two scenarios examined in
Sections~\ref{sec:scen1} and \ref{sec:scen2}.
Note that \eqs{Eq:bsgLO}{Eq:bsgNLO} are invariant under simultaneously transforming $a^U\to -a^U$ and $a^D\to -a^D$.\footnote{Changing the sign $a^F\to -a^F$ in the charged Higgs Yukawa couplings given in \eq{YUK5} corresponds to changing the sign of $\varepsilon$ [cf.~\eq{twosigns}] or equivalently changing the sign of the Higgs basis field $\mathcal{H}_2$, which has no physical consequence.
However, the sign of the product $a^U a^D$ is physical.}
For convenience, we shall henceforth take $a^U$ positive without loss of generality.
In our exploration of the A2HDM parameter space, we generously consider values of $\{a^U,a^D\}$ such that $0<a^U<2$ and $|a^D|<100$.

\begin{figure}[ht!]
\includegraphics[height=5.55cm,angle=0]{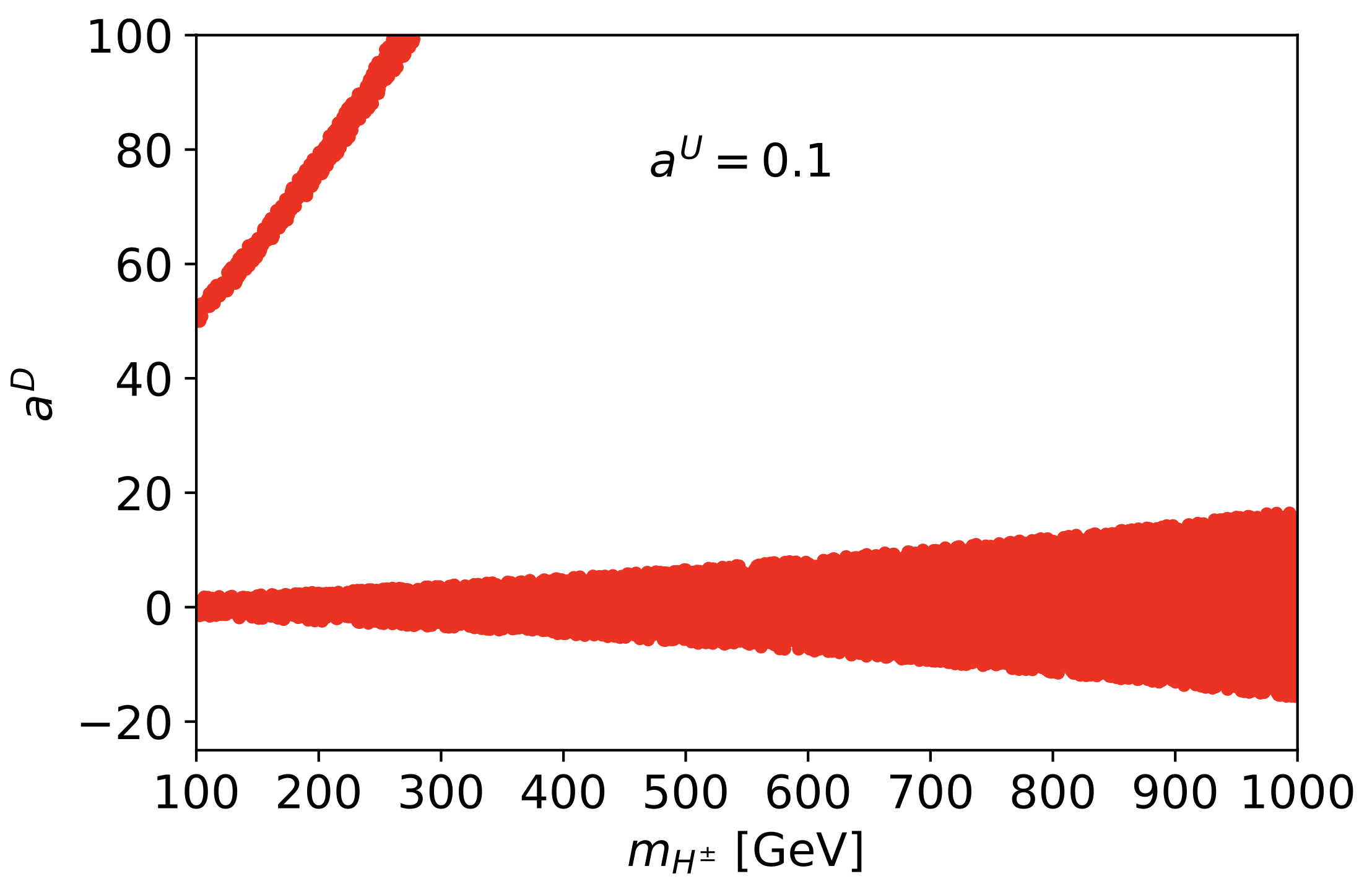}
\includegraphics[height=5.55cm,angle=0]{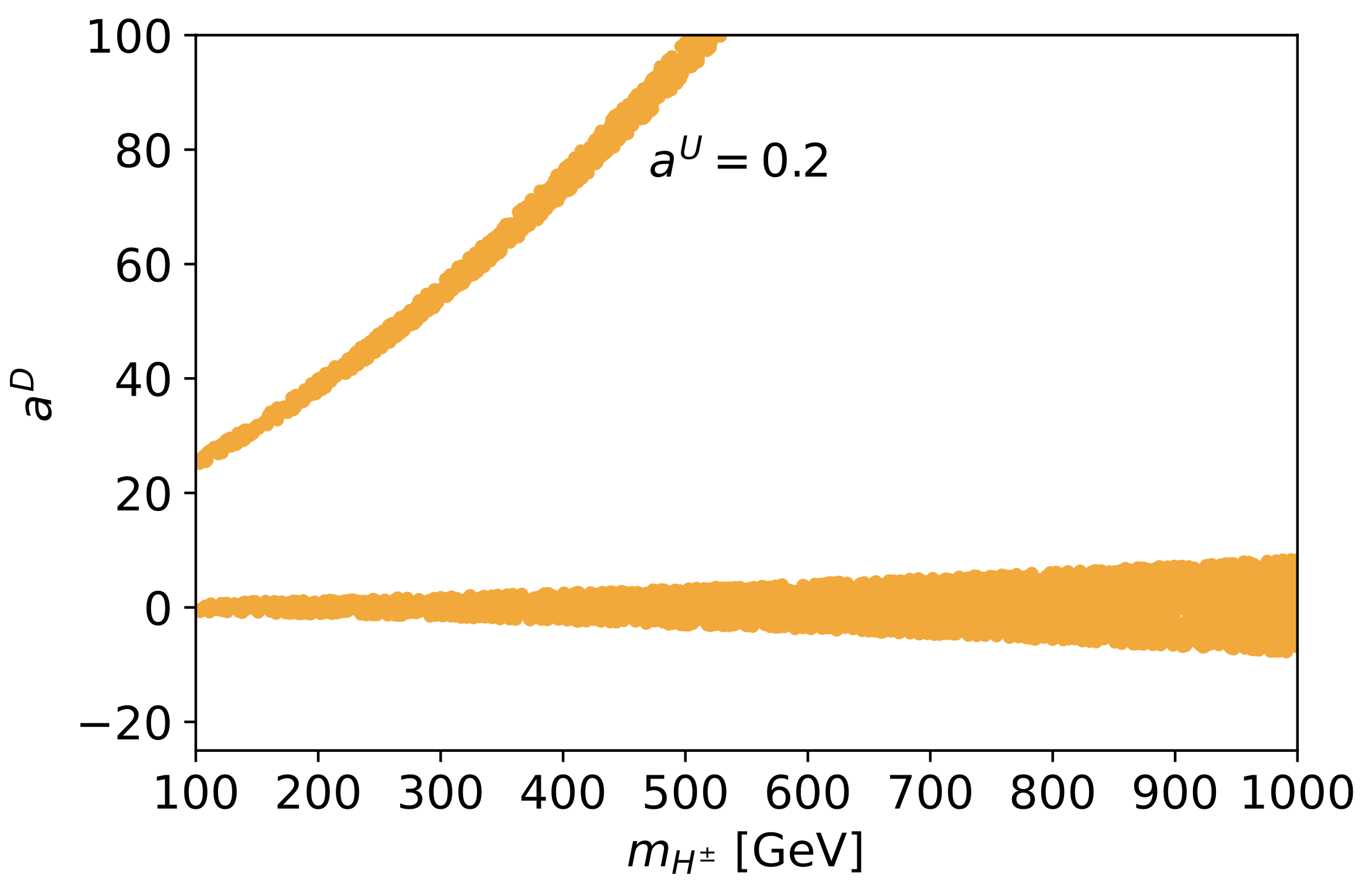}
\includegraphics[height=5.55cm,angle=0]{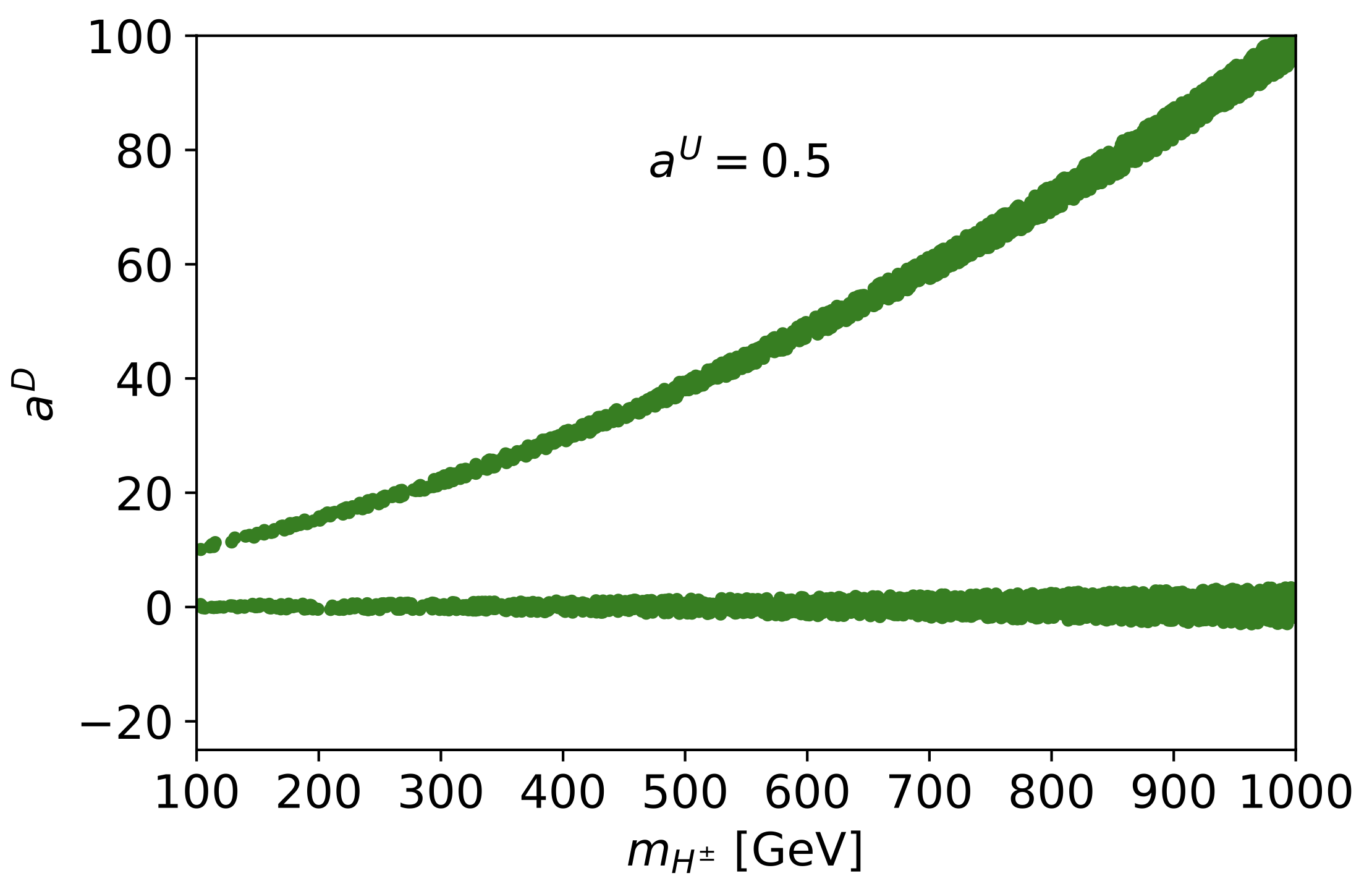}
\includegraphics[height=5.55cm,angle=0]{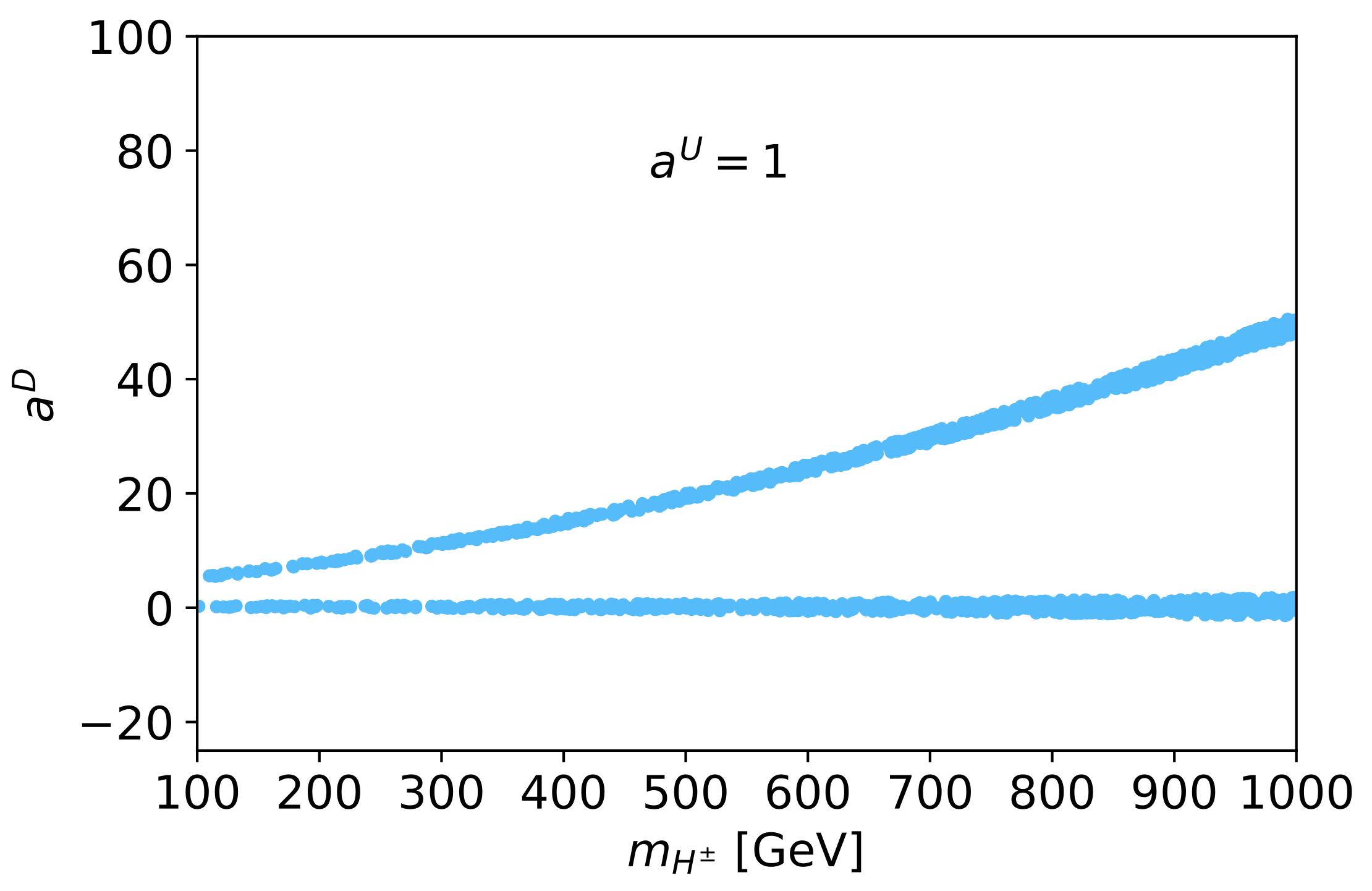}
\includegraphics[height=5.55cm,angle=0]{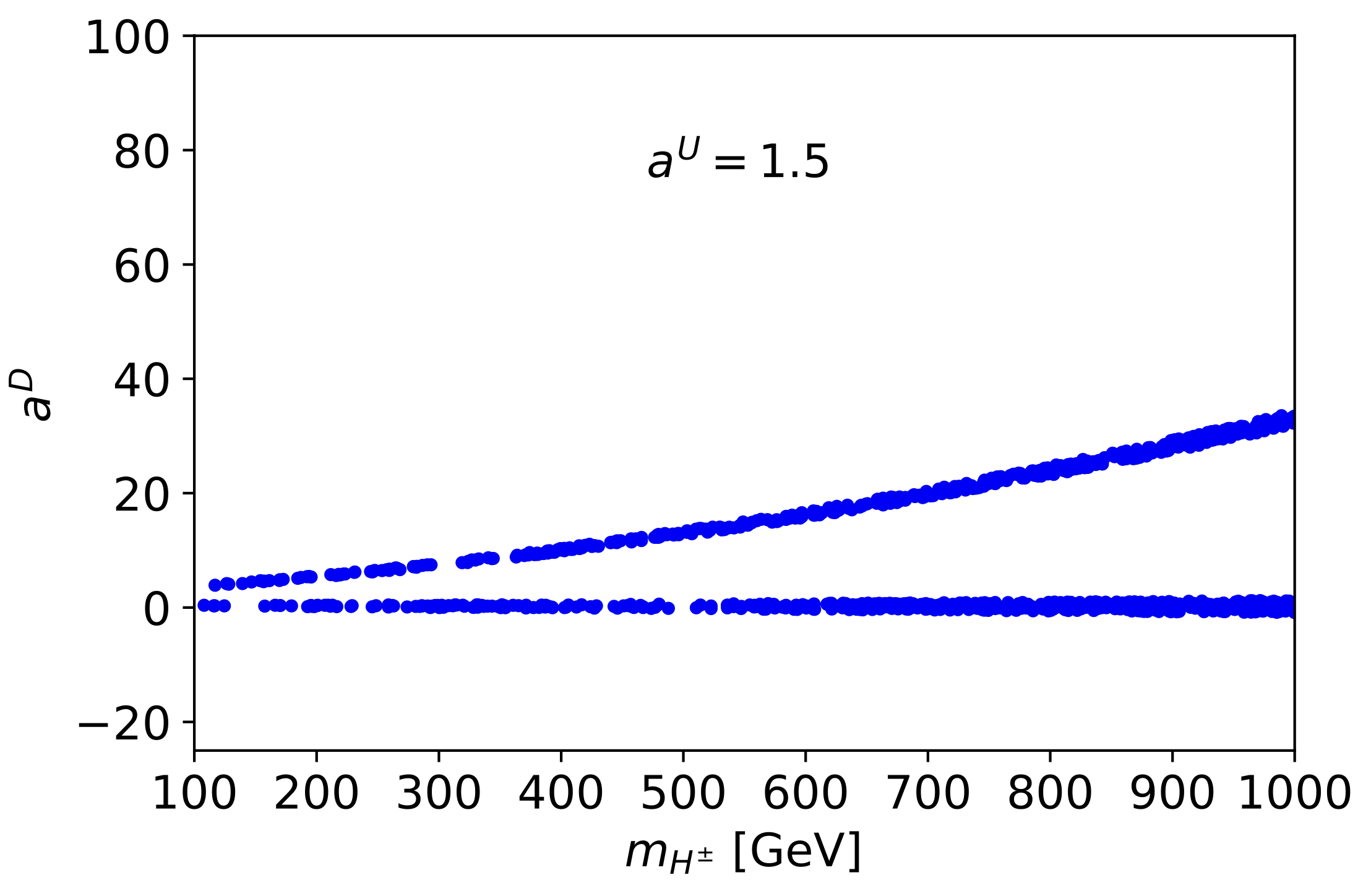}
\includegraphics[height=5.55cm,angle=0]{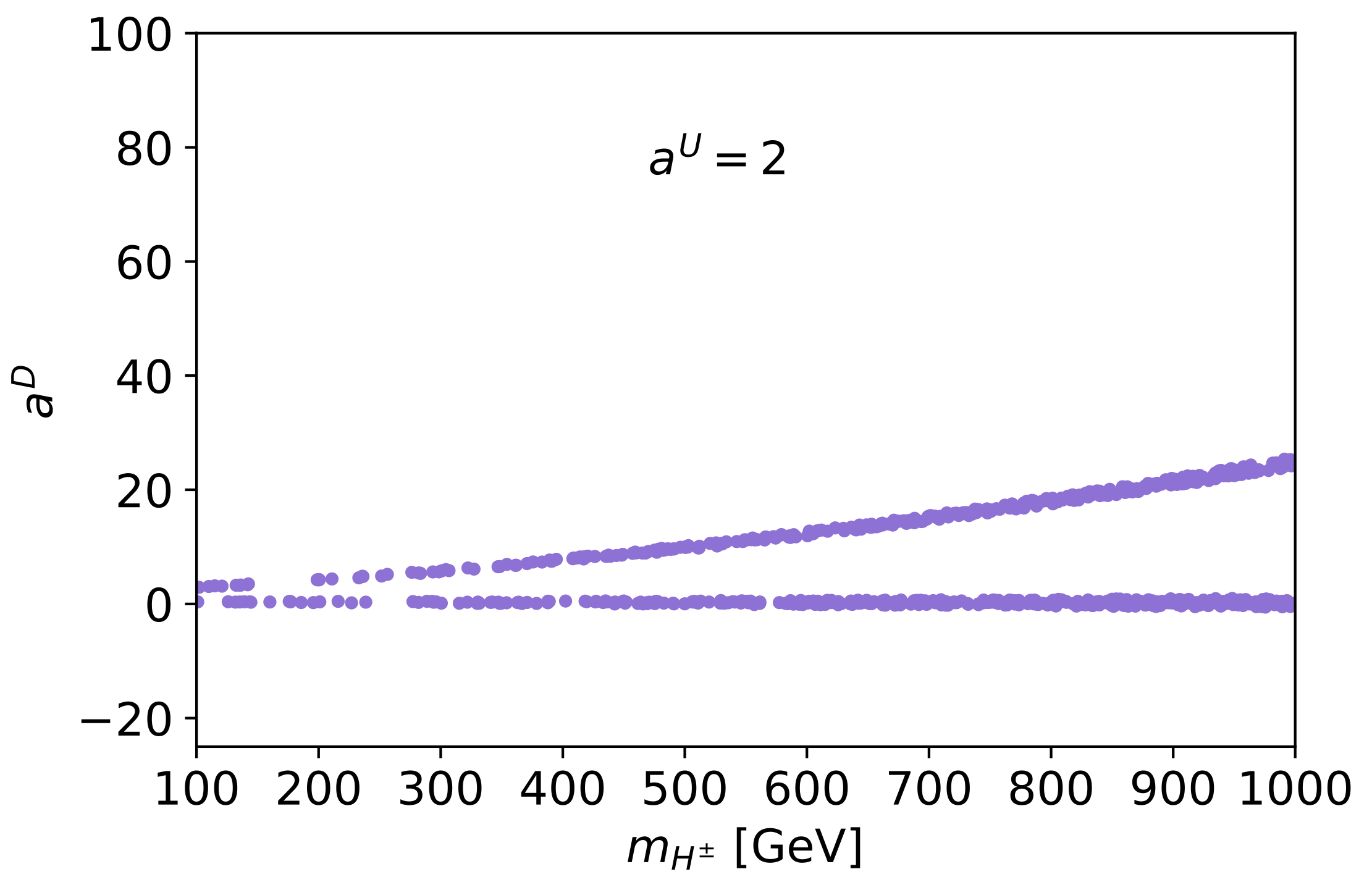}
  \caption{\small Regions of the A2HDM parameter space that satisfy $|\delta {\rm BR}(b\to s\gamma)|\leq 4\times 10^{-5}$
  for fixed values of $a^U=0.1$, 0.2, 0.5, 1.0, 1.5 and 2.0.  Combining the six panels above yields the plot shown in Fig.~\ref{fig:adautotal}.
  \label{fig:adau}}
\end{figure}

In Fig.~\ref{fig:adautotal}, we show the allowed regions of the A2HDM parameter space that satisfy the $b\to s\gamma$ constraint specified in \eq{deltalim}.
To better see the evolution of these results as $a^U$ increases, we have exhibited six separate panels in Fig.~\ref{fig:adau}, each one corresponding to a different fixed value of $a^U$.~~It is

\begin{figure}[t!]
\includegraphics[height=5.55cm,angle=0]{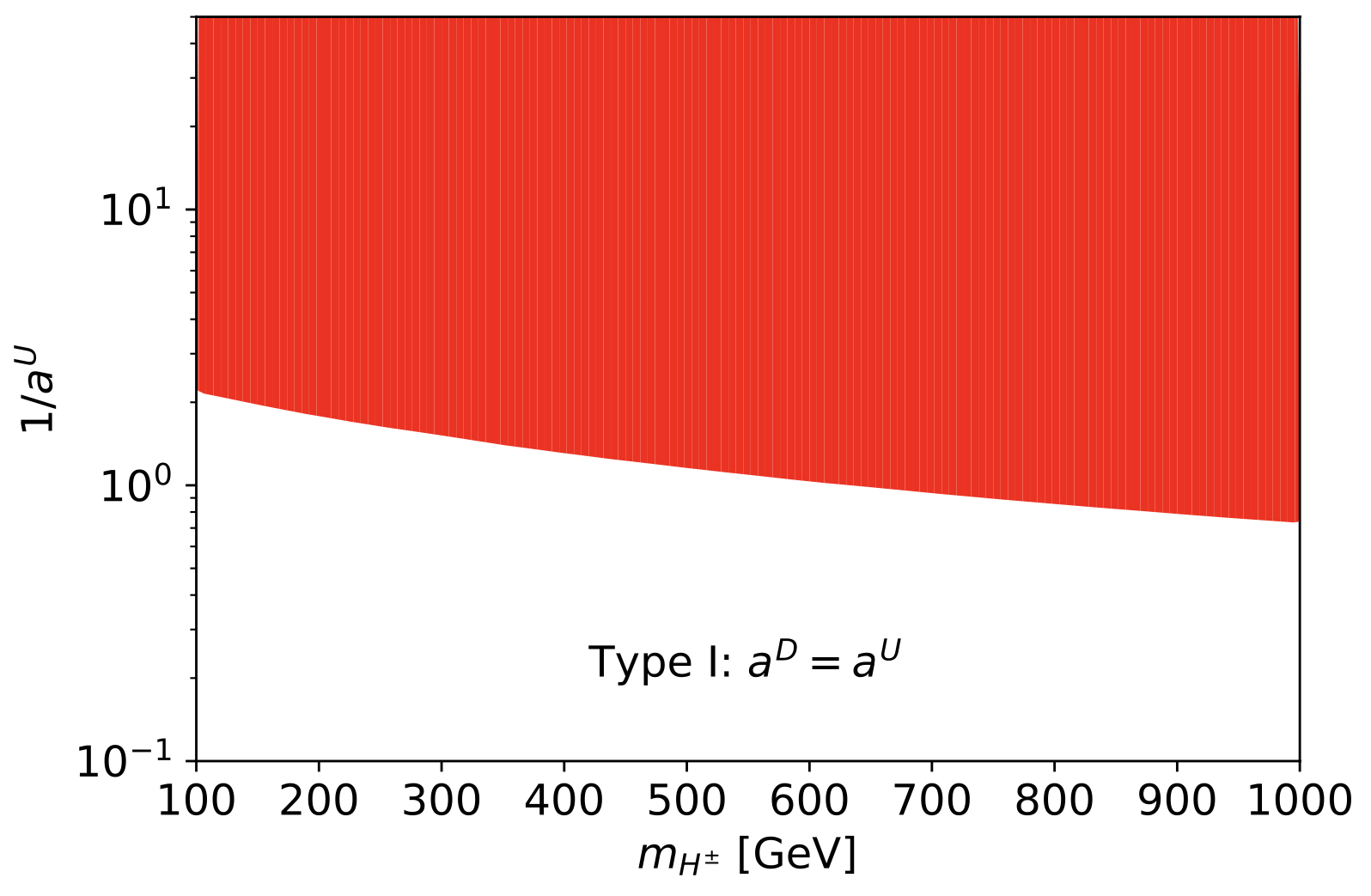}
\includegraphics[height=5.55cm,angle=0]{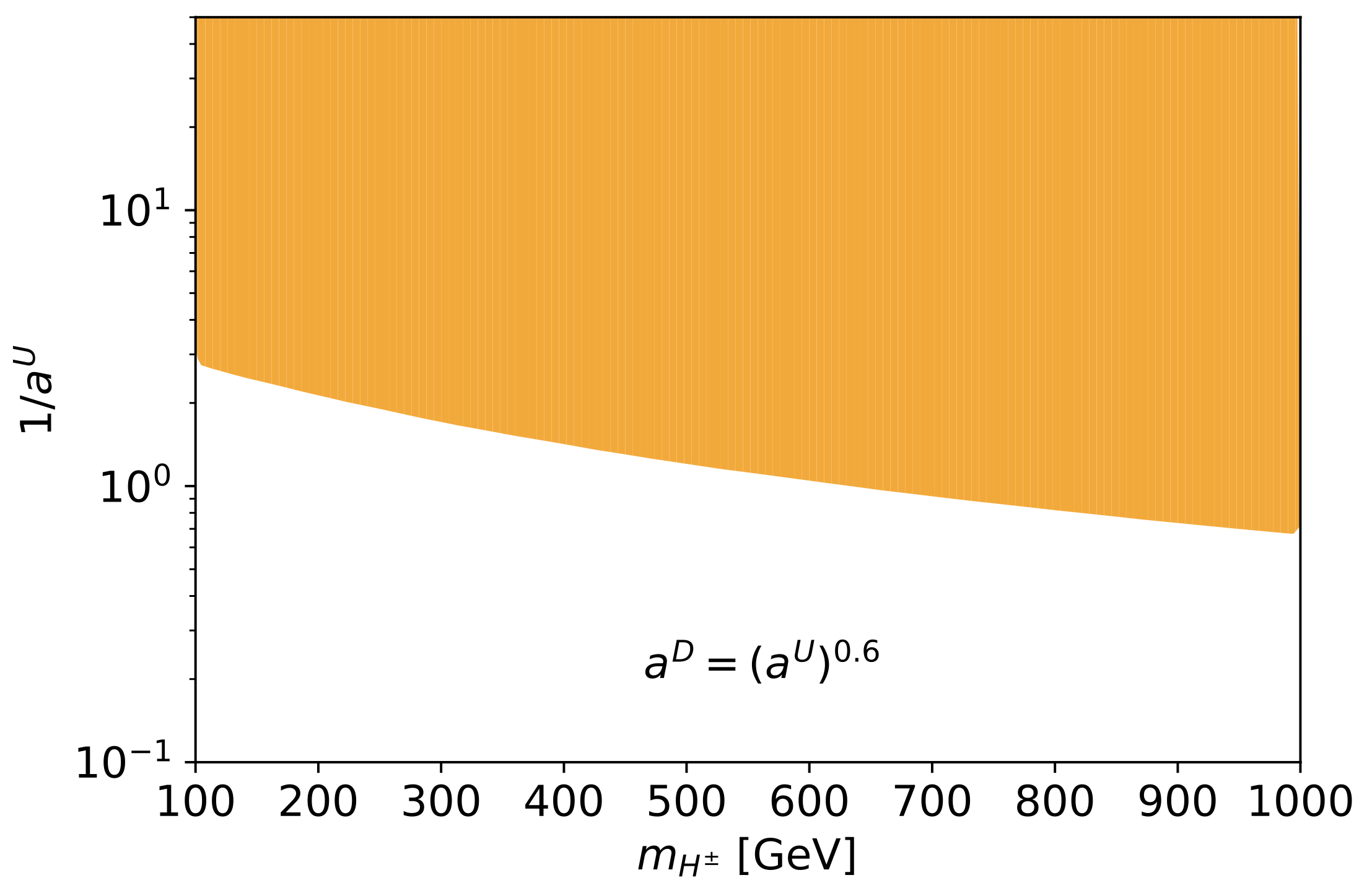}
\includegraphics[height=5.55cm,angle=0]{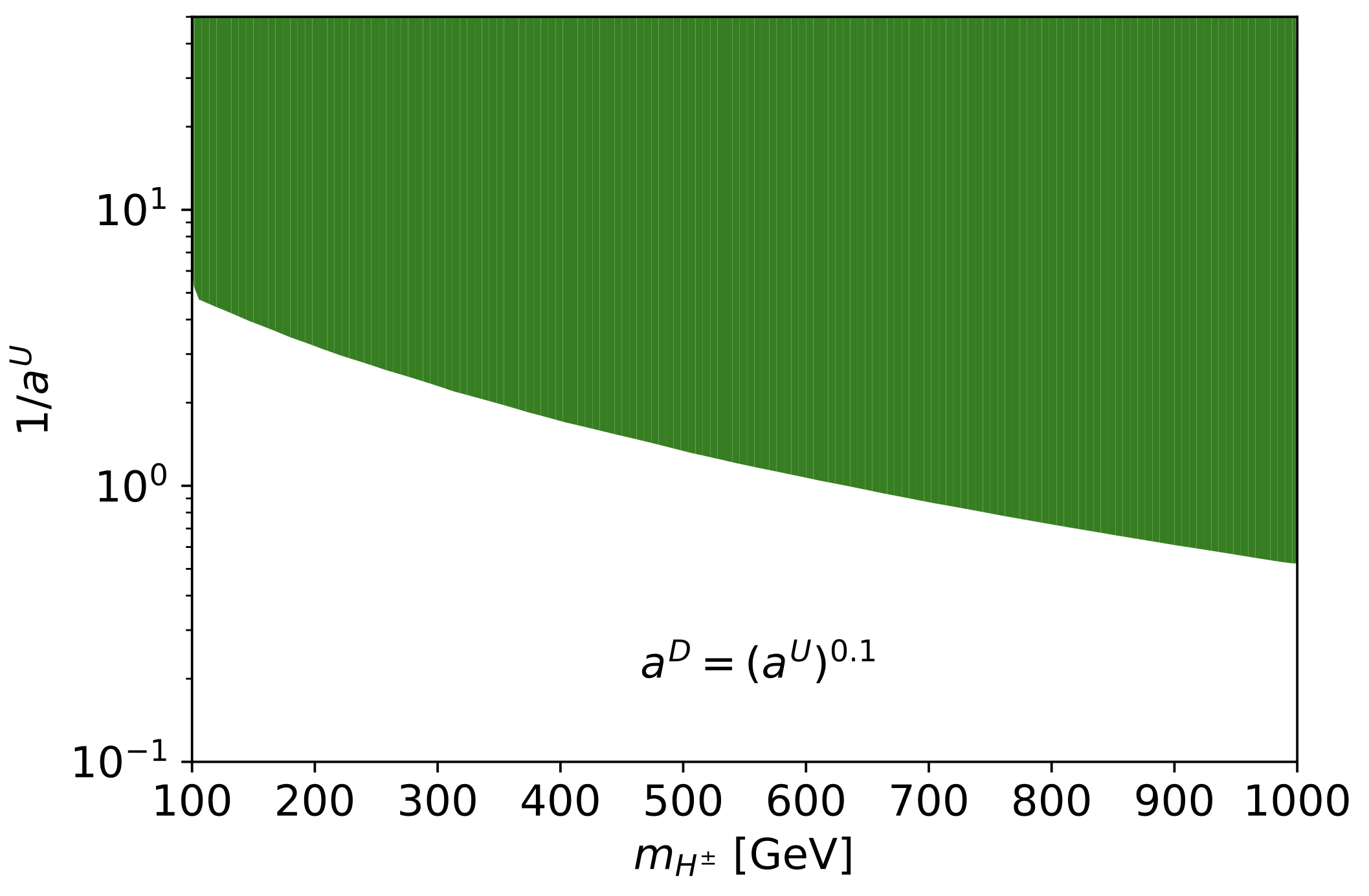}
\includegraphics[height=5.55cm,angle=0]{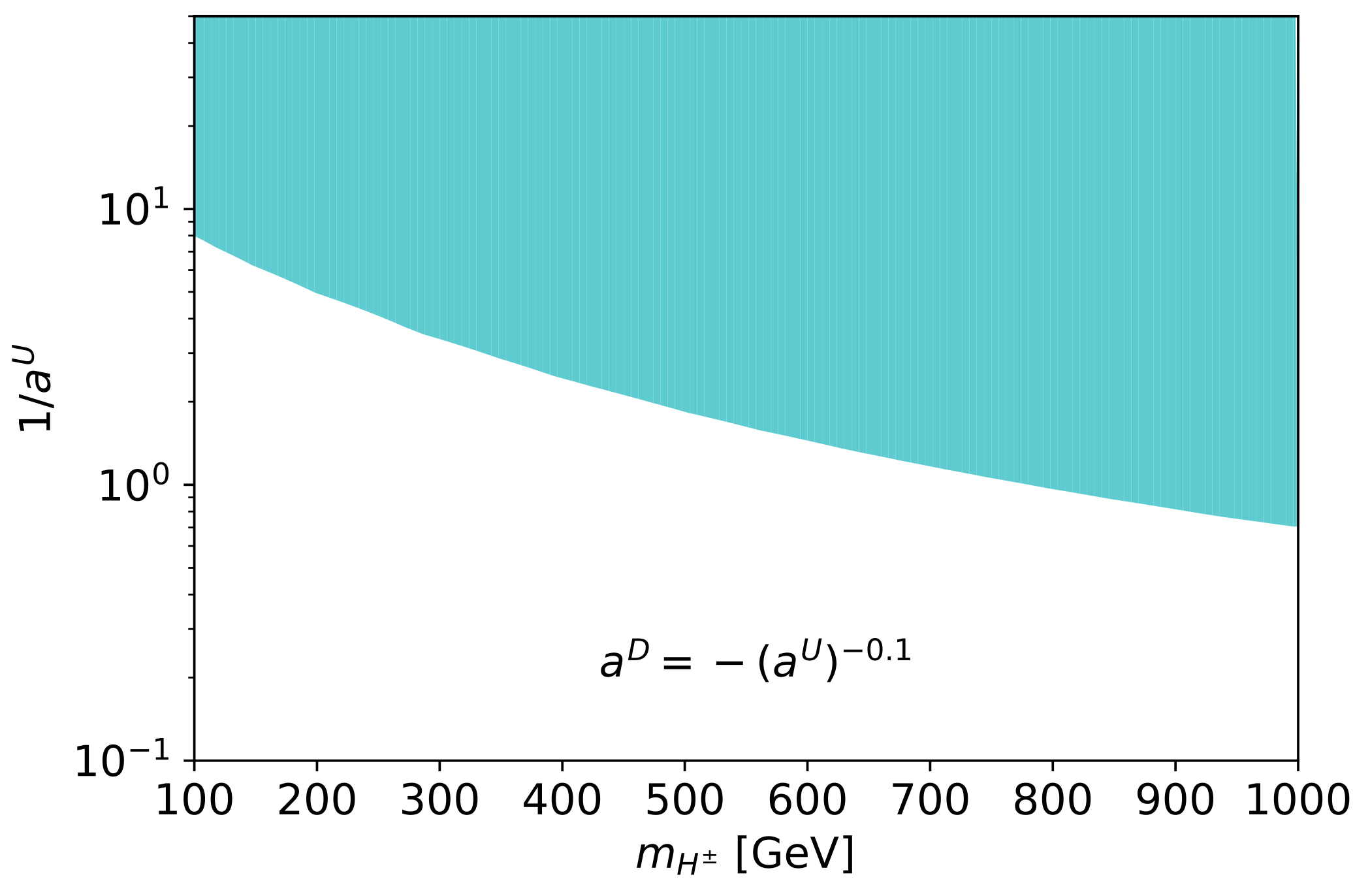}
\includegraphics[height=5.55cm,angle=0]{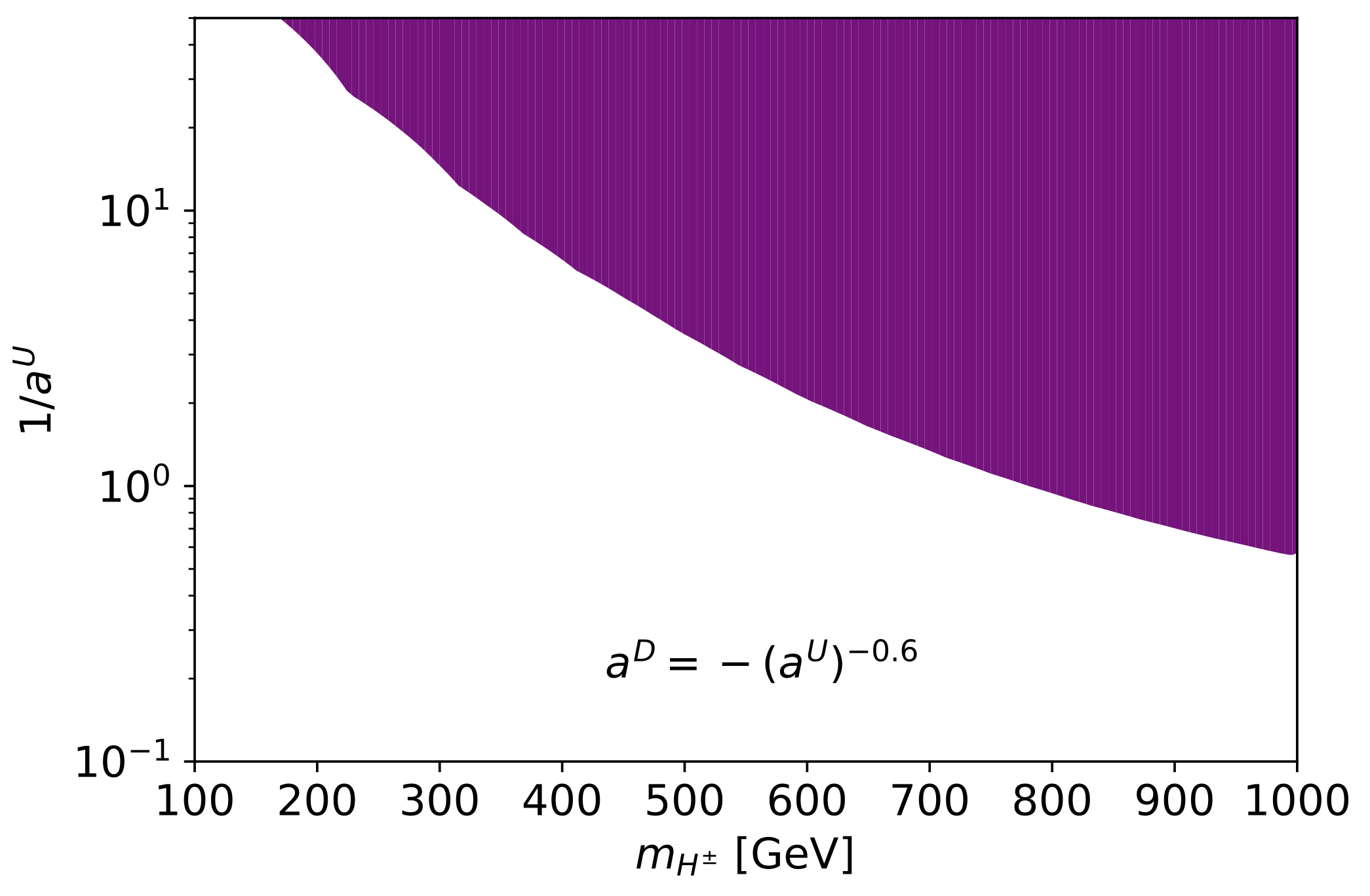}
\includegraphics[height=5.55cm,angle=0]{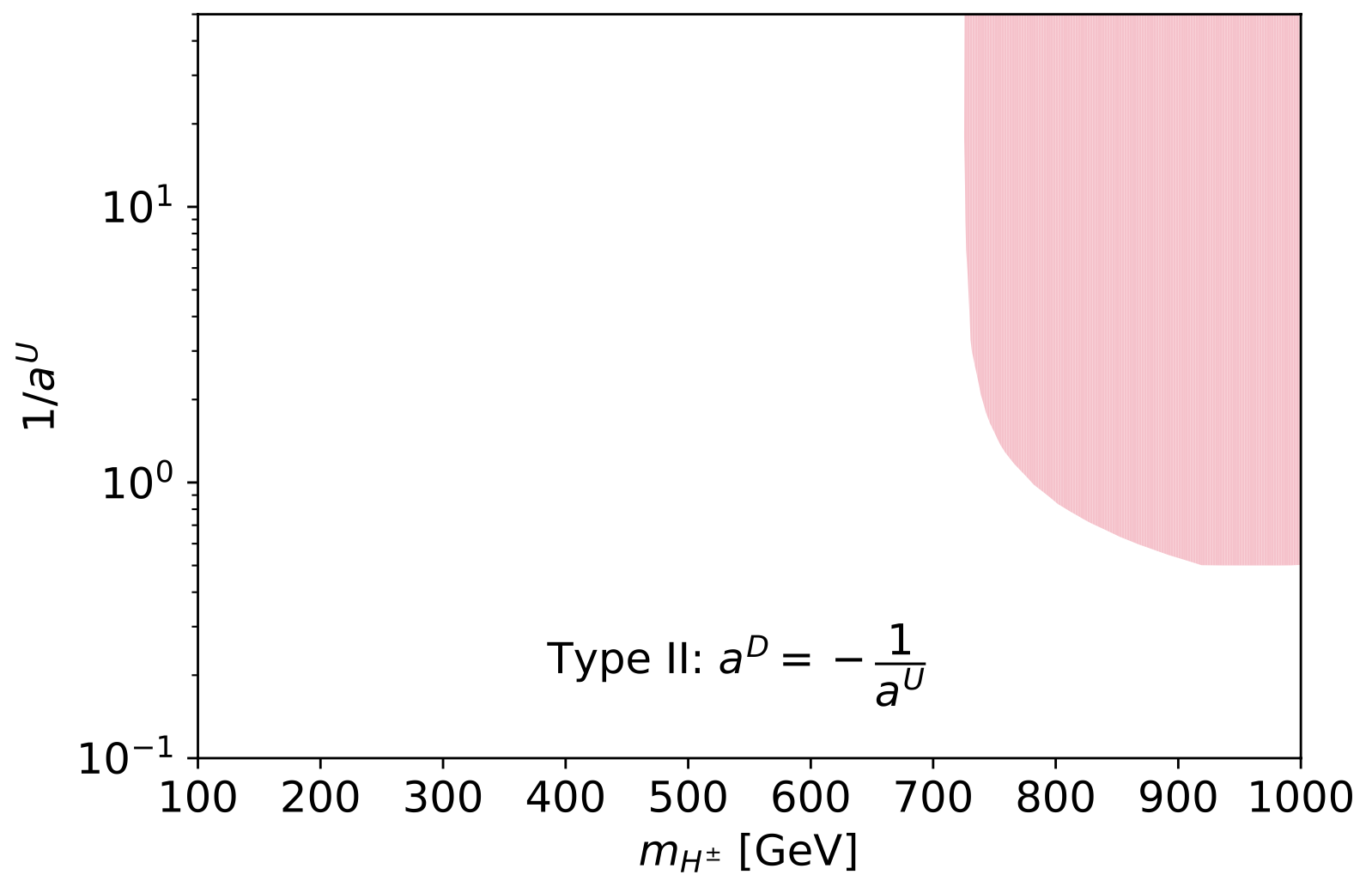}
   \caption{\small Regions of the A2HDM parameter space (indicated by the various colors)
   in the $m_{H^\pm}$ vs.~$1/a^U$ plane that satisfy $|\delta {\rm BR}(b\to s\gamma)|\leq 4\times 10^{-5}$.   The uncolored regions are excluded.  The value of $a^D$ is fixed by $a^D=(a^U)\rs{p}\sgn p$.   As $p$ varies, we include all parameter points in which $|a^D|<100$.   The sequence of panels correspond to
$p= 1, 0.6, 0.1, -0.1, -0.6$, and $-1$.  The case of $p=1$ [$p=-1$] corresponds to the Type I [Type II] 2HDM.  In these two cases,
in a convention where $\tan\beta$ is positive, one may identify $\tan\beta=1/a^U$.
  \label{fig:typesevolve}}
\end{figure}

\noindent
instructive to recover the constraints of the Type I and II 2HDM from the more general A2HDM constraints shown in Fig.~\ref{fig:adau}.
It is possible to illustrate the evolution from Type~I to Type II inside the A2HDM parameter space by employing the following parametrization (in a convention where $a^U>0$),
\beq \label{pee}
a^D=(a^U)\rs{p}\sgn p\,,\quad -1\leq p\leq 1\,.
\eeq
where $\sgn p=1$ for $p>0$ and $\sgn p=-1$ for $p<0$.
Note that $p=1$ corresponds to Type I Yukawa couplings whereas $p=-1$ corresponds to Type II Yukawa couplings.  By varying $p$, one can determine $a^D$ via \eq{pee} [subject to
$|a^D|<100$].  The evolution of the Type-I 2HDM constraints into the Type-II constraints as $p$ varies from $+1$ to $-1$ is shown in Fig.~\ref{fig:typesevolve}.  We show results in the $m_{H^\pm}$ vs.~$1/a^U$ plane, since in both the Type-I and Type-II 2HDM,
we can identify
$\tan\beta=1/a^U$.\footnote{More precisely, $\tan\beta=\varepsilon/a^U$ in light of \eqs{typeone}{typetwo}.  Having chosen $a^U>0$, it then follows that $\varepsilon=1$ in a convention where $\tan\beta$ is positive.}
Indeed, we see that the $a^D=a^U$ panel of Fig.~\ref{fig:typesevolve} is consistent with the
excluded parameter regime of the Type-I 2HDM, whereas
the $a^D=-1/a^U$ panel of Fig.~\ref{fig:typesevolve} is consistent with the
excluded parameter regime of the Type-II 2HDM (cf.~Ref.~\cite{Arbey:2017gmh}).
For values of $|p|\neq 1$, $1/a^U$ does not have the interpretation of $\tan\beta$ (as this parameter is no longer physical).  Nevertheless, the sequence of panels exhibited in Fig.~\ref{fig:typesevolve} provides some understanding on how the evolution between Type-I and Type-II occurs.\footnote{Strictly speaking, the evolution is not continuous, since at $p=0$, the sign of $p$ is undefined and one switches between positive and negative $p$ as one passes through zero.  Indeed, only half of the A2HDM parameter space is accessed in this way, since we do not consider parameter points where the sign of $a^D$ is $-\sgn p$ (in the convention of positive $a^U$).}

\begin{figure}[t!]
\includegraphics[height=5.85cm,angle=0]{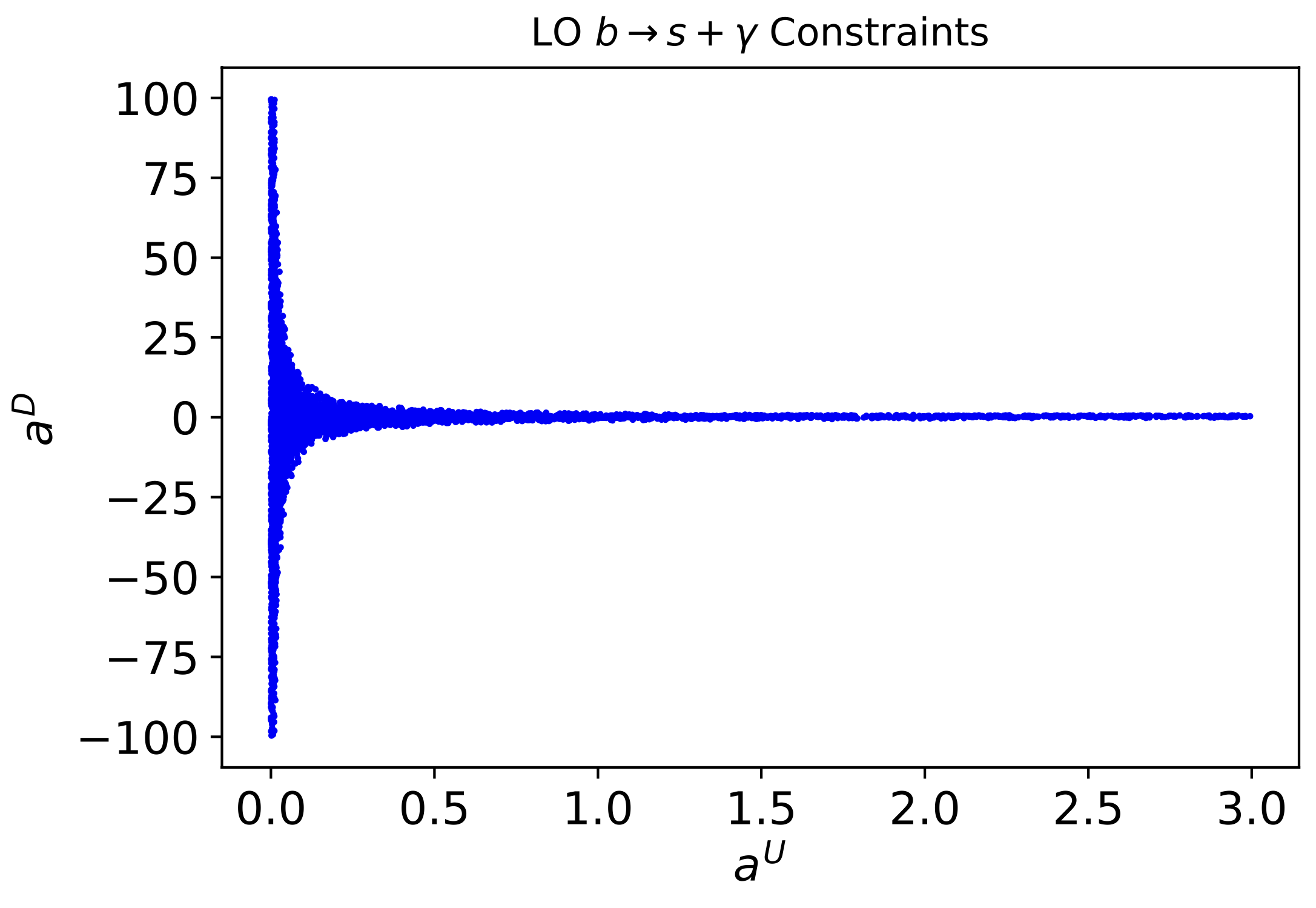}
\includegraphics[height=5.85cm,angle=0]{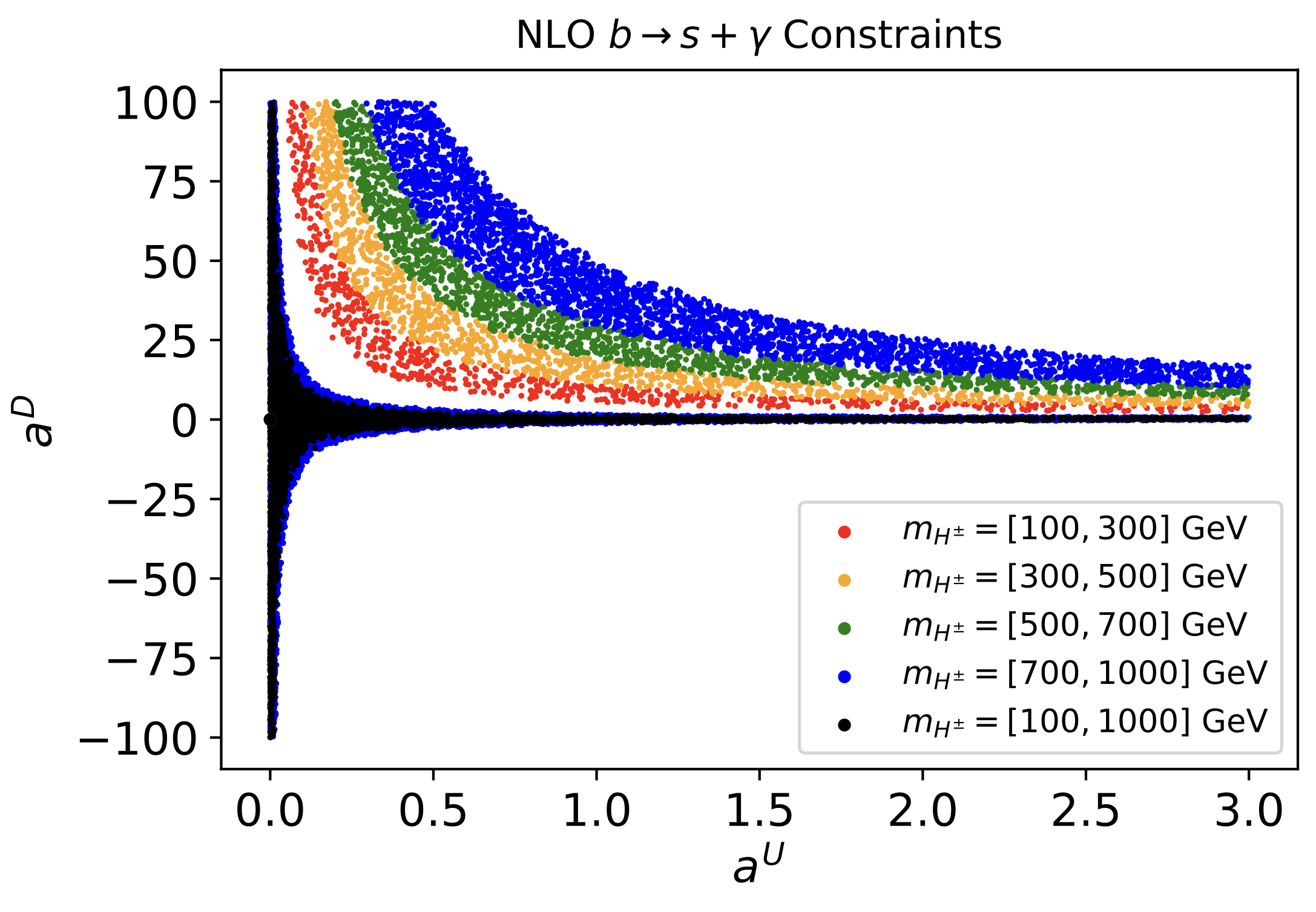}
  \caption{\small LO vs.~NLO constraints due to $b\to s\gamma$ in the $a^U$ vs.~$a^D$ plane, with charged Higgs masses separated by color.
  All points shown above satisfy $|\delta{\rm BR}(b\to s\gamma)|\leq 4\times 10^{-5}$.
   \label{fig:LOvsNLO}}
\end{figure}

In some earlier works, only the leading order (LO) corrections to $b\to s\gamma$ were included.   Although the LO results provide a fairly good representation of the excluded regions in some of the parameter regimes, there are noticeable differences with the more accurate NLO result.  In Fig.~\ref{fig:LOvsNLO}, we exhibit the regions of the $a^U$ vs.~$a^D$ parameter space in which $|\delta{\rm BR}(b\to s\gamma)|\leq 4\times 10^{-5}$ based on the LO computation (where $C_i^{\rm NLO}=0$ in \eq{Eq:bsgammaTHDM}) and the NLO computation, respectively.  A blue point is plotted
in Fig.~\ref{fig:LOvsNLO}(a) and (b) as long as the branching ratio inequality is satisfied for at least one value of the charged Higgs mass (which is allowed to vary between 100 and 1000 GeV).

\section{\texorpdfstring{$\Delta}{\uppercaseDelta}M_{B_s}$ constraints on the CP-conserving A2HDM parameter space}
\label{sec:MsubS}
\renewcommand{\theequation}{C.\arabic{equation}}
\setcounter{equation}{0}

Among the flavor observables discussed in Ref.~\cite{Enomoto:2015wbn}, the observation of $b\to s\gamma$, which is consistent with the prediction of the Standard Model,  provides the most important constraint  on the A2HDM parameter space.   However, there exists some regions of this parameter space where the consideration of $B_s$--$\overline{B}_s$ adds an additional constraint beyond what is excluded by $b\to s\gamma$.

In Ref.~\cite{UTfit:2022hsi}, the observed value of $\Delta M_{B_s}=17.241(20)~{\rm ps}^{-1}$ obtained from $B_s$--$\overline{B}_s$ oscillation data is compared with the Standard Model prediction, $17.94(69)~{\rm ps}^{-1}$ based on a global fit of flavor observables.  Since the error in the Standard Model prediction is still considerably larger than the precision of the measured value, we chose to identify the 2$\sigma$ error in the theoretical prediction as the upper limit to the contribution to $|\Delta M_{B_s}|$ of new physics beyond the Standard Model.
Using the results of Ref.~\cite{UTfit:2022hsi}, the contribution to $\Delta M_{B_s}$ due to the contributions of the charged Higgs boson arise through the effective operators
\beqa
\mathcal{O}_{V L L}&=&\bar{s}^\alpha \gamma_\mu\left(1-\gamma\ls{5}\right) b^\alpha \bar{s}^\beta \gamma^\mu\left(1-\gamma\ls{5}\right) b^\beta\,, \\
\mathcal{O}_{S R R}&=&\bar{s}^\alpha\left(1+\gamma\ls{5}\right) b^\alpha \bar{s}^\beta\left(1+\gamma\ls{5}\right) b^\beta\,, \\
\mathcal{O}_{T R R}&=&\bar{s}^\alpha \sigma_{\mu \nu}\left(1+\gamma\ls{5}\right) b^\alpha \bar{s}^\beta \sigma^{\mu \nu}\left(1+\gamma\ls{5}\right) b^\beta\,,
\eeqa
and is given by,
\beqa
\Delta M_{B_s} &\equiv & 2|\bra{B_s^0}H^{\Delta B=2}\ket{\overline{B}_s^0}|=(\Delta M_{B_s})^{\rm SM}+\delta\Delta M_{B_s}\,,
\eeqa
\beqa
\delta\Delta M_{B_s} &=& \frac{G_F^2 m_W^2 m_{B_s}}{24\pi^2} |V_{tq} V^*_{tb}|^2 f^2_{B_s}\bigl[\hat{B}_{B_s}\eta_B\mathcal{C}_V+\hat{B}^{ST}_{B_s}\eta_{B_s}^{ST}\mathcal{C}_{ST}\bigr]\,,\label{dms}
\eeqa
where the $\hat{B}_{B_s}$ [$\hat{B}_{B_s}^{ST}$] parametrize the nonperturbative effects in the hadronic matrix element of $\mathcal{O}_{V L L}$ [$\mathcal{O}_{S R R}$ and $\mathcal{O}_{T R R}$],  $\eta_B$ [$\eta^{ST}_{B_s}$] account for NLO QCD corrections~\cite{Buras}, and the corresponding Wilson coefficients
are given by\footnote{The Standard Model contribution to $\mathcal{C}_V$ has been omitted from \eq{Wilson1}.}
\beq
\begin{aligned}
\mathcal{C}_V & =x_t\left[2 x_t A_{W H}\left(x_t, x_b\right)+x_t A_{H H}\left(x_t, x_b\right)\right], \label{Wilson1}\\
\mathcal{C}_{S T} & =4 x_b x_t^2\left[A_{W H}^{S T}\left(x_t\right)+A_{H H}^{S T}\left(x_t\right)\right],
\end{aligned}
\eeq
with $x_q\equiv [m_q(m_q)]^2/m_W^2$ equal to the square of the $\overline{\rm MS}$ quark mass normalized to the $W$ boson mass.   The explicit expressions for $A_{WH}$, $A_{HH}$, $A_{WH}^{ST}$ and $A_{HH}^{ST}$ can be found in Appendix B.2 of Ref.~\cite{Enomoto:2015wbn}.
We have evaluated these functions employing the Standard Model parameters taken from the UT\textit{fit}
Collaboration global fit of flavor parameters~\cite{UTfit:2022hsi} and the parameters associated with the charged Higgs contributions given in Ref.~\cite{Enomoto:2015wbn}.\footnote{Values for $G_F$ and $m_W$ are taken from Ref.~\cite{ParticleDataGroup:2022pth} and we employ the value of $\eta_B=0.5510\pm 0.0022$ quoted in Ref.~\cite{Buras}.}

\begin{figure}[t!]
\begin{tabular}{cc}
\hspace{-0.3in}
\includegraphics[height=6.45cm,angle=0]{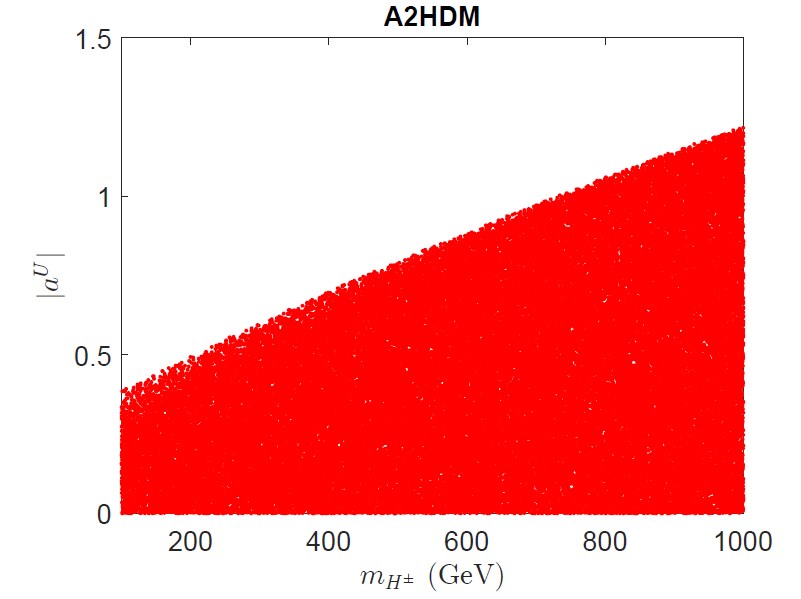}&
\includegraphics[height=6.45cm,angle=0]{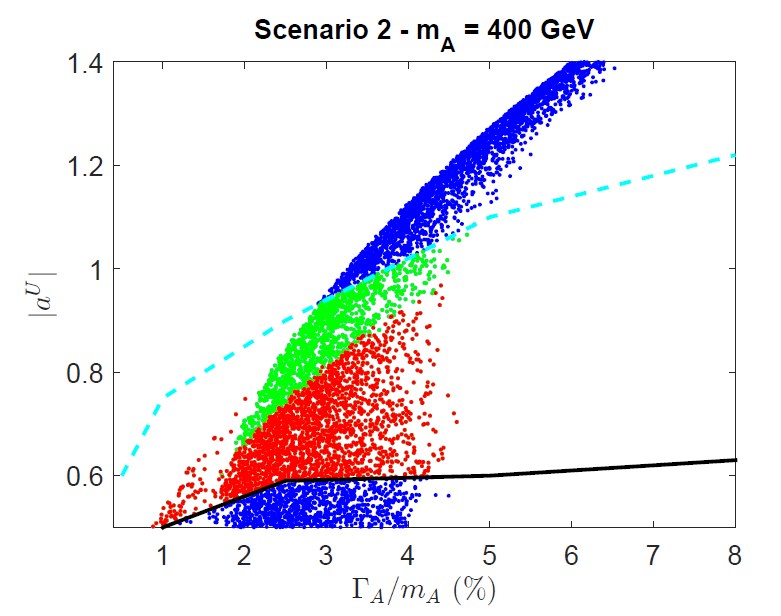}\\
 (a) & (b)
\end{tabular}
  \caption{\small Impact of the $\Delta M_{B_s}$ bound on the A2HDM parameter space. In panel (a) all points
  shown are
  from a general scan on $a^U$, $a^D$ and $m_{H^\pm}$ which pass the $\Delta M_{B_s}$ bound. In panel (b) we exhibit the
  ratio of the $A$ width to its mass (with $m_A=400$~GeV) as a function of the flavor-alignment parameter $|a^U|$ in Scenario 2.  The blue points are the result of
a scan over A2HDM parameters, subject to
  the theoretical and experimental constraints elucidated in Section~\ref{sec:scans} prior to imposing the
  $\Delta M_{B_s}$ bound.
  The dashed cyan (solid black) line correspond to the observed (expected) 95\% CL upper limit on the cross section for $gg\to A\to t\bar{t}$ reported by the CMS Collaboration in Ref.~\cite{CMS:2019pzc},
  translated into an upper limit for $|a^U|$ as a function of $\Gamma_A/m_A$.
  As for the remaining scan points that lie between the dashed cyan and solid black curve,
  the green points are eliminated
  after imposing the $\Delta M_{B_s}$ constraint.
The surviving red scan points constitute the proposed signal of Scenario~2.
  \label{fig:aU_mch}}
\end{figure}

Imposing the upper limit  to the contribution to $|\Delta M_{B_s}|$ of new physics beyond the Standard Model on the A2HDM parameter space when scanned over the values exhibited in Table~\ref{parms}, we have found no constraint on the flavor-alignment parameters $a^D$ and $a^E$ [of course, $a^E$ does not appear in \eq{dms}].  However, we do find an upper bound for $|a^U|$ shown in panel (a) of Fig.~\ref{fig:aU_mch}.   Note that the excluded region of the A2HDM parameter space due to the constraint on $\Delta M_{B_s}$ roughly coincides with the corresponding excluded region of the Type-I 2HDM
 in the $m_{H^\pm}$ vs.~$\tan\beta$ plane exhibited in Fig.~8 of Ref.~\cite{Arbey:2017gmh} after identifying $\tan\beta = 1/|a^U|$.   This result is not surprising in light of the small numerical contribution from terms that depend on $a^D$.

In panel (b) of
Fig.~\ref{fig:aU_mch}, we show the result of Fig.~\ref{fig:scen2_gamA} prior to imposing the constraint from $\Delta M_{B_s}$.   The red, green and blue points are all consistent with the $b\to s\gamma$ constraint.  Imposing the experimental constraint based on the 95\% CL upper limit on the cross section for $gg\to A\to t\bar{t}$ reported by the CMS Collaboration in Ref.~\cite{CMS:2019pzc}, we can eliminate the scan points that lie above the dashed cyan line.
Finally, the result of imposing the $\Delta M_{B_s}$ constraint is to remove the green points from the scan.  The remaining red scan points constitute the proposed signal of Scenario 2.

\end{appendices}

\end{document}